\documentclass[aps,pra,nofootinbib,twocolumn,longbibliography,citeautoscript,superscriptaddress]{revtex4-1}
\usepackage{amsmath, amsthm, amssymb,graphicx,color,bbm, dsfont, braket,bm}
\usepackage{mathtools} 
\usepackage{algorithm}
\usepackage{algpseudocode}
\usepackage{txfonts}
\usepackage{xcolor}
\usepackage{tikz}
\usepackage{float}
\usepackage{hyperref}
\hypersetup{
    colorlinks,
    filecolor=black,
    urlcolor=black
}

\algrenewcommand\algorithmicrequire{\textbf{Input:}}
\algrenewcommand\algorithmicensure{\textbf{Output:}}

\definecolor{sandiaBlue}{HTML}{00ACD5}

\newcommand{\stack}{%
  \mathop{%
    \vcenter{\hbox{%
      \tikz[baseline=-.2em,inner sep=0pt,outer sep=0pt,line width=.15mm]{
        \pgfresetboundingbox
        \path (0,0) rectangle (.4em,.4em);
        \draw (0,0) rectangle (.4em,.4em);
        \draw (0,.2em) -- (.4em,.2em);
      }%
    }}%
  }\displaylimits
}
\newcommand{\bigstack}{%
  \mathop{%
    \vcenter{\hbox{%
      \tikz[baseline=-.5em,inner sep=0pt,outer sep=0pt,line width=.37mm]{
        \pgfresetboundingbox
        \path (0,0) rectangle (1em,1em);
        \draw (0,0) rectangle (1em,1em);
        \draw (0,.5em) -- (1em,.5em);
      }%
    }}%
  }\displaylimits
}
\begin{document}

\newcommand{\gnF}{F}  
\newcommand{\nF}{\hat{F}}  
\newcommand{\nV}{\hat{V}}  
\newcommand{\vf}{\vec{f}}
\newcommand{\vfp}{\vec{f}'}
\newcommand{\vv}{\vec{v}}

\newcommand{\gss}{\Sigma}
\newcommand{\cM}{\mathcal{M}}
\newcommand{\cE}{\mathcal{E}} 
\newcommand{\cG}{\mathcal{G}} 
\newcommand{\cK}{\mathcal{K}} 
\newcommand{\cF}{\mathcal{F}} 
\newcommand{\cR}{\mathcal{R}} 
\newcommand{\cA}{\mathcal{A}} 

\newcommand{\pinv}[1]{\ensuremath{#1^{+}}}
\newcommand{\pvec}[1]{\vec{#1}\mkern2mu\vphantom{#1}}
\newcommand{\bvec}[1]{\bm{#1}}

\newcommand{\rrangle}{\rangle\!\rangle}                                     
\newcommand{\llangle}{\langle\!\langle}                                     
\newcommand{\rrpipe}{|}                                                     
\newcommand{\llpipe}{|}
\newcommand{\Id}{\mathds{1}}
\newcommand{\Tr}{\mathrm{Tr}}
\newcommand{\sket}[1]{\ensuremath{\llpipe#1\rrangle}}               
\newcommand{\sbra}[1]{\ensuremath{\llangle#1\rrpipe}}                
\newcommand{\sbraket}[2]{\ensuremath{\llangle#1|#2\rrangle}}         
\newcommand{\sketbra}[2]{\sket{#1}\!\sbra{#2}}                       
\newcommand{\ketbra}[2]{\ket{#1}\!\bra{#2}}   

\newcommand{\sbraopket}[3]{\ensuremath{\sbra{#1}#2\sket{#3}}}  
\newcommand{\proj}[1]{\ensuremath{|#1\rangle\langle#1|}}

\newcommand{\kcy}[1]{{\color{red}[#1]\textsubscript{kcy}}}

\newcommand{\jgm}[1]{{\color{red}[#1]\textsubscript{jgm}}}
\newtheorem{definition}{Definition}
\newtheorem{assumption}{Assumption}
\newtheorem{lemma}{Lemma}
\newtheorem{corollary}{Corollary}
\newtheorem{algo}{Algorithm}
\newtheoremstyle{breakpropnoparen}%
  {}{}%
  {\itshape}%
  {}%
  {\bfseries}%
  {.}%
  {\newline}%
  {\thmname{#1}\thmnumber{ #2}\thmnote{ --- #3}}

\theoremstyle{breakpropnoparen}
\newtheorem{proposition}{Proposition}

\title{A gauge-invariant theory of small Markovian errors in quantum gate sets}
\begin{abstract}
Noisy logic operations on a quantum computational register -- e.g., one or more qubits -- can be described by transfer matrices (a.k.a. CPTP maps or superoperators) that act linearly on the density matrix representing the register's quantum state.  Collectively, these operations form a \textit{gate set}.  Gate sets have a gauge freedom; many gate sets that appear different actually predict the same experimental outcomes.  A property of a gate set can be observable (and thus physically relevant) only if it is gauge-invariant.  Unfortunately, no good gauge-invariant parameterizations of gate sets are known.  We introduce the next best thing, a \textit{perturbative} gauge-invariant parameterization of small Markovian errors in gate sets. We construct vector spaces of properties that are first-order gauge-invariant (FOGI).  We show how to construct and understand FOGI properties, how to use them as coordinates to parameterize gate sets without gauge freedom, and how to extract approximately gauge-invariant error metrics.
\end{abstract}
\author{Juan Gonzalez de Mendoza}
\affiliation{Quantum Performance Laboratory, Sandia National Laboratories}
\affiliation{Department of Physics and Astronomy, Quantum New Mexico Institute,
University of New Mexico}
\author{Corey Ostrove}
\affiliation{Quantum Performance Laboratory, Sandia National Laboratories}

\author{Timothy Proctor}
\affiliation{Quantum Performance Laboratory, Sandia National Laboratories}

\author{Kevin Young}
\affiliation{Quantum Performance Laboratory, Sandia National Laboratories}
\author{Erik Nielsen}
\affiliation{IonQ}

\author{Robin Blume-Kohout}
\affiliation{Quantum Performance Laboratory, Sandia National Laboratories}

\date{\today}
\maketitle

\tableofcontents

\section{Introduction} \label{Introduction}

Quantum computing processors execute programs, described as a sequence of logic operations or gates, chosen by the user or programmer from a discrete set of available operations.  A \emph{gate set} \cite{Nielsen2021-nu, Hashim2025-rz} associates each logic operation with a dynamical map that describes how its execution transforms the quantum data register.  Gate sets are models that describe the execution of programs, including the effect of noise and errors in gates and SPAM (state preparation and measurement) operations.  Gate sets are central to the theory and practice of \emph{quantum characterization, verification, and validation} (QCVV) protocols like gate set tomography (GST) or randomized benchmarking (RB) \cite{Hashim2025-rz, Nielsen2021-nu, QCVV, RB_emerson, RB_Emerson2, RBKnill_2008}.

But gate sets have an inconvenient \textit{gauge freedom} \cite{Nielsen2021-nu, errorreconstructioncompiledcalibration}.  Two gate sets $\cG$ and $\cG'$ can appear very different, but actually describe exactly the same dynamics because they are \textit{gauge-equivalent}. This creates problems for estimating and interpreting gate sets  \cite{Nielsen2021-nu, Proctor2017-wc}.  Most notably, an estimated or predicted gate set $\cG$ that appears very different from an ideal \textit{target} gate set $\overline{\cG}$ is generally assumed to represent large errors (deviations from ideal behavior).  But if the discrepancy between $\cG$ and $\overline{\cG}$ corresponds mostly or entirely to a gauge transformation, then this is not true. The impacts of gauge freedom have been discussed in the context of GST \cite{Nielsen2021-nu,DiMatteo2020operationalgauge}, RB \cite{Proctor2017-wc, RBBeyond, RBConvolution, RBFramework}, Pauli noise learning and error mitigation \cite{chen_learnability_2023, senrui_efficient, chen2026disambiguatingpaulinoisequantum, BenchQI}, and cycle benchmarking / error reconstruction \cite{learnable_MCM, errorreconstructioncompiledcalibration, jordanmcmcycle}. Widely-used gate error metrics, such as process fidelity and diamond norm, vary with gauge and thus do not quantify observable phenomena \cite{Proctor2017-wc}. The use of \emph{gauge-invariant} error metrics and gate set models could avoid these problems, but few gauge-invariant metrics and no satisfactory gauge-invariant representations are known. Two notable exceptions are (1) the RB error rate $r$, which is gauge-invariant and can be measured experimentally or computed from transfer matrices \cite{Proctor2017-wc,  Carignan-Dugas2018-np, Lin2019-qx, RB_emerson, RB_Emerson2, RBKnill_2008}, and the (2) ``learnable'' degrees of freedom for Pauli noise channels \cite{chen_learnability_2023, senrui_efficient}. 

The most common way to mitigate gauge freedom is by \emph{gauge fixing} \cite{Nielsen2021-nu}. This means using some rule to select a representative gate set from each \emph{orbit} (equivalence class) defined by the gauge group. Two gate sets are on the same orbit if and only if they are related by a gauge transformation, and a gauge fixing procedure selects exactly one gate set in each orbit to represent the entire equivalence class.  There are many ways to fix a gauge. In theory and modeling, the representative is usually chosen to be easy to write down. Even when gauge fixing is not explicitly invoked in the construction of a protocol, the majority of commonly-used QCVV protocols do so implicitly. In gate set tomography, gauge fixing is also called ``gauge optimization,'' and a common choice is to select the gate set representation that is closest (by elementwise Frobenius distance) to an error-free target gate set \cite{Nielsen2021-nu}. 

But gauge fixing is fundamentally unsatisfactory for several reasons. It does not yield rigorously interpretable metrics, because it inherits the original gauge-variant metrics and parameters used to describe gauged gate sets. It also does not provide a gauge-free way to vary over gate sets, because each orbit's representative is defined implicitly rather than explicitly.

A more satisfying approach would be to construct a list of gauge-invariant properties (coordinates) whose values identify each gauge orbit uniquely. Ideally, these coordinates would also serve as error metrics, quantifying distance from the error-free ``target'' orbit containing $\overline{\cG}$ in an operationally relevant way.  Unfortunately, constructing such coordinates has proven hard.  Gauge orbits are curved and noncompact, and contain singular points analogous to the origin of polar coordinates. We'll later observe that gauge transformations act as similarity transforms on gates, and thus the eigenvalues of each logic gate $G_i$ are gauge-invariant. Gate eigenvalues serve as a standard metric in some protocols, like ACES \cite{ACES}, but gate spectra do not uniquely identify gauge orbits. A gate set can deviate from its target in observable ways that do not affect any gate's eigenvalues.  So eigenvalues' deviations from their target values can quantify \emph{some} errors, but not all. Construction of a fully gauge-invariant theory or representation of gate sets remains an open problem.  The gauge-invariant coordinates that have been constructed to date are either hard to interpret physically \cite{DiMatteo2020operationalgauge}, or only cover a subset of either operations or types of errors \cite{chen_learnability_2023, learnable_MCM}. 

In this paper, we introduce \textit{first-order gauge-invariant} (FOGI) properties of gate sets. FOGI properties quantify \textit{small} deviations from a target gate set, and a set of FOGI properties can quantify all small deviations from a target gate set. FOGI properties have been used as an analysis tool in previous papers \cite{JP,UNSWreference, Stemp_2024}, but without a complete derivation or explication. In this paper, we define a gate set's FOGI properties, construct a theory of them, and show how to construct FOGI properties systematically.  We explore and explain nontrivial consequences of gauge freedom, like the necessary existence of \textit{relational} FOGI errors that cannot be isolated to any single gate set operation. We also discuss how to algorithmically construct gauge-invariant bases for gate set errors, and we construct explicit FOGI bases using the elementary error generator representation \cite{Blume-Kohout2022-ln}.

\section{Gate sets and gauge} \label{GateSets}

\subsection{Gate sets}
A gate set is a collection of logic operations, including gates, state preparations, and measurements, that can be executed on a particular quantum processor. We consider gate sets that contain one initialization operation, one measurement operation, and one or more reversible logic operations.  In real-world use, these operations will be noisy and imperfect.  They transform the processor's quantum state, in various ways.  We describe that state by a density matrix $\rho$ on the processor's $d$-dimensional Hilbert space, where $d=2^r$ for an $r$-qubit\footnote{While the theory presented in this work applies to general qudit systems, we will focus on qubits for convenience.} processor.
A density matrix is also a vector $\sket{\rho}$ in the $d^2$-dimensional real \emph{Hilbert-Schmidt space} of Hermitian $d\times d$ matrices, and thus gates are described by linear \emph{superoperators} acting on that space. Specifically:
\begin{enumerate}
    \item The initialization operation, denoted by $\rho$, is represented by a $d^2 \times 1$ column matrix (vector) written as $\sket{\rho}$.
    \item The measurement operation, denoted by $M$, is represented by a $m \times d^2$ matrix, with $m=d$ in most cases, that maps a state into an $m$-element probability distribution over measurement outcomes according to Born's Rule.  The rows of $M$ are the effects $\sbra{E_i}$ of a POVM.
    \item Logic gates, denoted by $G_i$, are represented by $d^2\times d^2$ transfer matrices that map states to states. 
\end{enumerate}
A gate set therefore corresponds to a tuple,
\begin{equation}
\cG = \left(\sket{\rho},M,G_1,\ldots, G_N\right).
\end{equation}
\noindent We are especially interested in \emph{informationally complete} (IC) gate sets. A gate set is considered IC iff (1) the states resulting from all possible combinations of the available gates $G_i$ applied to $\sket{\rho}$ span the entire $d^2$-dimensional Hilbert-Schmidt space and (2) the measurement operations resulting from all possible combinations of the available gates $G_i$ applied to $M$ span the entire $d^2$-dimensional dual Hilbert-Schmidt space \cite{Nielsen2021-nu}. 

We usually represent each gate as a Pauli transfer matrix (PTM), with $\mathds{1}$ as the first element of the Pauli basis.  Physically valid quantum states $\rho$ must satisfy $\Tr(\rho)=1$, so that measurement probabilities add up to 1.  A PTM is \emph{trace preserving} (TP) iff its top row is $[1,0,0,\ldots]$.  Gate superoperators should also be \emph{completely positive} (CP) to ensure that all observable probabilities are non-negative, but we ignore this constraint throughout most of this document.

\begin{figure}[t!]
\centering
\includegraphics[width=8cm]{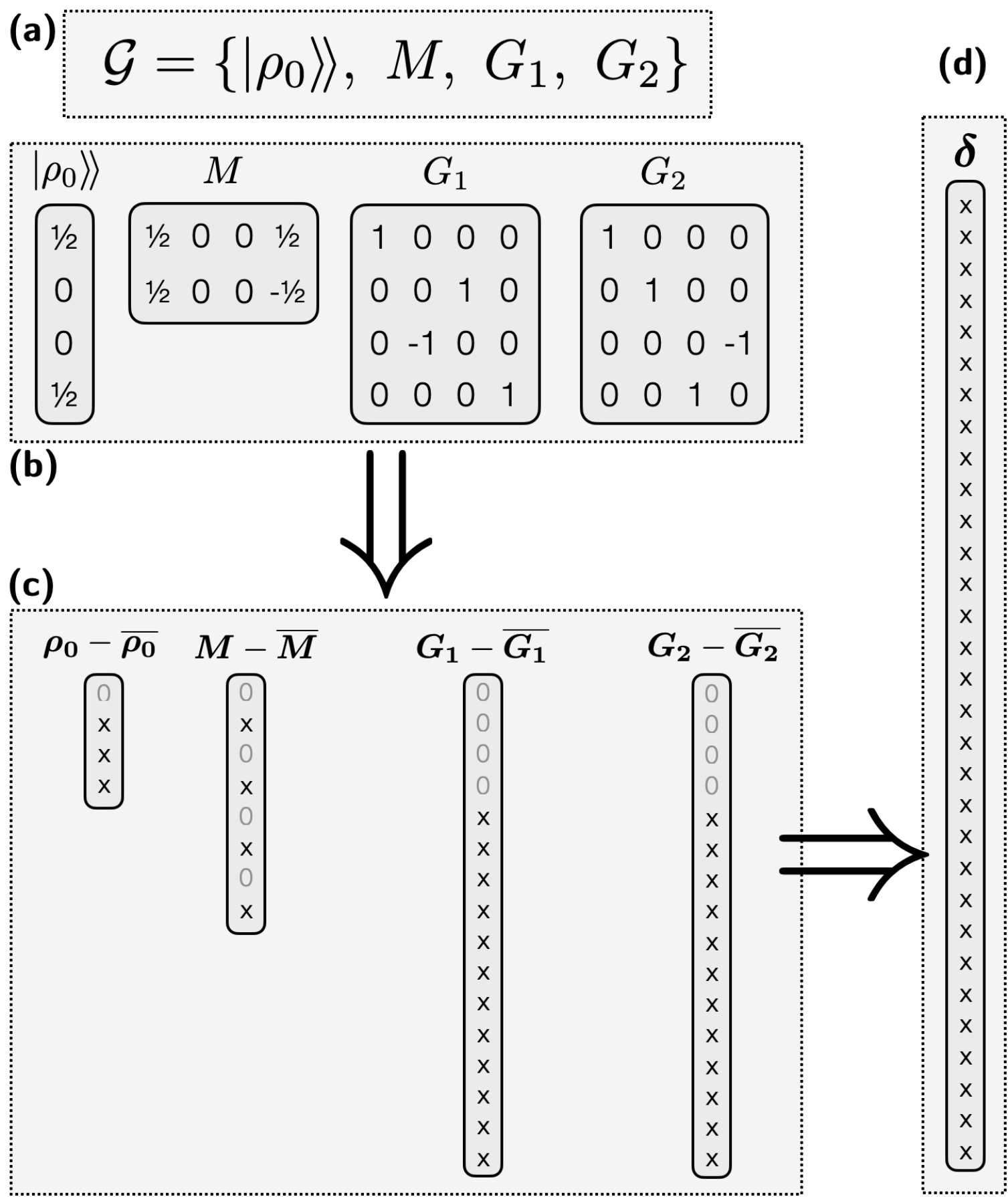}
\caption{We analyze gate sets $\cG = \left\{\sket{\rho},M,G_1,\ldots, G_N\right\}$ \textbf{(a)} containing operations that can be represented by Pauli transfer matrices \textbf{(b)}.  To analyze errors represented by small variations, we consider the difference $\delta = \cG - \overline{\cG}$ between the gate set and its target \textbf{(c)}, and stack the vectorized differences to form a single vector $\bvec{\delta}\in\Delta$ in \textit{error set space} \textbf{(d)}.  Zero entries in panel (c) represent unvarying (and thus discardable) matrix elements.  This figure illustrates a small 1-qubit gate set with $3+4+12+12=31$ degrees of freedom.}
\label{fig:gatesets}
\end{figure}

Each element $g$ of a gate set $\mathcal{G}$ (a state, gate, or measurement) can be described by a vector in a space of operators $\gss_g$.  We use bold type to denote such vectors $\bvec{g} \in \gss_g$. Stacking the vectors for $\rho$, $M$, and the $G_i$ together yields a vector $\bvec{\mathcal{G}}$ that describes the entire gate set in the \emph{gate set space} $\gss$ formed as the direct sum of all the operator spaces 
\begin{equation}
\gss = \gss_{\rho} \oplus \gss_{M} \oplus \gss_{G_1} \oplus \cdots \oplus \gss_{G_N}.
\end{equation}
We use the $\stack$ symbol to denote vertical stacking of both matrices and vectors, so, e.g.,
\begin{equation}
    A \stack B \equiv \left(\begin{array}{c}A \\ B \end{array}\right).
\end{equation}
Therefore,
\begin{equation}
   \mathrm{vec}(\cG)=\bvec{\mathcal{G}} = \bvec{\rho} \stack \bvec{M} \stack \bvec{G}_1 \stack \cdots \stack \bvec{G}_N .
\end{equation}

 Throughout this paper we identify several different vector spaces whose elements take a variety of forms such as tuples and matrices. All of which have their corresponding (bold) vector representation. As we later show, some results are greatly simplified when represented in vector form, while other results rely on exploiting the structure of the tuples or matrices. Thus, we freely move within the original and vectorized representations, and refer to both representations as part of the same space e.g. $\cG \in \Sigma$ and $\bvec{\cG} \in \Sigma$.
\subsection{Gauge}

The primary purpose of a gate set is to predict outcome probabilities of quantum circuits. A circuit comprises initialization ($\rho$), a sequence of $L$ gates ($G_i:\ i=1\ldots L$), and a terminating measurement ($M$).  Circuit outcome probabilities are given by a vector $\vec{P}$:
\begin{equation} \label{eq:Born}
\vec{P} = M G_L \cdots G_1 \sket{\rho}.
\end{equation}
Every directly observable probability can be written in this form. This fact is responsible for the gauge freedom we are concerned with.

Gauge freedom is a property of parameterized theories.  A parameterized theory has a gauge if (and only if) there exist multiple distinct points (i.e., values of the theory parameters) that predict \emph{exactly} the same observable consequences for all possible experiments.  Gate set space is a parameterized theory -- each gate set is a theory about how the noisy quantum computer works.  The gauge freedom for TP gate set space corresponds to
\begin{eqnarray}\label{gaugetransformation}
  G_i &\rightarrow& T G_i T^{-1} \nonumber\\
  \sket{\rho} &\rightarrow& T \sket{\rho} \label{eq:gaugetransform1}\\
  M &\rightarrow& M T^{-1}, \nonumber
\end{eqnarray}
where $T$ is any invertible TP matrix\footnote{For non-informationally-complete gate sets there exist additional \emph{effective} gauge degrees of freedom not captured by Eq. (\ref{eq:gaugetransform1}); see Appendix \ref{app:unobservableFOGIs}.}. Direct substitution into Eq. \eqref{eq:Born} reveals that this transformation leaves every probability unchanged:

\begin{align}
     \vec{P} & \to MT^{-1} TG_LT^{-1} \cdots TG_1T^{-1} T\sket{\rho} \\&=M G_L \cdots G_1 \sket{\rho}= \vec{P}.
\end{align}

\subsubsection{Small gauge transformations}
\label{sec:smallgauge}
A TP gauge transformation is ``small'' if it is close to the identity
\begin{equation}
    \|T-\Id\|=O(\epsilon),
\end{equation}where $\epsilon \ll 1$. A small gauge transformation can be written in exponential form as 

\begin{equation}
T = e^{\epsilon K},
\end{equation}

\noindent where $K$ is an $\mathcal{O}(1)$ matrix that we call a \textit{gauge generator}. Expanding the exponential to leading order gives a linear approximation that we will use extensively,
\begin{equation}\label{eq:gaugegenerator}
T = \mathds{1} + \epsilon K + O\left(\epsilon^2\right).
\end{equation}
Gauge generators can be represented in the Pauli basis as real matrices whose top row is constrained to be zero by trace preservation (TP). Thus, gauge generators can be viewed as elements of a vector space $\mathbb{R}^{d^2(d^2-1)}$ that we denote $\cK$.  We use $K\in\cK$ interchangeably with the vector $\bvec{k}\in\cK$, depending on whether we are treating the gauge generator as a Pauli transfer matrix or an abstract vector.

\subsection{Error set space}
\label{sec:errorsspace}
Typically, we are interested in how a gate set deviates from the intended  (error-free) operations described by a target gate set $\overline{\cG} = \left(\overline{\rho},\overline{M},\overline{G}_1\ldots \overline{G}_N\right)$. 
The \textit{error} in $\cG$ is its deviation from $\overline{\cG}$, 
\begin{align}
    \delta &= \cG - \overline{\cG}\label{eq:error}\\
     &= \left(\sket{\rho}-\sket{\overline{\rho}} , M-\overline{M} , G_1-\overline{G}_1  , \ldots  ,G_N-\overline{G}_N \right)\\
     &=\left(\sket{\delta_\rho}, \delta_M, \delta_{G_1}, \ldots, \delta_{G_N} \right)\label{eq:errortuple},
\end{align}
where  $\delta_g = g - \overline{g} \in \Delta_g$ is the error in an individual operation $g \in \mathcal{G}$. Its vectorized form is
\begin{align}
\bvec{\delta} &= \left(\bvec{\rho}-\overline{\bvec{\rho}}\right) \stack \left(\bvec{M}-\overline{\bvec{M}}\right) \stack \left(\bvec{G}_1-\overline{\bvec{G}}_1 \right) \stack \ldots \stack \left(\bvec{G}_N-\overline{\bvec{G}}_N \right)\notag\\
&=\bvec{\delta_\rho} \stack \bvec{\delta_M} \stack \bvec{\delta_{G_1} \stack}\ldots \bvec{\delta_{G_N}}.\label{eq:linearform}
\end{align}

Errors uniquely identify gate sets as $\cG(\delta)=\overline{\cG} + \delta$. They form a vector space $\Delta$ that we call \textit{error set space}, with the same dimensions and block structure as $\Sigma$:

\begin{equation}
\Delta = \Delta_{\rho} \oplus \Delta_{M} \oplus \Delta_{G_1} \oplus \cdots \oplus \Delta_{G_N}.
\end{equation}

\noindent In this paper, we are only concerned with \emph{small} errors.

\begin{definition}\label{def:neighborhood}
A gate set $\mathcal{G}$ is within an $\epsilon$-neighborhood of a target gate set $\overline{\cG}$ if $\| \delta\| =\|\cG-\overline{\cG}\| \leq \epsilon$, where $\| \cdot \|$ is some fixed norm on $\Delta$.
\end{definition} \noindent

\begin{assumption}\label{ass:epsilon}
    Throughout the remainder of this paper, all gate sets are assumed to be within an $\epsilon$-neighborhood of their target gate set, for some $\epsilon\ll 1$ such that $\epsilon^2$ is negligible.
\end{assumption} 

\subsection{Gate set properties}
\label{sec:properties}
A \emph{property} of a gate set $\cG$ is a scalar-valued function $f(\cG)$.  Errors describe gate sets relative to a known target gate set, and properties \emph{quantify} errors. Since we have assumed that $\cG$ is close to $\overline{\cG}$, any property $f(\cG)$ can be accurately approximated by a \textit{linear} functional defined by $f$'s first-order Taylor expansion around $\overline{\cG}$,
\begin{equation}
f(\cG) = f\left(\overline{\cG}\right) + \varphi(\delta) + O\left(\epsilon^2\right),\notag
\end{equation}
where $\varphi$ is a linear functional on elements of $\Delta$, i.e. an element of the dual space $\Delta^*$. The property's value on the target gate set, $f\left(\overline{\cG}\right)$, is a constant that does not depend on $\delta$, so we will drop it. Hereafter, ``property'' will mean a linear functional $\varphi: \Delta \to \mathbb{R}$.

Any such property $\varphi$ can be represented, like the error vector $\delta$ in Eq. (\ref{eq:errortuple}), as a tuple over individual gate set elements,

\begin{equation}\label{eq:tuple}
    \varphi = (\varphi_\rho, \varphi_M, \varphi_{G_1}, \ldots, \varphi_{G_N}),
\end{equation}

\noindent where each element $\varphi_g$ is a property intrinsic to gate $g$, and thus

\begin{equation}
    \varphi(\delta) = \sum_{g\in\cG}\varphi_g(\delta_g).\label{eq:propsum}
\end{equation}

\noindent It is also useful to vectorize a property $\varphi$, following Eq. (\ref{eq:linearform}), as a covector (row vector) $\bvec{\phi}$. Any $\bvec{\phi}=\mathrm{vec}(\varphi)$ can be written as a stack of properties $\bvec{\phi}_g$ that are each intrinsic to a single operation $g$. Thus Eq. (\ref{eq:tuple}) becomes
\begin{equation}\label{eq:phistack}
    \bvec{\phi} = \begin{pmatrix}
        \bvec{\phi}_{g_1}&
        \bvec{\phi}_{g_2}&
        \cdots&
        \bvec{\phi}_{g_N}
    \end{pmatrix}=\left(\bigstack_{g\in\cG} \bvec{\phi}_g^\top\right)^\top,
\end{equation}

and

\begin{equation}
\varphi(\delta)=\bvec{\phi}\bvec{\delta}=\sum_{g\in\cG}\bvec{\phi}_g\bvec{\delta}_g.
\end{equation}

\subsection{Inner products on $\Delta$ and $\cK$}\label{sec:innerprod}
When we represent abstract error vectors and properties ($\delta$, $\varphi$) as concrete  column and row vectors ($\bvec{\delta}$, $\bvec{\phi}$), we do so with caution because there is no natural, satisfying inner product on $\Delta$ or $\cK$.  Both spaces contain several qualitatively different kinds of directions that have different ``units'' -- i.e., represent fundamentally different variations in the gate set.  For example, coherent and stochastic errors are not directly comparable. Neither are errors associated with gates and SPAM operations (respectively). An inner product implicitly makes these incomparable dimensions comparable, by combining them all into a quadratic form with specific weights. So we view error set space as \textit{not} equipped with a canonical inner product.  Our analysis should properly proceed \textit{without} using an inner product, and therefore without asserting a canonical isomorphism between $\Delta$ and $\Delta^*$, or between $\cK$ and $\cK^*$.

Operating in a completely coordinate-free formalism is possible (as shown in Sec.~\ref{canonical}), but it is inconvenient. For computations, it is useful to choose bases and represent errors, gauge generators, and properties by concrete vectors and matrices (e.g., as in Fig.~\ref{fig:spaces}). Furthermore, several constructions in this paper \textit{require} specifying a basis and an inner product. For instance, it will be helpful to iteratively partition $\Delta$ into subspaces. A subspace of $\Delta$ has many complements, and singling out an orthogonal complement requires an inner product. So do defining pseudoinverses, normalizing the basis vectors, and identifying each property in the dual space $\Delta^*$ with an error vector in the primal space $\Delta$. 

Choosing a basis and choosing an inner product are not entirely independent. Once a basis is fixed, it is tempting to compute inner products as dot products of coordinate vectors $\bvec{v}^\top\bvec{w}$. This action defines the unique inner product for which the chosen basis is orthonormal. But if the basis is changed without changing the formula $\bvec{v}^\top\bvec{w}$, the inner product on the underlying abstract space has changed as well. For example, the dot product of PTM-element coordinates (Fig.~\ref{fig:gatesets}) and the dot product of elementary error generator rates (Sec.~\ref{EEG}) define different inner products on $\Delta$, because the elementary error generators are not orthonormal under the dot product (which is why extracting their rates requires dual generators). Two scientists making different, individually sensible choices of basis will therefore disagree about which errors are orthogonal and how big they are. 

Therefore, for convenience and to enable numerical calculations, we sometimes equip $\Delta$ and $\cK$ with a specific inner product. We use this inner product only when necessary, and recognize that its consequences (e.g. orthogonality of subspaces) are not fundamental.  

\begin{assumption}\label{ass:innerprod}
     Whenever an inner product on $\Delta$ and $\cK$ is required, we use the inner product induced by the basis in use at that point (the unique inner product under which that basis is orthonormal) computed as the dot product of coordinate vectors:

     \begin{equation}
         \langle \bvec{u},\bvec{v} \rangle=\bvec{u}^\top\bvec{v}
         \quad\text{for } \bvec{u},\bvec{v}\in\Delta \text{ or } \bvec{u},\bvec{v}\in\cK.
 \end{equation}
\end{assumption}
This inner product defines a correspondence (the Riesz map) between properties and errors, which in coordinates is simply the transpose. It associates every property $\bvec{\phi}\in\Delta^*$ with a corresponding unique error vector $\bvec{\phi}^\top = \bvec{\delta}_{\bvec{\phi}} \in \Delta$ such that
\begin{equation}
    \varphi(\delta)=\bvec{\delta}_{\bvec{\phi}}^\top\bvec{\delta}.
\end{equation} 
The same reasoning extends to gauge generator space, relating $\cK^*$ to $\cK$.  This correspondence is well defined, but it is not canonical. It depends on the chosen basis through the inner product induced by the dot product.
The basis similarly induces an inner product on the dual space $\Delta^*$. When we refer to two properties as orthogonal, it is with respect to this induced inner product. 

We could ask if any particular inner product on the abstract space $\Delta$ or $\cK$ is preferable to any other. One reasonable criterion is that it should not depend on unitary changes of basis. This essentially singles out the Hilbert-Schmidt (trace) inner product, up to arbitrary weights on the various sectors of $\Delta$. The trace inner product coincides with the dot product of coordinate vectors in \textit{any} normalized Pauli basis, independent of the unitary reference frame. However, we can identify no canonical choice for the relative weighting of qualitatively different error sectors (eg., coherent versus stochastic rates, or gate versus SPAM errors). Fortunately, most of our results are entirely agnostic to which inner product is used. The FOGI space and the intrinsic FOGI subspaces admit fully canonical definitions, as  Sec.~\ref{canonical} shows without the assumption of an inner product at all. What will inherit the choice of basis (and thus inner product) is the orthogonal decomposition of FOGI space into purely relational subspaces in Sec.~\ref{Taxonomy}. 

\subsection{FOGI Properties}

Intuitively, a first-order gauge-invariant (FOGI) property is one whose derivative with respect to \textit{any} small gauge transformation vanishes.  A FOGI property's value may not be strictly gauge-invariant, but it doesn't vary to leading order.  This is formalized as follows:
\begin{definition} \label{def:FOGI}
A property $\varphi(\delta)$ is FOGI if and only if, for any two gate sets described by errors $\delta$ and $\delta'$ that are (1) related by a gauge transformation and (2) both within an $\epsilon$-neighborhood of the target gate set, so that $\|\delta\| \leq \epsilon$ and $\|\delta'\| \leq \epsilon$, it holds that
\begin{equation}
\varphi(\delta) - \varphi(\delta') = O\left(\epsilon^2\right).
\end{equation}
\end{definition}

Definition \ref{def:FOGI} allows us to associate FOGI properties with a vector space, and to characterize it.  If two informationally complete gate sets are both close to the target, then they are also close to each other. If they are also gauge-equivalent, then they are related by a \textit{small} gauge transformation $T = e^{\epsilon K} = \Id + \epsilon K + O\left(\epsilon^2\right)$, and the $O\left(\epsilon^2\right)$ term can be neglected (See Appendix \ref{app:smallgauge}). Applying this linearized gauge transformation to a gate set $\cG$ as described in Eq. (\ref{gaugetransformation}) yields
\begin{align}
    \cG \to \cG' &=
     \left( e^{\epsilon K}\sket{\rho}, Me^{-\epsilon K}, e^{\epsilon K} G_1 e^{-\epsilon K}, \dots, e^{\epsilon K} G_N e^{-\epsilon K}\right) \\
    &= \cG + \epsilon\left(  K \sket{\rho}  , -MK, [K,G_1], \dots, [K,G_N]\right)  + O\left(\epsilon^2\right),\label{eq:linearaction}
\end{align}

\noindent where, here and throughout, adding tuples denotes elementwise addition of their elements. Therefore, because $\delta = \cG - \overline{\cG}$, we have that 
\begin{align}
    \delta' - \delta 
    &= \epsilon\left(  K\sket{\rho}, -MK, [K,G_1], \dots, [K,G_N]\right)  + O\left(\epsilon^2\right). \label{eq:delta-dif-1}
\end{align}
Since all errors are assumed to be small (Assumption \ref{ass:epsilon}), $\cG = \overline{\cG} + O(\epsilon)$.  Substituting this into Eq.~\eqref{eq:delta-dif-1} yields
\begin{equation} \label{eq:affineshift}
    \delta' - \delta = \epsilon \left(K\sket{\overline{\rho}}, - \overline{M}K, [K,\overline{G}_1], \dots, [K,\overline{G}_N] \right) + O\left(\epsilon^2\right).
\end{equation}

\noindent To first order, the right hand side of Eq.~\eqref{eq:affineshift} is linear in $K$ and independent of the error $\delta$.  Thus, small gauge transformations produce an \emph{affine shift} in $\delta$ that is linear in $K$. We call this linear relationship the \emph{gauge action map}, $\mathcal{A}_\cG:\mathcal{K}\rightarrow \Delta$:
\begin{align}
\delta' &= \delta + \epsilon \mathcal{A}_{\cG}(K) + O\left(\epsilon^2\right)\\
\Rightarrow &\delta' - \delta = \epsilon \mathcal{A}_{\cG}(K) + O\left(\epsilon^2\right).\label{eq:gaugeaction}
\end{align}  
$\mathcal{A}_\cG$ captures the linear relationship between a small gauge transformation and the shift that it induces in error set space. It has the same operational block structure as $\Delta$ and can be written as

\begin{equation}\label{eq:Atuple}
    \mathcal{A}_{\cG} = \left(\mathcal{A}_{\rho} , \mathcal{A}_{M}  , \mathcal{A}_{G_1} , \cdots , \mathcal{A}_{G_N} \right),
\end{equation}

\noindent where the subscript on an $\mathcal{A}$ map indicates the subspace on which it acts (e.g., $\mathcal{A}_{\mathcal{G}}$ acts on the full error set space), not a dependency. The explicit action of each $\mathcal{A}_g$ is given, following Eq. \eqref{eq:affineshift}, by
\begin{equation}\label{eq:ALinMap}
    \mathcal{A}_g(K)=\begin{cases}
        K\sket{\overline{\rho}}& \text{if } g=\rho,\\
        -\overline{M}K & \text{if } g=M,\\
        [K, \overline{G}_i]& \text{if } g=G_i.
    \end{cases}
\end{equation}

Using $\mathcal{A}_\cG$, we can identify the subspace of FOGI properties within $\Delta^*$.  By construction, $\delta$ and $\delta'$ are within an $\epsilon$-neighborhood of the target gate set. Therefore, from Definition \ref{def:FOGI}, a property $\varphi$ is FOGI if and only if $\varphi(\delta - \delta') = O\left(\epsilon^2\right)$ for any two error sets $\delta$ and $\delta'$ related by a gauge transformation $T = e^{\epsilon K}$. If we substitute $\delta'-\delta$ from Eq. \eqref{eq:gaugeaction}, we see that this holds if and only if the $O(\epsilon)$ term vanishes for all $K\in\cK$. It follows immediately that:
\begin{lemma}\label{lem:FOGI}
A property $\varphi(\delta)$ is FOGI if and only if $\varphi\left(\mathcal{A}_\cG(K)\right) = 0$, for all gauge generators $K\in\cK$.
\end{lemma}

The abstract map $\mathcal{A}_\cG$ can be represented by a matrix $A_\cG$ that maps vectorized gauge generators $\bvec{k} \in \cK$ to vectorized errors as defined in Eq.~\ref{eq:linearform}, and then Eq.~\ref{eq:gaugeaction} can be rewritten as
\begin{equation}\label{eq:linearA}
\bvec{\delta}' - \bvec{\delta} =  \epsilon A_{\cG}\bvec{k} + O\left(\epsilon^2\right).
\end{equation}

\noindent We now write $A_\mathcal{G}$ as a stack of gate-specific matrices, following Eq.~\ref{eq:Atuple}:
\begin{equation}\label{eq:Astack}
A_{\cG} = A_{\rho} \stack A_{M}  \stack A_{G_1} \stack \cdots \stack A_{G_N},
\end{equation}
where $A_g:\cK \to \Delta_g$. By Lemma \ref{lem:FOGI}, a property $\bvec{\phi}\cdot\bvec{\delta}$ is FOGI if and only if 
\begin{equation}\label{eq:matrixFOGIcond}
\bvec{\phi} \cdot A_\cG\bvec{k} = 0, \forall \bvec{k}\in\cK.
\end{equation}

 The range of $A_\cG$, which we denote $\Gamma_\cG$, has dimension at most $d^2(d^2-1)$, so it cannot span $\Delta$.  Therefore, there must be a subspace of $\Delta^*$ that annihilates all errors in $\Gamma_\cG$, i.e. a space of properties such that $\bvec{\phi}\cdot\bvec{\delta}=0$ for all $\bvec{\delta}\in\Gamma_\cG$, and invariant to leading order in $\epsilon$ under the application of any small gauge transformation.  This subspace $\cF_\cG\subseteq \Delta^*$ corresponds to FOGI properties. 

From Eq. \ref{eq:matrixFOGIcond}, all vectors $\bvec{\phi}\in\cF_\cG$ must satisfy $\bvec{\phi} A_\cG=0$ or equivalently $A_\cG^\top\bvec{\phi}^\top=0$. Expressed formally:
\begin{center}
\fbox{%
\begin{minipage}{0.95\linewidth}
\begin{lemma}\label{lem:kernel}
The FOGI space of a gate set is isomorphic to the kernel of the transpose of the gauge action matrix:
\begin{equation}
    \cF_\cG\cong\mathrm{kernel}\left(A_\cG^\top\right).
\end{equation}
\end{lemma}
\end{minipage}%
}
\end{center}

Explicitly constructing a matrix $A_{\cG}$ and obtaining FOGI properties from it requires choosing a basis. In Sec. \ref{sec:Detailed} we provide an example construction of gauge action matrices in the elementary error generator basis \cite{Blume-Kohout2022-ln}, but for now we analyze their properties abstractly.

\section{Taxonomy and Construction of FOGI Properties} \label{Taxonomy}

\begin{figure}[t!]
\centering
\includegraphics[width=8cm]{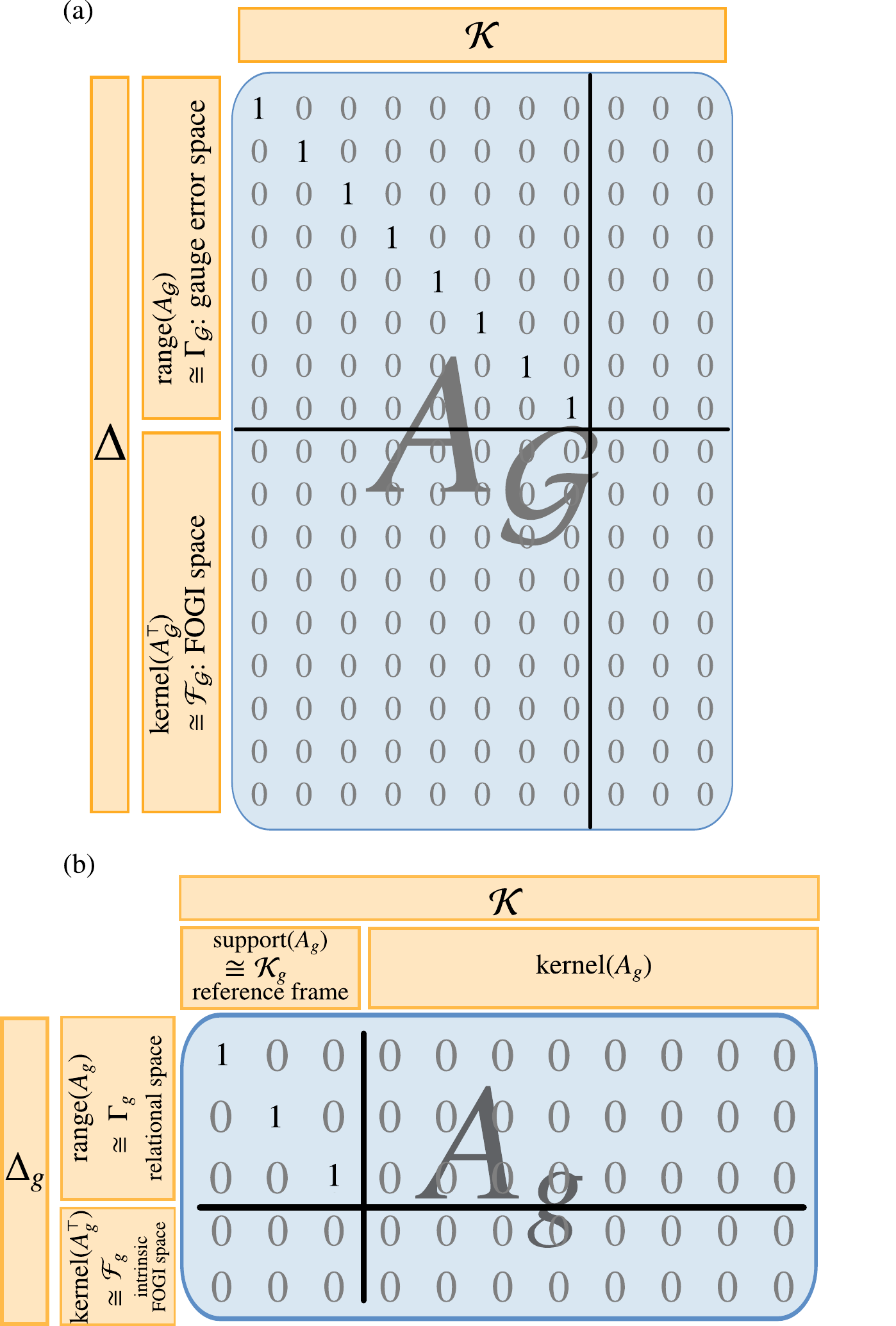}
    \caption{Small gauge transformations, described by $\bvec{k}\in\cK$, apply affine shifts to error set vectors $\bvec{\delta}\in\Delta$. \textbf{(a)} Gauge-induced shifts in a gate set $\cG$ are described by a gauge action matrix $A_\cG: \cK \to \Delta$.  As seen in the analysis that yields Lemma \ref{lem:FOGI}, the row space of $A_\cG$  is isomorphic to two fundamental subspaces. $A_\cG$ is shown in reduced row echelon form (RREF) to illustrate $\cF_\cG \cong \mathrm{kernel}\left(A_\cG^\top\right)$, the gate set's \textit{FOGI space}, and $\Gamma_\cG \cong \operatorname{range}(A_\cG)$, the \textit{gauge error space}. The gauge action matrix for the entire gate set $A_\cG$, in its original non-RREF form, is composed of stacked operation-specific gauge action matrices $A_{g}$. \textbf{(b)} For any $A_g$, shown here in its RREF, we focus on three fundamental subspaces: $\cF_g \cong \mathrm{kernel}\left(A_g^\top\right)$, the \textit{intrinsic FOGI space} of gate $g$; $\Gamma_g \cong \mathrm{range}(A_g)$, the \emph{relational space} of gate $g$; and $\cK_g \cong \mathrm{support}(A_g)$, the \emph{reference frame space} of gate $g$.}
\label{fig:spaces}
\end{figure}

To identify and make use of specific physically-motivated FOGI properties, we need to break down the space of \textit{all} FOGI properties $\cF_\cG \cong \mathrm{kernel}\left(A_{\mathcal{G}}^\top\right)$ into smaller subspaces of properties with useful characteristics.  We can do much of this work in an abstract coordinate-free way, without imposing \textit{a priori} structure on $\cF_\cG$ or $\Delta^*$, but we can't explicitly partition $\Delta^*$ into orthogonal subspaces without relying on an (arbitrary) inner product as discussed in Sec. \ref{sec:innerprod}.  We state clearly when and why we do this.  Our analysis leverages the decomposition $A_\cG = A_{g_1} \stack A_{g_2}\ldots$ (Eq. (\ref{eq:Astack})). We also make use of four fundamental subspaces (Fig.~\ref{fig:spaces}) of a gauge action matrix $A$:
\begin{enumerate}
\item Its \emph{kernel}, or null space, which is the space of all $\bvec{k}$ annihilated by $A$---i.e., all $\bvec{k}$ such that $A\bvec{k}=0$.
\item Its \emph{corange}, or support, which is the span of its row vectors, the orthogonal complement of its kernel, and the space of all $\bvec{k}$ such that $\bvec{k} = A^\top\bvec{\delta}$ for some $\bvec{\delta}$.
\item Its \emph{range}, or image, which is the span of its column vectors, and the space of all $\bvec{\delta}$ such that $\bvec{\delta} = A\bvec{k}$ for some $\bvec{k}$. 
\item Its \emph{cokernel}, or the kernel of its transpose, which is the orthogonal complement of its range, and the kernel of $A^\top$.
\end{enumerate}

Each matrix in Eq. (\ref{eq:Astack}) can be explicitly constructed by choosing a basis for $\Delta$ and $\cK$. Then, each basis element can be plugged into the linear maps defined in Eq. (\ref{eq:ALinMap}) to obtain matrix elements. For many of the algorithms we present in this section we will work under the assumption that all the matrices $A_g$ have already been constructed.

\subsection{Intrinsic FOGI properties} \label{sec:intrinsicFOGI}

A property $\bvec{\phi}$ can depend on the variations in a single operation $g \in \cG$, e.g.~ as
\begin{equation}
    \bvec{\phi}=\left(\bvec{\phi}_{g_1},0,\cdots,0\right),\notag
\end{equation}
or it can depend on variations in multiple operations within $\cG$. We call a property that depends only on variations in one gate $g$ an \textit{intrinsic property of $g$}. Intrinsic FOGI properties correspond to the kernels of submatrices $A_g^\top$ of $A_\cG^\top$ as defined in Eq. (\ref{eq:Astack}).

\begin{definition} \label{def:intrinsicFOGI}
 A property $ \bvec{\phi}$ is both FOGI and intrinsic to gate $g$ if and only if $\bvec{\phi}^\top \in  \mathrm{kernel}\left(A_g^\top\right)\cong \cF_g$. 
\end{definition}

Algorithm \ref{algo:intrinsic} below can be used to find all the intrinsic FOGI quantities within $\mathcal{F}_\mathcal{G}$. 

\begin{algorithm}[H]
  \caption{Construct a basis for the intrinsic FOGI space $\cF_g$}\label{algo:intrinsic}
  \begin{algorithmic}[1]
    \State Choose a basis $\{\bvec{b}_i\}$ for $\Delta_g$
    \State Construct $A_g^\top$ in that basis
    \State Compute the kernel of $A_g^\top$
\end{algorithmic}
\end{algorithm}

\subsection{Relational FOGI properties}\label{relationalFOGI}

There exist FOGI properties that are \textit{not} intrinsic to a single gate $g$.  For example, if $\bvec{\phi}_1\in\cF_{G_1}$ and $\bvec{\phi}_2\in\cF_{G_2}$, then the linear combination $\bvec{\phi}_1 + \bvec{\phi}_2$ is supported on both, and thus not intrinsic to any gate.  

But there are also FOGI properties that \textit{cannot} be written as linear combinations of intrinsic properties.  The span of all of a gate set's intrinsic properties, 
\begin{equation}
    \cF_{\mathrm{intrinsic}} \equiv \bigoplus_{g\in\cG}{\cF_g},
\end{equation}
is generally a strict subspace of $\cF_\cG$.  This implies the existence of \textit{relational} FOGI properties that quantify errors in the relationship between two or more operations.  Figure \ref{fig:pairwise} illustrates such an error.  We refer to any property that does not lie in $\cF_{\mathrm{intrinsic}}$, and therefore cannot be reduced to a sum of intrinsic properties, as \textit{irreducibly relational}.

Adding an intrinsic property to an irreducibly relational property yields another irreducibly relational property that is different from the first, but not in an interesting way.  They have the same relational content.  We wish to isolate it.  We can do so formally by defining an equivalence relation -- $\bvec{\phi} \sim \bvec{\phi}'$ iff $\bvec{\phi} - \bvec{\phi}' \in \cF_{\mathrm{intrinsic}}$ -- and using it to construct equivalence classes of properties. These equivalence classes define the quotient space $\cF_\cG / \cF_{\mathrm{intrinsic}}$.  But since a relational property is an element of $\cF_\cG$, and the quotient space defined above is not a subspace of $\cF_\cG$, we identify the subspace of purely relational properties with a \textit{complement} of $\cF_{\mathrm{intrinsic}}$ within $\cF_\cG$, which is a subspace of $\cF_\cG$ that is isomorphic to the quotient space defined above.  This complement is not unique, so to define a unique subspace of relational properties, we resort to Assumption \ref{ass:innerprod} and define it as the \textit{orthogonal} complement of $\cF_{\mathrm{intrinsic}}$ in $\cF_\cG$ according to an (arbitrary) inner product on $\Delta^*$.

\begin{definition}\label{def:relational}
    A property $\bvec{\phi}$ is purely relational if and only if (1) is FOGI and (2) is orthogonal\footnote{This requires choosing an inner product on $\Delta^*$ (See Assumption \ref{ass:innerprod}).} to all intrinsic FOGI properties: \begin{equation}
        \bvec{\phi}\in\cF_\cG \cap \cF_{\mathrm{intrinsic}}^\perp
    \end{equation}
\end{definition}

Hereafter, by ``relational FOGI property'', we will mean \textit{purely} relational FOGI properties.

\subsubsection{The structure of [purely] relational properties}
\label{sec:relprops}
In this section, we treat $\cK$ and $\Delta^*$ explicitly as inner product spaces (Assumption \ref{ass:innerprod}). We also rely on the resulting isomorphism $\cK^*\cong\cK$. This allows for the derivation of a convenient characterization of relational FOGI properties using the gauge space $\cK$.  

Recall that a relational FOGI property $\bvec{\phi}$ must be both (1) FOGI and (2) orthogonal to every gate's intrinsic FOGI space $\cF_g$.  Recall from Eq. (\ref{eq:phistack}) that any $\bvec{\phi}$ can be written as a linear combination (stack) of properties that are intrinsic to individual gates,
\begin{equation}
\bvec{\phi} = \left(\bigstack_{g\in\cG} \bvec{\phi}_g^\top\right)^\top,
\end{equation}
so that
\begin{equation}
    A_\cG^\top\bvec{\phi}^\top = \sum_{g\in\cG}A_g^\top\bvec{\phi}^\top_g.
\end{equation}
Per Lemma \ref{lem:FOGI}, $\bvec{\phi}$ is FOGI if and only if

\begin{equation}\label{eq:relcon}
    \sum_{g\in\cG}A_g^\top\bvec{\phi}_g^\top = 0.
\end{equation}

\noindent But in order to satisfy condition (2) in Definition \ref{def:relational}, no individual $\bvec{\phi}_g$ can itself be FOGI. Thus, for every non-trivial $\bvec{\phi}_g$, $A_g^\top\bvec{\phi}_g^\top$ must be non-zero because $\bvec{\phi}_g^\top\notin\mathrm{kernel}\left(A_g^\top\right)$ by Definition \ref{def:intrinsicFOGI}.

Recall that $A_g$ maps $\cK \to \Delta$. The support of $A_g$ is the space of gauge transformations that act faithfully on $g$ and we denote it $\cK_g\subseteq \cK$. On the other hand, its transpose by definition acts on the corresponding dual spaces, i.e. $A_g^\top:\Delta^*\to\cK^*$. But because $\cK\cong\cK^*$,  the range of $A_g^\top$ is isomorphic to $\cK_g$ and so
\begin{equation}
    A_g^\top\bvec{\phi}_g^\top=\bvec{k}_g \in \cK_g. 
\end{equation}
Substituting this relation into Eq.~\eqref{eq:relcon} yields a simple condition for $\bvec{\phi}^\top$ to be FOGI:
\begin{equation} \label{eq:ZeroSum}
\sum_g{ \bvec{k}_g} = 0.
\end{equation}

The space of relational FOGI properties of $\cG$ is linearly equivalent to the space of solutions to Eq.~\eqref{eq:ZeroSum}, subject to $\bvec{k}_g \in \cK_g$. 
For each operation \(g\), the kernel of the map \(A_g^\top : \Delta_g^* \to \cK\) is the intrinsic FOGI space of gate \(g\), \(\cF_g\), while \(A_g^\top\)'s image \(\cK_g \subseteq \cK\) contains the gauge directions that act faithfully and nontrivially on \(g\). Therefore, \(A_g^\top\) induces an isomorphism between the quotient space \(\Delta_g^*/\cF_g\) and \(\cK_g\).
Thus, each gate-local component \(\bvec{\phi}_g\) of a relational FOGI property is
specified by the gauge vector \(\bvec{k}_g = A_g^\top \bvec{\phi}_g^\top \in \cK_g\). Given \(\bvec{k}_g\), a convenient representative for \(\bvec{\phi}_g\) can be obtained using the Moore--Penrose pseudoinverse,
\[
\bvec{\phi}_g^\top = \left(A_g^\top\right)^+ \bvec{k}_g .
\]
As a corollary, any solution $\{\bvec{k}_g:\ g\in\cG\}$ can be immediately mapped to a relational FOGI property by stacking all the $\bvec{\phi}$ components
\begin{equation} \label{eq:InverseSolution}
\bvec{\phi}^\top = \bigstack_g{\left(A_g^\top\right)^{+}\bvec{k}_g} .
\end{equation}

\noindent This characterization of relational FOGI properties can be summarized in the following lemma: 
\begin{lemma}
  Any set of gauge vectors $\left\{\bvec{k}_g \in \cK_g\right\}$ that satisfies Eq.~\eqref{eq:ZeroSum} defines a relational FOGI property by Eq.~\eqref{eq:InverseSolution}, and any relational FOGI property can be expressed in this way. 
\end{lemma}

Note that the transformation of Eq.~\eqref{eq:InverseSolution} preserves linear independence, but not orthogonality.  So solutions to Eq. \eqref{eq:ZeroSum} that are orthogonal in $\cK$ will yield vectors in $\Delta^*$ that are linearly independent, but not necessarily orthogonal. Furthermore, the relationship  $\Delta_g^*/\cF_g \cong \cK_g$ is not uniquely defined without the isomorphism that accompanies the inner product assumption (Assumption \ref{ass:innerprod}). For example,  the Moore-Penrose pseudoinverse is undefined without an inner product, making Eq. (\ref{eq:InverseSolution}) directly dependent on the inner product choice.

\subsubsection{Pairwise relational properties} \label{sec:pairwise}

The simplest relational FOGI properties are \textit{pairwise} relational FOGI properties that depend only on variations in two operations $g_1$ and $g_2$.  

\begin{figure}[th!]
\centering
\includegraphics[width=8cm]{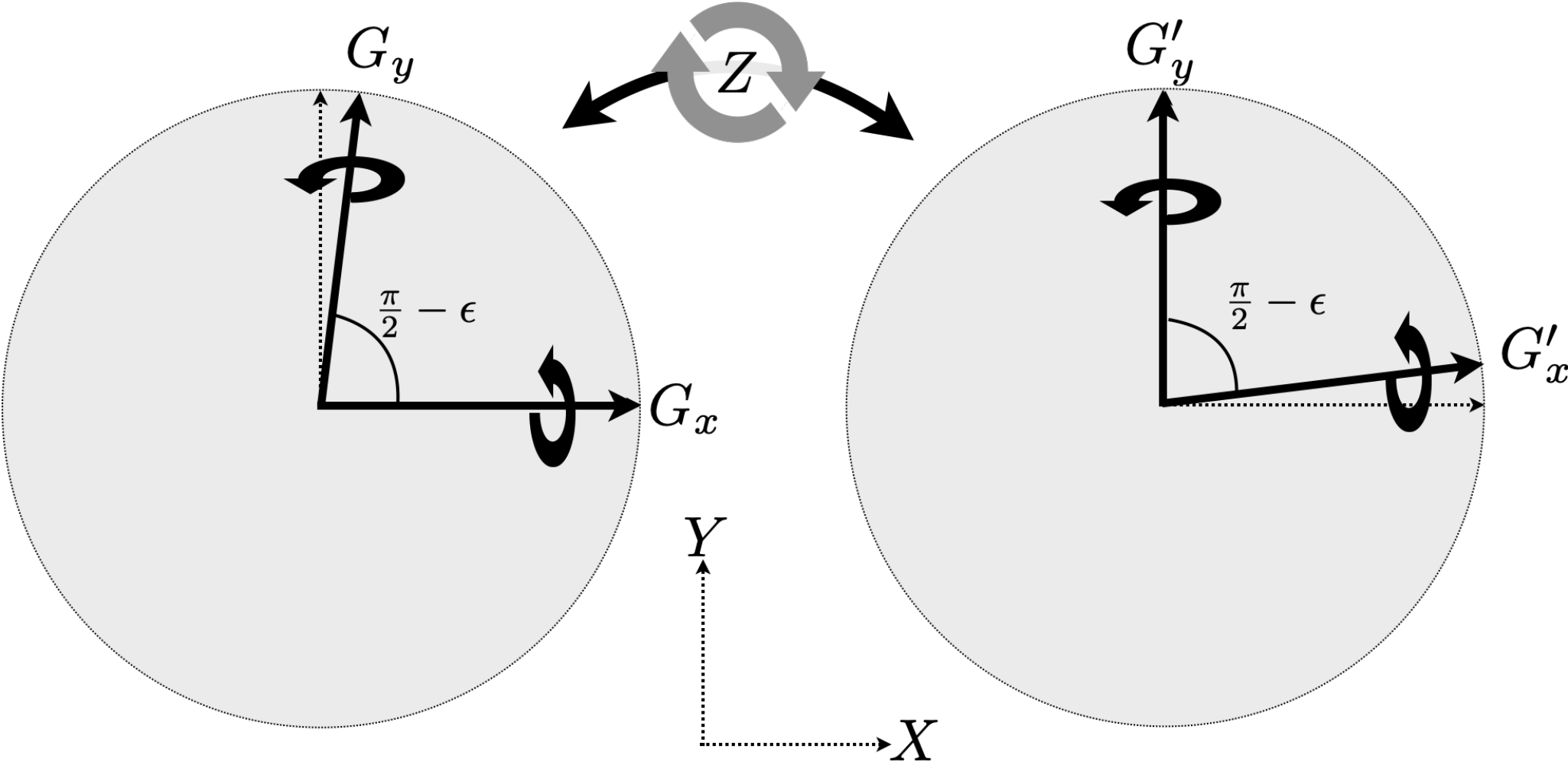}
\caption{If a gate set contains $\pi/2$ rotations around the Pauli X ($G_x$) and Pauli Y ($G_y$) axes, but the actual rotation axes make an angle of $\pi/2 - \epsilon$ with each other instead of $\pi/2$, this is a pairwise relational error.  Either gate in isolation behaves perfectly, but there is an observable error that can be shifted between them by a gauge $Z$ rotation. }
\label{fig:pairwise}
\end{figure}

Figure \ref{fig:pairwise} illustrates a simple example.  Consider a single-qubit processor whose target gate set $\overline{\cG} = \left\{\overline{G_x},\overline{G_y}\right\}$ contains $\pi/2$ rotations around the Pauli $X$ and $Y$ axes, and consider a noisy gate set $\cG$ in which $G_x = \overline{G_x}$ but $G_y$ is a unitary rotation by $\pi/2$ around the wrong axis $\sin(\epsilon)X + \cos(\epsilon) Y$. As described, the $G_x$ gate appears perfect, while $G_y$ is flawed. But there is an alternative description $\cG'$---a gauge-equivalent gate set, related to $\cG$ by a gauge rotation around the $Z$ axis---in which $G_y' = \overline{G_y}$, but $G_x'$ rotates by $\pi/2$ around the axis $\cos(\epsilon)X-\sin(\epsilon)Y$.  This is physically indistinguishable, just described in a different gauge. But in this description, the error is all in the $G_x$ gate.

The error in this gate set is observable, but neither gate is intrinsically flawed.  Experiments that use \emph{only} $G_x$ or \emph{only} $G_y$ will display Rabi oscillations at exactly the correct frequency. The error can be described in a gauge-invariant way as ``The angle between the rotation axes of $G_x$ and $G_y$ is $\pi/2-\epsilon$ instead of $\pi/2$.'' This describes a relational FOGI error.  Relational errors can always be shifted, by gauge transformations, between the gates they relate.

The pairwise relational FOGI properties between gates $g_1$ and $g_2$ are precisely the FOGI relational properties that are supported entirely on $\Delta^*_{g_1} \oplus \Delta^*_{g_2}$ and are orthogonal to both (1) all FOGI properties supported entirely on $\Delta^*_{g_1}$ and (2) all FOGI properties supported entirely on $\Delta^*_{g_2}$.  We extend the definition of ``intrinsic'' from Sec.~\ref{sec:intrinsicFOGI}, and say that such a property is \textit{intrinsic to $\{g_1,g_2\}\subseteq\cG$}.  To define the pairwise relational properties between $g_1$ and $g_2$ precisely, we first identify the space of \textit{all} FOGI properties that are intrinsic to $\{g_1,g_2\}$,
\begin{equation}    
\cF_{g_1,g_2} \cong \mathrm{kernel}\left(A_{g_1,g_2}^\top\right) = \mathrm{kernel}\left(\left(A_{g_1}\stack A_{g_2}\right)^\top \right),
\end{equation}
and then identify pairwise relational FOGIs between $g_1$ and $g_2$ with elements of the subspace
\begin{equation}
    \cF_{g_1,g_2}^{rel} = \cF_{g_1,g_2} \cap (\cF_{g_1})^\perp \cap (\cF_{g_2})^\perp.
\end{equation}
Algorithm \ref{algo:pairwise}, below, defines a systematic procedure that uses \emph{blocking matrices} B to construct a basis for $\cF_{g_1,g_2}^{rel} $:

\begin{algorithm}[H]
  \caption{Construct a spanning set for the relational FOGI space $\cF^{rel}_{g_1,g_2}$}\label{algo:pairwise}
  \begin{algorithmic}[1]
    \Require Subset of operations $S = \{g_1, g_2\}$
    \Require $A_{g_1},A_{g_2}$
    \Ensure  Basis for $\cF^{rel}_{g_1,g_2}$
    \State $B=\emptyset$ \Comment{Initialize empty blocking matrix}

    \State $\cF_{g_1} = \mathrm{kernel}\left(A_{g_1}^\top\right)$ \Comment{Compute lists of intrinsic FOGI properties}
    \State $\cF_{g_2} = \mathrm{kernel}\left(A_{g_2}^\top\right)$ 
    
    \For{$\bvec{\phi}_{g_1}^\top \text{ in } \cF_{g_1}$}
        \State $B$ = $B \stack \bvec{\phi}_{g_1}$ \Comment{Add intrinsic FOGIs to blocking matrix}
    \EndFor
    \For{$\bvec{\phi}_{g_2}^\top \text{ in } \cF_{g_2}$}
        \State $B$ = $B \stack \bvec{\phi}_{g_2}$\Comment{Add intrinsic FOGIs to blocking matrix}
    \EndFor
    \State $A_{g_1,g_2} = A_{g_1} \stack A_{g_2}$
    \State$\cF_{g_1,g_2}^{rel}=\mathrm{kernel}\left(A_{g_1,g_2}^\top\stack B\right)$

    \State \Return $\cF_{g_1,g_2}^{rel}$
  \end{algorithmic}
\end{algorithm}
\noindent This algorithm works because stacking a matrix $B$ inside a kernel computation effectively blocks the span of the rows of $B$. To see this, consider the following example. Let $H$ be an arbitrary matrix, and $h$ an arbitrary vector. Now consider a vector $h'$ that lies within the kernel of $H \stack h^\top$. By definition, then

\begin{equation}
    \begin{pmatrix}
        H\\h^\top
    \end{pmatrix} h'= 
    \begin{pmatrix}
        H h'\\h^\top h'
    \end{pmatrix}=0,
\end{equation}
thus $h'$ satisfies both $H h'=0$ and $h^\top h'=0$. In other words, any vector $h'$ in the kernel of $H \stack h^\top$ is both (1) in the kernel of $H$, and (2) orthogonal to $h$ under the dot product\footnote{Although we construct the blocking matrices to work specifically with the dot product, the same procedure can be adapted to any choice of inner product by scaling the rows of B appropriately.} 
\begin{equation}
h\cdot h' = h^\top h'=0.
\end{equation}

\noindent By construction, the rows of $B$ span $\cF_{g_1}$ and $ \cF_{g_2}$ .  Thus, the kernel of $A_{g_1,g_2}^\top\stack B$ contains vectors that are both in (1) in the kernel of $A^\top_{g_1,g_2}$ and (2) orthogonal to all FOGI properties intrinsic to $g_1$ and $g_2$. These vectors are by definition in $\cF_{g_1,g_2}^{rel}$.

Algorithm \ref{algo:pairwise} provides an existence proof for constructing pairwise relational FOGIs, but there's a much more elegant and intuitive way to construct $\cF_{g_1,g_2}^{rel}$ by identifying it with gauge vectors.  We observe that Eq.~\ref{eq:ZeroSum} provides a simple mathematical characterization of pairwise relational properties.  Every pairwise relational property between operations $g_1$ and $g_2$ is described by a pair of gauge vectors $\bvec{k}_1\in\cK_1$ and $\bvec{k}_2\in\cK_2$, where $\cK_g = \mathrm{support}(A_g)$, that satisfy
\begin{align}
    \bvec{k}_1 + \bvec{k}_2 &= 0 \\
    &\Downarrow \notag\\
    \bvec{k}_2 &= -\bvec{k}_1.
\end{align}
It follows that every pairwise relational FOGI property $\bvec{\phi}$ between $g_1$ and $g_2$ can be identified with a gauge vector $\bvec{k} \in \cK_1 \cap \cK_2$.  These properties form the subspace $\cF_{g_1,g_2}^{rel}$, and Eq.~\ref{eq:InverseSolution} implies that it is given by
\begin{equation}
    \cF_{g_1,g_2}^{rel} = \mathrm{span}\left(\left[\left(A_1^\top\right)^+ \stack \left(-A_2^\top\right)^+\right]\bvec{k}\right)\ \forall \bvec{k} \in \cK_1 \cap \cK_2.
\end{equation}
So, every pairwise relational FOGI property between $g_1$ and $g_2$ can be found by (1) identifying a gauge transformation $\bvec{k}$ that acts \textit{faithfully} on both operations -- i.e., it is orthogonal to all the gauge transformations in $\mathrm{kernel}\left( A_g\right)$ that have no effect at all on $g$, for $g=g_1$ \textit{and} $g=g_2$ -- and then (2) computing a vector 

\begin{equation}\bvec{\phi}^\top = \left[\left(A_1^\top\right)^+\stack\left(-A_2^\top\right)^+\right]\bvec{k}\notag\end{equation} 

\noindent that measures how much $g_1$ and $g_2$ have been \textit{relatively} shifted in the gauge direction corresponding to $\bvec{k}$.  

If the same gauge transformation acts faithfully on 3 or more operations, then there will exist a pairwise relational FOGI property for \textit{each} pair of operations, but they will not be linearly independent.  For example, if there exists a $\bvec{k} \in \cK_1 \cap \cK_2 \cap \cK_3$, then three pairwise FOGI properties can be constructed corresponding to
\begin{equation}
    \left\{ (\bvec{k},-\bvec{k},0),\ (0,\bvec{k},-\bvec{k}),\ (-\bvec{k},0,\bvec{k})\right\}.
\end{equation}
Because each can be written as a linear combination of the other two, these three quantities only span a 2-dimensional space.  If the gauge transformation defined by $\bvec{k}$ acts faithfully on $n$ operations, then $n(n-1)/2$ distinct pairwise FOGI properties will share an $(n-1)$-dimensional space.  There is no obvious, natural way to break this symmetry and choose a unique linearly independent basis.

\subsubsection{$n$-wise relational properties}
\label{sec:n-wise-relational}

It is often the case that the intrinsic and pairwise relational FOGI properties of a gate set do not span the entire space of FOGI properties.  That is\footnote{The use of $+$ (or equivalently $\Sigma$) describes summing subspaces that may overlap. On the other hand,  \(U\oplus V\) denotes a direct sum which describes a sum of spaces with a trivial overlap $U\cap V=\{0\}$ and decompositions are unique.
},
\begin{equation}
    \bigoplus_{g\in\cG}{\cF_g} \oplus \left( \sum_{g,g'\in\cG}{\cF_{g,g'}^{rel}} \right) \subset \cF_\cG.
\end{equation}
This implies the existence of \textit{irreducibly} $n$-wise  FOGI properties that depend on variations in $n>2$ operations, and cannot be written as linear combinations of intrinsic or pairwise relational properties.  To analyze these properties, we follow the convention introduced previously and say that a FOGI property is irreducibly $n$-wise if it cannot be reduced to ``smaller'' properties -- i.e., written as a sum of $n'$-wise FOGI properties for $n'<n$, with $n'=1$ corresponding to intrinsic FOGI properties -- and that it is \textit{purely} $n$-wise if it is orthogonal (under some chosen inner product) to all $n'$-wise FOGI properties for $n'<n$.

\begin{definition}
    A FOGI property $\bvec{\phi}$ is a purely $n$-wise FOGI property if and only if it is (1) intrinsic to a set of $n$ operations, and (2) orthogonal to all $n'$-wise relational FOGI properties for all $n'<n$.
\end{definition}

Let the set of all collections of $n$ elements of a gateset be denoted  
\begin{equation}
    \mathcal{P}_n(\cG) =\{\,x \mid x\in \mathcal{P}(\cG) \land |x|=n\} \label{eq:combs}
\end{equation}
where $\mathcal{P}$ denotes a power set, and $|x|$ denotes the cardinality of $x$. For example, for the gate set $\{M,\sket{\rho},G \}$ 
\begin{align}
    &\mathcal{P}_1(\{M,\sket{\rho},G \}) = \{\{M\},\{\sket{\rho}\},\{G\} \},\\
    &\mathcal{P}_2(\{M,\sket{\rho},G \})  = \{\{M,\sket{\rho} \}, \{M,G\}, \{\rho,G \} \}, \text{ and }\\
    &\mathcal{P}_3(\{M,\sket{\rho},G \})  = \{\{M,\sket{\rho},G \} \}.
\end{align}
The purely $n$-wise FOGI properties between a subset of $n$ operations $S=\{g_1,\ldots,g_n\}\subseteq \cG$ form a subspace of $\cF_\cG$ that we denote $\cF_{S}^{rel}$ and is defined as

\begin{equation}\label{eq:orthogonal_irreducible}
    \cF_{S}^{rel} = \cF_S\cap \left(\sum_{\{g_i\}\in \mathcal{P}_{n-1}(\cG)} \cF_{\{g_i\}}\right)^\perp.
\end{equation}

Notice that the sum goes over $\mathcal{P}_{n-1}(\cG)$, i.e. all combinations of $n-1$ operations in the entire gate set. This is because purely $n$-wise FOGI properties need to be orthogonal to \emph{all} FOGI properties with support on a smaller set of operations, even if they include some operations outside of S.

\begin{equation}
    \cF_S = \mathrm{kernel}\left(\left(\bigstack_{g\in{S}}A_g\right)^\top\right)
\end{equation}

\noindent is the space of all the FOGI properties intrinsic to the subset of gates in $S$.

To preemptively address a potential confusion, we observe that $\cF_{S}^{rel}$ could plausibly indicate \textit{either} (1) the space of \textit{all} purely relational FOGI properties that are intrinsic to $S$, or (2) the space of purely $n$-wise FOGI properties intrinsic to $S$.  These coincide for the $n=2$ (pairwise) case analyzed previously, but for $n>2$ they do not because the former also includes (e.g.) pairwise relational properties intrinsic to 2-element subsets of $S$.  We therefore emphasize that $\cF_{S}^{rel}$ indicates the (smaller) space of purely $n$-wise FOGI properties, and we do not discuss or define a notation for the space of all relational FOGI properties on $S$.
\begin{figure}[th!]
\centering
\includegraphics[width=5cm]{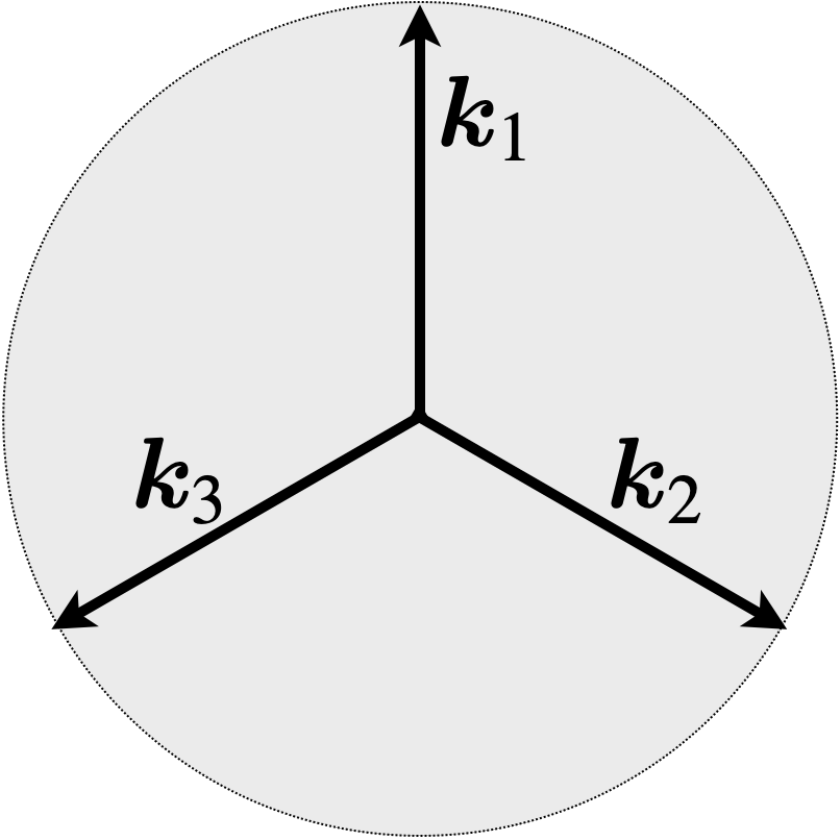}
\caption{If three operations $g_1,g_2,g_3$ have distinct 1-dimensional gauge spaces $\bvec{k}_1,\bvec{k}_2,\bvec{k}_3$ that span a 2-dimensional space, like the trine vectors illustrated here, then no pairwise FOGI relational operations can exist because no pair of gauge spaces intersects and so the equation $\bvec{k}_i + \bvec{k}_j = 0$ has no nonzero solutions.  But a 3-wise relational property exists, because $\bvec{k}_1 + \bvec{k}_2 + \bvec{k}_3 = 0$.}
\label{fig:trine}
\end{figure}

Once more we observe that Eq.~\ref{eq:ZeroSum} provides a good mathematical characterization of purely n-wise relational properties.  Every purely n-wise property of a set of operations $S=\{g_1,\ldots,g_n\}$ is described by $n$ gauge vectors $\{\bvec{k}_1,\ldots,\bvec{k}_n\}$, such that $\bvec{k}_i\in\cK_{g_i}$ where $\cK_{g_i} = \mathrm{support}(A_{g_i})$.  Figure \ref{fig:trine} offers an intuitive explanation for why $n$-wise relational properties are necessary, using only a 2-dimensional gauge space $\cK = \mathbb{R}^2$.  Consider three operations $g_1,g_2,g_3$.  Suppose that each operation's gauge space $\cK_i$ is 1-dimensional, and that they form a trine as in Fig.~\ref{fig:trine}.  Each pair of gauge spaces is linearly independent, and so no pairwise FOGI relational properties exist, but because the triple $\{\cK_1,\cK_2,\cK_3\}$ is linearly dependent, a solution to Eq.~\ref{eq:ZeroSum} exists and so a 3-wise relational FOGI property exists.

We do not yet know an elegant mathematical characterization of a gate set's purely $n$-wise relational FOGI properties.  However, they are easy to construct explicitly (when they exist) by a straightforward modification of Algorithm \ref{algo:pairwise}. As an example, here is a procedure to compute all purely 3-wise FOGI properties between three operations $g_1,g_2,g_3$:

\begin{algorithm}[H]
  \caption{Construct a spanning set for a purely 3-wise FOGI space $\cF^{rel}_{g_1,g_2,g_3}$}\label{algo:3-wise}
  \begin{algorithmic}[1]
  \Require The gate set $\cG$
    \Require Subset of operations $S = \{g_1, g_2, g_3\}\subseteq\cG$
    \Require $A_{g}$ for all $g\in \cG$
    \Ensure  Basis for $\cF^{rel}_{g_1,g_2,g_3}$
    \State $B=\emptyset$ \Comment{Initialize empty blocking matrix}

    \For{$\{g,g'\}$ in $\mathcal{P}_2(\cG)$} \Comment{See Eq. (\ref{eq:combs}) for $\mathcal{P}_2$ definition}
        \State $\cF_{g,g'}= \mathrm{kernel}\left(\left(A_{g} \stack A_{g'}\right)^\top\right)$ 
        \For{$\bvec{\phi}^\top \text{ in } \cF_{g,g'}$}
        \State $B$ = $B \stack \bvec{\phi}$ \Comment{Add 2-wise properties to blocking  matrix}
    \EndFor
    \EndFor

    \State $A_{g_1,g_2,g_3} = A_{g_1} \stack A_{g_2} \stack A_{g_3}$
    \State$\cF_{g_1,g_2,g_3}^{rel}=\mathrm{kernel}\left(A_{g_1,g_2,g_3}^\top\stack B\right)$

    \State \Return $\cF_{g_1,g_2,g_3}^{rel}$
  \end{algorithmic}
\end{algorithm}
We first construct the \textit{pairwise} FOGI properties for each pair of operations in $\cG$.  Then, we compute $\mathrm{kernel}(A_{g_1,g_2,g_3}^\top\stack B)$  where $B$ blocks all the pairwise properties constructed in the previous step.  This yields purely 3-wise properties.  This algorithm can be generalized to compute any purely $n$-wise subspace $\cF_{g_1,\ldots,g_n}^{rel}$ by blocking the FOGI spaces of all subsets of $n-1$ operations.

Every purely $n$-wise FOGI space spans a unique subspace of $\cF_{\cG}$. By collecting bases for all of the purely $n$-wise FOGI spaces (including intrinsic properties), we construct a spanning set for the entire FOGI space
\begin{align}
    \cF_\cG =& \left(\bigoplus_{g\in\cG}\cF_g\right) \oplus \left(\sum_{g_1,g_2\in\cG}\cF_{g_1,g_2}^{rel} \right) \oplus \notag \\
    & \left(\sum_{g_1,g_2,g_3\in\cG}\cF_{g_1,g_2,g_3}^{rel} \right)\oplus\cdots\oplus\left(\sum_{g_1,\ldots,g_N\in\cG}\cF_{g_1,\ldots,g_N}^{rel} \right).
\end{align}

Although this set is guaranteed to span $\cF_\cG$, it will generally be overcomplete. For example, in Sec \ref{sec:pairwise} we showed that if there exists a gauge generator $\bvec{k}$ such that $\bvec{k}\in\cK_1\cap\cK_2\cap\cK_3$, then three pairwise relational FOGI properties corresponding to\begin{align}\label{eq:lindep}
    &(\bvec{k}, -\bvec{k},0) \in \cF_{g_1,g_2}^{rel},\notag\\
    &(0, \bvec{k},-\bvec{k}) \in \cF_{g_2,g_3}^{rel}, \\
    &(\bvec{k}, 0,-\bvec{k}) \in \cF_{g_1,g_3}^{rel},\notag
\end{align} 
can be defined.  These three pairwise FOGI properties are intrinsic to distinct pairs of gates, but they overlap because these pairs are not disjoint. Nevertheless, with some work it is possible to construct a basis for $\cF_\cG$ from elements of these spaces.

\subsection{Constructing FOGI bases}
\label{sec:constructingFogiBases}
The preceding subsections provided a procedure to construct FOGI properties that span any purely $n$-wise FOGI subspace. A sufficiently large set will span $\cF_\cG$, making it possible to construct a complete basis — even with some work, an orthonormal basis — for FOGI properties. Unfortunately, we have not found any way to identify a \emph{uniquely good} basis. Despite this degeneracy, FOGI bases have useful applications such as parameterizing gate sets (e.g. for optimization) and uniquely distinguishing gauge orbits.

It is possible to compute $\mathrm{kernel}\left(A_\cG^\top\right)$ using any computational linear algebra library to obtain a FOGI basis, but this approach does not take into account the interpretability of the basis elements. FOGI properties with support on fewer operations tend to be more amenable to interpretation. Nonetheless, properties with multiple operations are required to span most relational FOGI spaces. Algorithm \ref{algo:basis} presents an iterative (generalized) implementation of Algorithm \ref{algo:pairwise} and \ref{algo:3-wise} which yields a basis for $\cF_\cG$ containing only properties that are purely $n$-wise for some $n$.

Given a gate set with operations $\{\rho, M, G_1,\dots, G_N\}$ and their corresponding gauge action matrices $\{A_{\rho}, A_M, A_{G_1},\dots,A_{G_N}\}$, we construct a FOGI basis $\{\bvec{\phi}_i\}$ through the following procedure:

\begin{algorithm}[H]
  \caption{Construct a FOGI basis for gate set $\cG$}\label{algo:basis}
  \begin{algorithmic}[1]
    \Require Gate set $\cG = \{\rho, M, G_1,\dots,G_N\}$
    \Require $A_{\rho},A_M,A_{G_1},\cdots,A_{G_N}$
    \Ensure  Basis for $\cF_\cG$
    \State $\cF_\cG = \emptyset$ \Comment{Initialize empty list}
    \State $B = \emptyset$ \Comment{Initialize empty matrix}
    \For{$n \text{ in } (1,2,\dots,N+2)$}
        \ForAll{$S \text{ in } \mathcal{P}_n(\cG)$} \Comment{$\mathcal{P}_n(\cG)$ defined in Eq.\ref{eq:combs}}
        \State $A_S =  \bigstack_{g\in S}\; A_g$ 
        
      \If{n == 1}\Comment{Intrinsic properties}
        \State $\cF_S^{pure} = \mathrm{ker}\left(A_S^\top\right)$
      \Else \Comment{Relational Properties}
        
        \State $\cF_S^{pure} = \mathrm{ker}\left(A_S^\top\stack B\right)$
      \EndIf
          \ForAll{$\bvec{\phi}^\top \text{ in }\cF_S^{pure}$ }
          \If{$\bvec{\phi}^\top \notin \mathrm{span}(\cF_\cG)$}
        \State $\cF_\cG = \cF_\cG \cup \bvec{\phi}^\top$ 
        \State $B = B \stack \bvec{\phi}$ \label{line:blocking}
        \EndIf
        \EndFor
        \EndFor
      \EndFor
    \State \Return $\cF_\cG$
  \end{algorithmic}
\end{algorithm}

Algorithm \ref{algo:basis} considers all possible combinations of operations in ascending order; single operations, pairwise combinations, 3-wise combinations, and so on. At every step it constructs all $n$-wise FOGI properties that are orthogonal to every $(n-1)$-wise FOGI property by blocking the basis elements for these subspaces found in previous steps. Furthermore, it only adds properties to the basis if they are linearly independent from the other properties already found (to avoid situations like Eq. (\ref{eq:lindep})). This ensures that all the FOGI properties within the basis are linearly independent.

It is important to note that this procedure does not yield a deterministic basis for $\cF_\cG$. The order iterated over $\mathcal{P}_n(\cG)$, which we leave unspecified, affects the output of Algorithm \ref{algo:basis}. We have not yet found any motivation to prefer a specific ordering.

\section{Understanding FOGI properties} \label{Understanding}
The analysis in Section \ref{Taxonomy} treated the error set space $\Delta$ as a generic vector space and the gauge action $A_\cG$ as an arbitrary matrix.  But these objects have additional structure.  Operations and gauge generators are represented by superoperators (PTMs), and gauge transformations can be usefully described using these (much smaller) matrices rather than the generic $A_\cG$.  We now apply this analysis to reveal more about the structure and nature of FOGI properties.

\subsection{PTM representations of properties and gauge actions}

Recall from Sec. \ref{GateSets} that
\begin{enumerate}
    \item Errors and (linear) properties of gate sets are described by tuples $\delta\in\Delta$ and $\varphi\in\Delta^*$, respectively.
    
    \item Gauge transformations are described by gauge generators $K\in\cK$ and act affinely as $\delta \to \delta + \epsilon A_\cG(K) + O\left(\epsilon^2\right)$.

    \item A property $\varphi$ is FOGI iff $\varphi(\delta-\delta')=O\left(\epsilon^2\right)$, for any two errors $\delta$ and $\delta'$ related by a gauge transformation $K$.
\end{enumerate}
To analyze FOGI properties as functionals of the Pauli transfer matrices usually used to represent quantum operations, we revert from representing errors and properties by abstract vectors back to representing them as tuples of PTMs:
\begin{align}
    \delta &= \cG - \overline{\cG} \\
    &= \left( \sket{\rho} - \sket{\overline{\rho}},\ M-\overline{M},\ G_1-\overline{G}_1,\ldots,\ G_N - \overline{G}_N \right).
\end{align}
Now any linear functional $\varphi(\delta)$ can be represented by a tuple of objects of the same size and degrees of freedom as $\delta$, $\Phi = \left( \sbra{\Phi_\rho},\ \Phi_M,\ \Phi_1,\ldots \Phi_N \right)$, and computed as
\begin{equation} \label{eq:PTMproperty}
    \varphi(\delta) =  \sbraket{\Phi_\rho}{\rho-\overline{\rho}} + \Tr\left(\Phi_M (M-\overline{M})\right) + \sum_i{\Tr\left(\Phi_i (G_i-\overline{G}_i)\right)}.
\end{equation}

Recall from Eq. (\ref{eq:affineshift}) that the gauge action induced by a small gauge transformation generated by a gauge generator $K$ is given by 
\begin{equation*}
    \delta' - \delta = \left(\epsilon K\sket{\overline{\rho}}, -\epsilon \overline{M}K, \epsilon[K,\overline{G}_1], \dots, \epsilon[K,\overline{G}_N] \right) + O\left(\epsilon^2\right).
\end{equation*}

\noindent We can use this representation to write
\begin{align}
    \varphi(\delta'-\delta) =&\notag\\ &\epsilon\left( \sbraopket{\Phi_\rho}{K}{\overline{\rho}} - \Tr\left(\Phi_M \overline{M} K\right) + \sum_i{ \Tr \left(\Phi_i \left[K,\overline{G}_i\right]\right)}\right)\notag\\ &+ O\left(\epsilon^2\right).\label{eq:deltadiff}
\end{align}

\noindent Therefore $\varphi$ is FOGI iff the $O(\epsilon)$ term vanishes for all $K$ in Eq. (\ref{eq:deltadiff}). By using the cyclic property of the trace and the fact that \begin{equation}
\Tr CB=0\ \forall\ C\notag, 
\end{equation} 
where $C$'s top row and $B$'s left column are zero, implies $B=0$, we obtain
\begin{center}
\fbox{%
\begin{minipage}{0.95\linewidth}
\begin{proposition}[FOGI criterion in PTM form]\label{prop:fogi-ptm}
A property $\Phi = \left( \sbra{\Phi_\rho},\ \Phi_M,\ \Phi_1,\ldots \Phi_N \right)$ is FOGI iff 
\begin{equation}
     \sketbra{\overline{\rho}}{\Phi_\rho} - \Phi_M \overline{M} + \sum_i{ \left[ \overline{G}_i,\Phi_i\right]}=0. \label{eq:GaugeChange}
\end{equation}
\end{proposition}
\end{minipage}%
}
\end{center}

\noindent Eq.~\ref{eq:GaugeChange}  is simply $ A_\cG^\top\bvec{\phi}^\top=0$ written in terms of PTMs.  Several important properties of intrinsic and relational FOGI properties can be derived straightforwardly from this expression.

\subsection{Intrinsic FOGI properties}
\label{sec:intrinsic}
A property $\varphi(\delta)$ of a gate set $\cG = (\sket{\rho},M,G_1,\ldots, G_N)$ is an intrinsic FOGI property of operation $g$ iff it is (1) FOGI and (2) dependent only on variations in operation $g$, i.e. there is only one non-trivial term $\Phi_g\neq0$ in Eq. (\ref{eq:GaugeChange}).  By Proposition \ref{prop:fogi-ptm}, an intrinsic property $\Phi$ is FOGI if and only if the non-trivial $\Phi_g$ satisfies Eq. (\ref{eq:GaugeChange}), so

\begin{align}
\mathrm{if\ }g = \rho &\mathrm{\ then\ } \sketbra{\overline{\rho}}{\Phi_\rho}=0, \\
\mathrm{if\ }g = M &\mathrm{\ then\ } \Phi_M \overline{M}=0,\ \mathrm{and}\\
\mathrm{if\ }g = G_i &\mathrm{\ then\ }  \left[\Phi_i, \overline{G}_i\right]=0.
\end{align}

\subsubsection{State preparation operations have no intrinsic FOGI properties}
\label{sec:SPAMHavenoIntrinsics}
A property intrinsic to state preparation is FOGI iff
$\sketbra{\overline{\rho}}{\Phi_\rho}=0$, which is not the case for any $\sbra{\Phi_\rho} \neq 0$. Therefore:

\begin{lemma}\label{lem:stateFOGI}
    State preparation operations $\sket{\rho}$ do not have any nontrivial intrinsic FOGI properties.
\end{lemma}

\subsubsection{(Most) measurement operations have no intrinsic FOGI properties}

An intrinsic property $\Phi=(0,\Phi_M,0,0,\cdots,0)$ is FOGI if and only if (Eq.~\ref{eq:GaugeChange})
\begin{equation} \label{eq:int-M}
    \Phi_M \overline{M}=0.
\end{equation}

\noindent Here, $\overline{M}$ is an $m\times d^2$ PTM describing an $m$-outcome POVM.  We will assume that the rows of $\overline{M}$, which represent the \textit{effects} of the POVM, are linearly independent.  This is not true for all POVMs, but it holds for projective measurements (even slightly noisy ones).

For Eq. (\ref{eq:int-M}) to be satisfied all the (linearly independent) columns of $\overline{M}$ must be contained within the kernel of $\Phi_M$, so
\begin{equation}
    \mathrm{rank}\left(\overline{M}\right) \leq \mathrm{dim}\left(\mathrm{kernel}(\Phi_M)\right).
\end{equation}
Because the rows of $\overline{M}$ are linearly independent, $\mathrm{rank}(\overline{M})=m$, so 
\begin{align}
    m & \leq \mathrm{dim}\left(\mathrm{kernel}(\Phi_M)\right) \\
    \Rightarrow& m - \mathrm{dim}\left(\mathrm{kernel}(\Phi_M)\right) \leq 0.
\end{align}
But since $\Phi_M$ has shape $d^2 \times m$, the rank-nullity theorem states that 
\begin{equation}
m = \mathrm{rank}(\Phi_M) + \mathrm{dim}\left(\mathrm{kernel}(\Phi_M)\right).
\end{equation}
Combining the last two equations implies that $\mathrm{rank}\left(\Phi_M\right)=0$ and hence $\Phi_M = 0$.  Therefore:

\begin{lemma}\label{lem:MFOGIs}
    Measurement operations $M$ with linearly independent effects have no intrinsic FOGI properties.
\end{lemma}

\subsubsection{Intrinsic FOGI properties of gates correspond to commutants}\label{sec:commutant}

From Eq. (\ref{eq:GaugeChange}) an intrinsic property \begin{equation}
\Phi = (0,0,\cdots,0, \Phi_i,0,\cdots,0)\notag\end{equation} of gate $G_i$ is FOGI if and only if 
\begin{equation} \label{eq:gateFOGIcomm}
        \left[ \overline{G}_i,\Phi_i\right] = 0.
\end{equation}

Unlike the previous two cases, this equation \textit{does} have obvious solutions.  Therefore, $\Phi$ is FOGI iff $\Phi_i$ commutes with $\overline{G}_i$.

The space of $G_i$'s intrinsic FOGI properties therefore corresponds almost directly to the \textit{commutant} of $\overline{G}_i$\footnote{The intrinsic FOGI space does not directly correspond to the commutant of a gate. Because gates have their top row constrained due to trace preservation,  linear functionals $\Phi_i$ have their left column constrained to all 0. Thus the space of all functionals $\Delta^*_{G_i}$ is a strictly smaller space than the commutant of $G_i$. Appendix \ref{app:intrinsicFOGIdim} rigorously delineates this subspace.}.  One consequence of this is the following.  Suppose that $G_i = \mathcal{E} \overline{G}_i$, where $\mathcal{E}$ is an unknown error process that commutes with $\overline{G}_i$.  Then $\mathcal{E}$ can be reconstructed and identified uniquely just by measuring intrinsic FOGI properties.  This follows from the fact that $\overline{G}_i$'s commutant is a vector space that is self-dual with respect to the Hilbert-Schmidt inner product, and so any element of it (e.g. $\mathcal{E}$) is uniquely identified by its inner products with a finite set of basis elements (each of which is a FOGI property).  So the part of $G_i$'s error process that commutes with $\overline{G}_i$ is FOGI.

\subsubsection{FOGI vs fully gauge-invariant}

Most FOGI properties are not fully gauge-invariant.  But gate sets \textit{do} have fully gauge-invariant properties.  Eigenvalues of individual gates' PTMs are the canonical example, since gauge transformations $G_i \to TGT^{-1}$ are similarity transformations.  Some FOGI properties can be matched up 1:1 with corresponding gauge-invariant properties.  

If a gate $\overline{G}_i$ is diagonalizable and has non-degenerate eigenvalues, then its $d^2$-dimensional commutant is precisely equal to the span of the projectors onto its $d^2$ distinct eigenvectors.  So the intrinsic FOGI properties of $G_i$ (coefficients of error processes that commute with $\overline{G}_i$) are in 1:1 correspondence with its intrinsic \textit{fully} gauge-invariant properties (eigenvalues).  These FOGI properties are linearized gauge-invariant properties -- or, more precisely, the space of intrinsic FOGI properties is simply the tangent space to the manifold of fully gauge-invariant properties.

However, intrinsic FOGI properties cannot always be identified with fully gauge-invariant properties. The correspondence breaks whenever $\overline{G}_i$ has degenerate eigenvalues.  The most extreme example is an idle or identity gate for which $\overline{G}_i = \Id$.  \textit{Every} perturbation commutes with $\overline{G}_i$, so every property of $G_i$ is FOGI.  In contrast, only the eigenvalues of $G_i$ are truly gauge-invariant -- the fact that $G_i$ is close to $\overline{G}_i = \Id$ does not in any way change the mathematical fact that its eigenvectors can be changed by gauge transformations.  So when $\overline{G}_i = \Id$, $G_i$ has $\sim d^4$ intrinsic FOGI properties but (typically) only $\sim d^2$ truly gauge-invariant ones.

This happens because an \textit{approximately gauge-invariant property} is not the same as an \textit{approximation to a gauge-invariant property}.  The eigenvectors of perturbations to $\overline{G}_i = \Id$ \textit{are} approximately invariant under small gauge transformations.  They just don't have any nice properties under arbitrarily large gauge transformations.  We conclude that FOGI and true gauge invariance are distinct, related, and both useful.  In many practical circumstances, we find FOGI to be more operational and useful, because the gauge is effectively ``known'' up to small perturbations.  This is directly analogous to local cartography on Earth's surface, where Cartesian coordinates work very well in practice (even though they can't describe the \textit{entire} globe), as long as their dependence on a specific origin point is accounted for.

\subsection{Relational FOGI properties and generalized commutation}

Equation \ref{eq:GaugeChange} can be used to investigate relational FOGI properties as well as intrinsic ones, although we have not (yet) found this approach to yield concrete useful results.  

Restricting Eq.~\ref{eq:GaugeChange} to just two gate operations $G_1,G_2$ yields the condition

\begin{equation}
    \left[\Phi_1, \overline{G}_1\right] = - \left[\Phi_2, \overline{G}_2\right].
\end{equation}
This equation defines an interesting generalization of commutation, since it is solved not by a single operator $\Phi$ whose commutator with the target gate vanishes, but a pair of operators whose commutators sum to zero.  It can be solved, in principle, by defining the commutation superoperator
\begin{equation}
    [\cdot, G]( \Phi ) \equiv [\Phi,G],
\end{equation}
and inverting it on its support (the orthogonal complement of $G$'s commutant) to get
\begin{equation}
    \Phi_2 = -\left[\cdot,\overline{G}_2\right]^{-1}\circ \left[\cdot,\overline{G}_1\right]( \Phi_1 ).
\end{equation}

If instead we restrict to the state preparation and measurement (SPAM) operations, or to one SPAM operation and one gate operation, we obtain the following equations that pairwise relational FOGI properties must satisfy:
\begin{align}
    \Phi_M \overline{M} - \sketbra{\overline{\rho}}{\Phi_\rho} &= 0 \\
    \left[\Phi_i, \overline{G}_i\right] - \sketbra{\overline{\rho}}{\Phi_\rho} &= 0 \\
    \left[\Phi_i, \overline{G}_i\right] + \Phi_M \overline{M} &= 0 
\end{align}

\subsection{Operations as Reference Frames}
\label{sec:referenceframes}
\begin{figure}[th!]
\centering
\includegraphics[width=8cm]{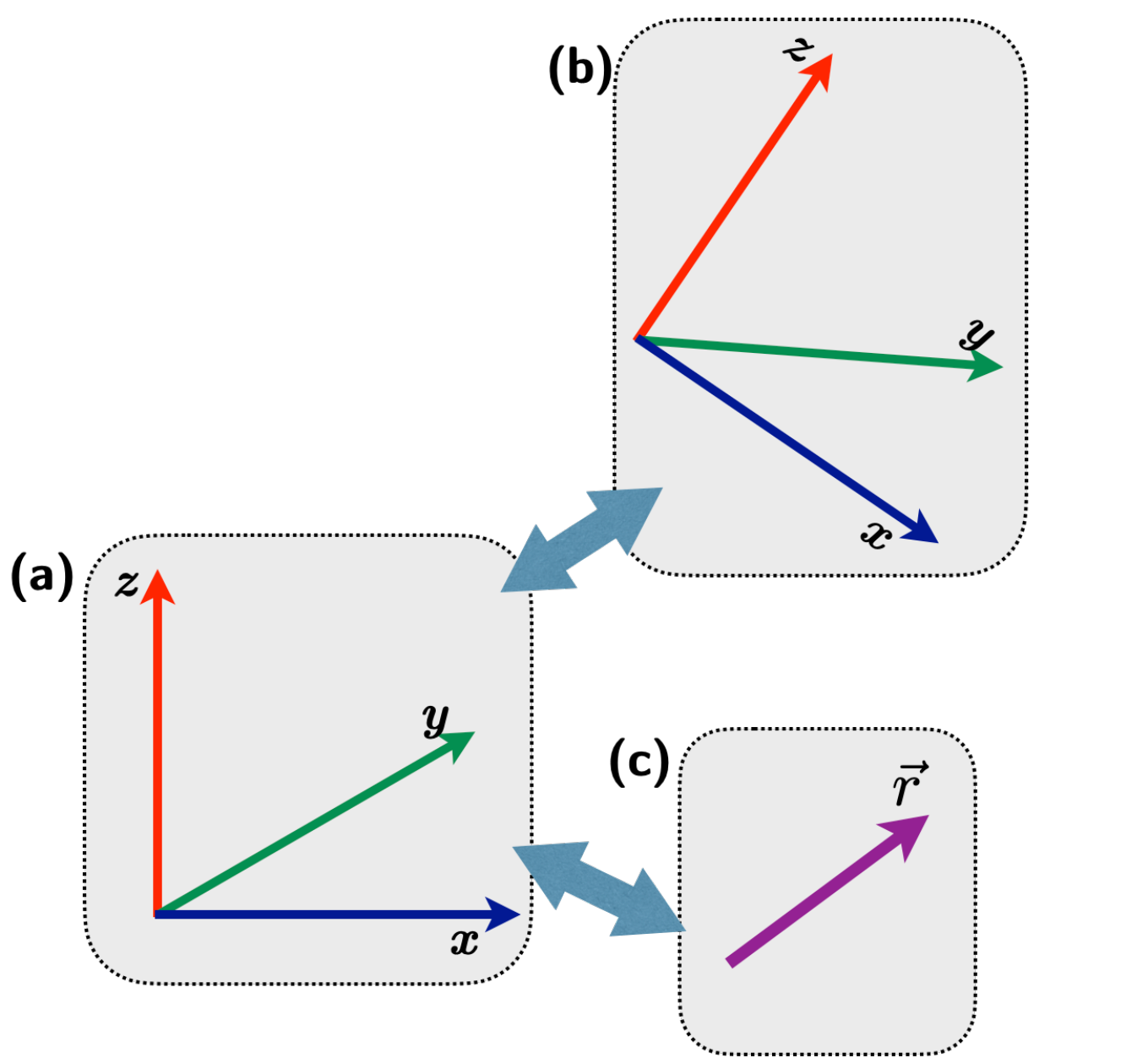}
\caption{An orthogonal basis for $\mathbb{R}^3$ constitutes a reference frame, relative to which other objects can be measured and coordinatized.  Here, panel (a) shows such a frame.  Panel (b) shows another object of the same type.  Panel (c) shows a simpler object, a single vector in $\mathbb{R}^3$.  The coordinates of an object [e.g., (b)]  \textit{relative} to another object that serves as a reference frame [e.g., (a)] are relational properties between the two objects.  The objects in (a) and (b) serve as \textit{complete} reference frames for each other, because their relational properties suffice to detect \textit{any} local gauge change [$\mathrm{SO}(3)$ rotation] on either of them.  But the object in (c) only provides a \textit{partial} reference frame for the object in (a), because it is invariant under rotations around its own axis.}
\label{fig:frame}
\end{figure}

We conclude this section by connecting relational FOGI quantities to the general physical concept of \textit{reference frames} \cite{bartlett2007reference}.
A reference frame (Fig.~\ref{fig:frame}) is a physical or mathematical object used to define the direction, magnitude, or nature of other objects of the same type.  Physical reference frames include meter sticks (used to define length) and the star Polaris (used to define navigational coordinates).  An orthogonal basis for a vector space is an example of a mathematical reference frame.  Generally speaking, when an object is subject to gauge freedom, its coordinates can only be defined \textit{relative} to some reference frame.

We can view each operation $g\in\cG$ as providing a reference frame that can be used to measure a property of \textit{another} operation that would otherwise be gauge-dependent.  This relationship is mutual.  If operation $g_1$ provides a reference frame for certain properties of $g_2$, then $g_2$ also provides a reference frame for $g_1$.  For example, in Fig.~\ref{fig:frame}, the objects in panels (a,b) can each be coordinatized relative to the other, but neither one constitutes an absolute frame of reference.  When we reference $g_1$ against $g_2$, we measure a relational property \textit{between} $g_1$ and $g_2$.  

The mathematics of relational FOGI quantities (Sec.~\ref{sec:relprops}) can be framed in this way.  Recall (Eq.~\ref{eq:ZeroSum}) that relational FOGI properties correspond to lists of gauge vectors $\{\bvec{k}_g\in \mathcal{K}_g \subseteq \mathcal{K}\}$ that add up to zero, where $\mathcal{K}_g = \mathrm{support}(A_g)$, and that pairwise FOGI relational quantities between two operations $g_1,g_2$ correspond to
\begin{equation}
\bvec{k}_1 + \bvec{k}_2 = 0\ \mathrm{with}\ \bvec{k}_i\in\cK_i = \mathrm{support}(A_{g_i}).
\end{equation}
For each operation $g$, $\Gamma_g$ is the space of variations that can be produced by local gauge transformations on $g$.  We call $\Gamma_g$ the \textit{relational space} of $g$.  Properties of $g$ in $\Gamma_g$ cannot be measured in isolation -- but they can be measured \textit{relative} to reference frames provided by other operations.  The pre-image $\cK_g$ of $g$'s relational space, in the space $\cK$ of gauge transformations, defines the set of gauge variations for which that gate can serve as a reference frame.  We refer to $\cK_g$ as $g$'s \textit{reference frame space}.  When two gates share a reference frame -- i.e., there exists $\bvec{k}\in\cK_1\cap\cK_2$ -- their \textit{relative} variations induced by that (local) gauge transformation define a pairwise relational FOGI property.

Operations almost never serve as \textit{complete} reference frames for other operations, because they are typically invariant under a subgroup of gauge transformations corresponding to $\mathrm{kernel}(A_g)$.  An informationally complete measurement operation $M$ with $d^2$ linearly independent effects is the only exception to this rule -- state preparations, gates, and projective measurements are all invariant under some gauge transformation.  The most extreme case is an idle or identity gate, which commutes with \textit{every} gauge transformation and therefore cannot serve as a reference frame for anything ($\cK_\Id = \emptyset$).  In general, reference frames of two operations $g_1$ and $g_2$ may be:
\begin{itemize}
    \item coincident ($\cK_1 = \cK_2$),
    \item independent ($\cK_1 \cap \cK_2 = \emptyset$), or
    \item partially overlapping ($\cK_1 \cap \cK_2 \subset \cK_1 \cup \cK_2$)
\end{itemize}
If their shared reference frame space $\cK_1 \cap \cK_2$ is $d$-dimensional, then $d$ linearly independent pairwise relational properties can be defined between them.  These properties can be interpreted as measuring a $d$-dimensional subspace of $g_1$'s properties \textit{relative to} $g_2$ -- or, they can be interpreted the other way around.

The preceding example involving $G_x$ and $G_y$ gates whose rotation axes are ``tilted'' so that the angle between them is not $\pi/2$ serves as a good illustration.  We can consider just unitary gauge transformations, which are generated by the 3-dimensional Lie algebra $\mathrm{so}(3)$.  Neither $G_x$ nor $G_y$ provides us a reference frame for the \emph{entire} unitary gauge, because $G_x$ is invariant under $X$ rotations and $G_y$ is invariant under $Y$ rotations. $G_x$ provides a 2-dimensional reference frame for $Y$- and $Z$-rotations, while $G_y$ provides a 2-dimensional reference frame for $X$- and $Z$-rotations.  Their intersection is 1-dimensional, containing just rotations around $Z$.  Thus, the \emph{difference} between local $Z$-axis gauge rotations on $G_x$ and $G_y$ defines a gauge-invariant property.  The angle between their rotation axes is fully invariant (not just FOGI) under unitary gauge transformations, and the corresponding FOGI property is the linearized error in that angle, evaluated at $\overline{\cG}$.

The reference frame paradigm also helps illuminate $n$-wise relational properties.  Although the gauge action on a single operation $g$ is almost never faithful because $A_g$ has a nontrivial kernel, its action on a \textit{set} of operations can easily be faithful. Such a set provides a \textit{faithful reference frame}, against which \textit{any} property in a gate $g$'s relational space can be referenced to define a relational FOGI property. In contrast to pairwise FOGI properties, irreducible \(n\)-wise relational properties arise not from a single shared direction, but from a nontrivial closed polygonal chain of reference-frame directions in gauge space. Concretely, they correspond to sets of nonzero vectors \(\bvec{k}_i \in \cK_{g_i}\) such that
\[
\sum_{g\in\cG} \bvec{k}_g = 0,
\]
where this relation cannot be decomposed into a sum of lower-order relations. Geometrically, the vectors \(\{\bvec{k}_i\}\) form the edges of a closed polygonal chain in gauge space, with the \(i\)-th edge drawn from the non-overlapping portion of the reference-frame space of the \(i\)-th operation. The trine example of Fig.\ref{fig:trine} illustrates the simplest such case: three one-dimensional reference-frame spaces in a two-dimensional gauge space that share no common pairwise direction, yet form a closed triangle and therefore define an irreducible 3-wise relational property.

\section{A canonical (inner product free) approach to FOGI properties}
\label{canonical}
In Sec.~\ref{GateSets} we introduced several vector spaces and then, in Assumption~\ref{ass:innerprod}, equipped them with an inner product. Choosing an inner product amounts to choosing how to compare and weight distinct error directions. Since there is no natural way to do this in full generality, this choice is not canonical. We made it purely for convenience and to enable numerical calculations. It significantly simplified the analysis underlying the previous results. In particular, we could not have partitioned FOGI space into orthogonal subspaces (e.g. intrinsic and relational FOGIs) without an inner product to define orthogonality, nor could we have made a connection between gauge-generator vectors and relational FOGI properties

Nevertheless, the choice of inner product is arbitrary, and many of the key ideas presented so far admit canonical, inner-product-independent formulations. These may be useful to readers interested in applying FOGI methods across multiple inner-product choices; other readers may wish to skip this section on a first pass. Here, we present a canonical characterization of FOGI space and a canonical taxonomy of FOGI subspaces.

\subsection{Canonical FOGI properties}

Lemma \ref{lem:FOGI} states that a property $\varphi(\delta)$ is FOGI if and only if $\varphi\left(\mathcal{A}_\cG(K)\right) = 0$, for all matrices $K\in\cK$.  In other words, a property is FOGI iff it \emph{annihilates} every vector in the range of $\cA_\cG$, which we denote $\Gamma_\cG$, and refer to it as \emph{gauge error space}. The space of linear functionals $\varphi$ that map a vector subspace $\Gamma_\cG$ to 0 is called its \emph{annihilator}, denoted $\Gamma_\cG^0$.
Thus, the FOGI space of a gate set ($\cF_\cG$) is the subspace of $\Delta^*$ that annihilates the gauge space, i.e.

\begin{equation}
    \cF_\cG  = \Gamma_\cG^0 = \left\{ \varphi \in \Delta^* : \varphi\left(\mathcal{A}_\cG(K)\right) = 0, \forall K \in \mathcal{K} \right\}.
\end{equation}
Furthermore, this can be re-written in terms of the gauge action's dual map $\mathcal{A}_\cG^*:\Delta^{*}\rightarrow\mathcal{K^*}$ (where $\mathcal{K}^*$ is the space of linear functionals on gauge generators, i.e. \emph{the space of dual gauge generators}) defined as
\begin{equation}\label{eq:dualmap}
    \left(\mathcal{A}_\cG^*(\varphi)\right)(K) :=\varphi\left(\mathcal{A}_\cG(K)\right),
\end{equation}

\noindent by noticing the following equivalent conditions

\begin{align}
    &\varphi\left(\mathcal{A}_\cG(K)\right) = 0, \forall K \\&\Leftrightarrow \left(\mathcal{A}_\cG^*(\varphi)\right)(K) = 0, \forall K \\ & \Leftrightarrow \mathcal{A}_\cG^*(\varphi)=0
    \\ &\Leftrightarrow \varphi \in  \mathrm{kernel}\left(\mathcal{A}_\cG^*\right).
\end{align}

\begin{figure*}[t!]
\centering
\includegraphics[width=16cm]{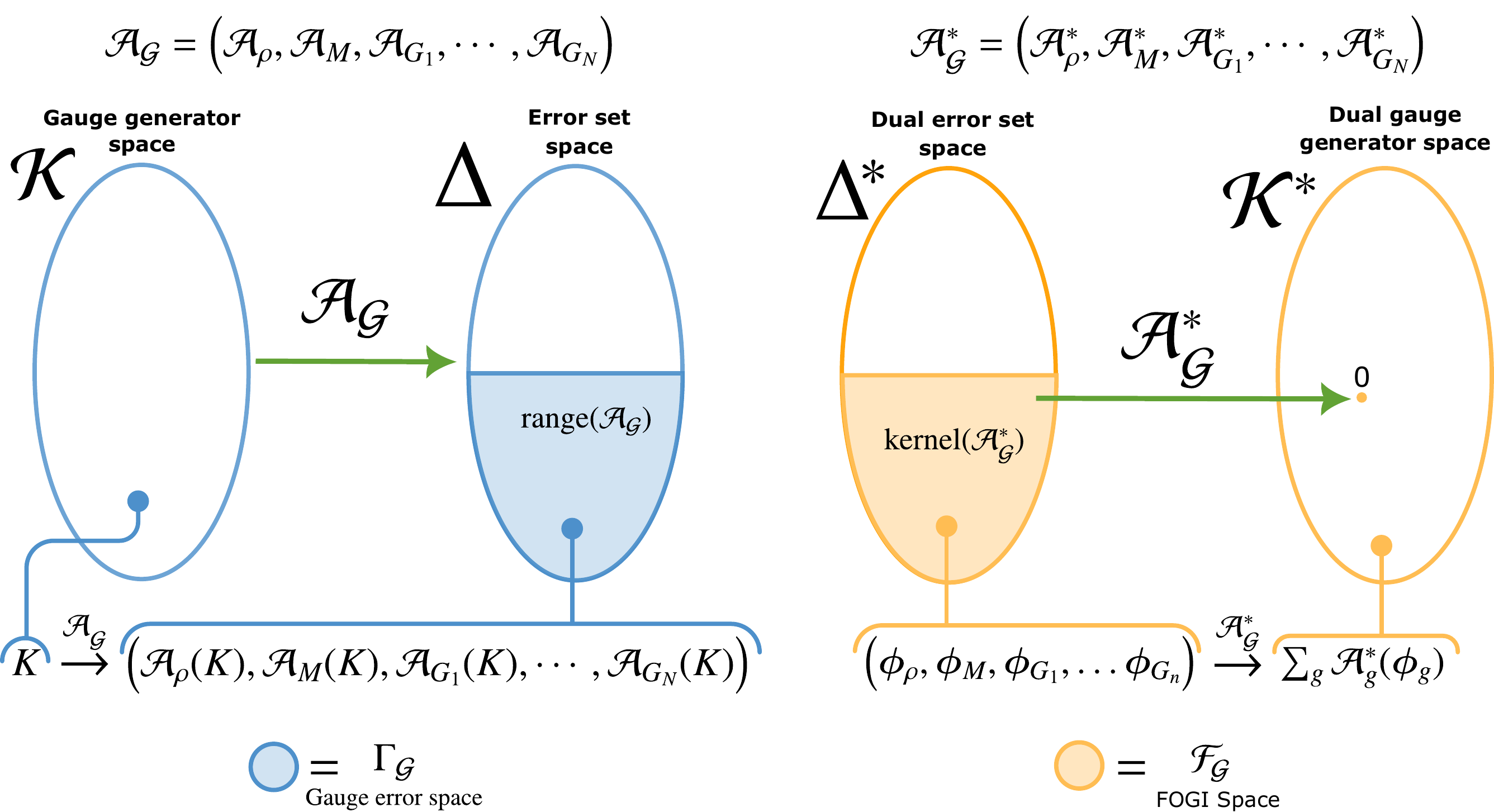}
\caption{The gauge action map $\cA_\cG$ captures the affine shift in gate set errors produced by gauge generators. It takes gauge generator matrices $K$ to error tuples $\delta$. Its range is solely composed of error directions which are physically unobservable (to first order), and thus we call it the \emph{gauge error space} $\Gamma_\cG$. Its dual map $\cA^*_\cG$ maps property tuples into gauge generator functional matrices. The kernel of the dual map is the FOGI space of the gate set.}
\label{fig:canonical_spaces}
\end{figure*}

\noindent Therefore, as illustrated in Fig. \ref{fig:canonical_spaces}, and analogous to Lemma \ref{lem:kernel}, the (canonical) space of FOGI properties of a gate set is
\begin{equation}\label{eq:FOGIspace}
    \boxed{\cF_\cG = \Gamma_\cG^{0} = \mathrm{kernel}\left(\mathcal{A}_\cG^*\right) \subseteq \Delta^*.}
\end{equation}

\subsection{Canonical FOGI subspaces}

Recall that in Eq. (\ref{eq:Atuple}) we decomposed the gauge action map into operator-wise maps

\begin{equation}
    \mathcal{A}_{\cG} = \left(\mathcal{A}_{\rho} , \mathcal{A}_{M}  , \mathcal{A}_{G_1} , \cdots , \mathcal{A}_{G_N} \right).\notag
\end{equation}
 Recall from Eq. (\ref{eq:propsum}) that any property can be expressed as a sum of intrinsic properties $\varphi_g\in\Delta_g^*$ 
\begin{equation}
    \varphi(\delta) = \sum_{g\in\cG}\varphi_g(\delta_g).\notag
\end{equation}
Thus, the action of a property on the output of the gauge action map can also be decomposed into a sum of operator-wise elements
\begin{align}
    \varphi(\cA_\cG(K))
     = \sum_g\varphi_g(\cA_g(K)).\label{eq:gaugeActionDecomp}
\end{align}
\noindent The dual map $\cA_\cG^*:\Delta^*\to\cK^*$, which acts on a property ($\varphi$) and outputs a linear functional on gauge generators, was defined as the left hand of the expression above (Eq. (\ref{eq:dualmap})), therefore \begin{equation}
    \left(\mathcal{A}_\cG^*(\varphi)\right)(K) =\varphi(\mathcal{A}_\cG(K))= \sum_g\varphi_g(\cA_g(K)).\notag\end{equation} From this, we observe that the dual map $\cA^*$ can also be decomposed into operator-wise elements defined as

\begin{equation}
    \left(\cA_g^*(\varphi_g)\right)(K) := \varphi_g(\cA_g(K)).
\end{equation}
Thus, we can represent gate set dual maps by the sum of operator-wise dual maps
\begin{equation}\label{eq:sumduals}
    \cA_\cG^*(\varphi)
    =
    \sum_g \cA^*_g (\varphi_g).
\end{equation}

\noindent This representation allows us to isolate FOGI properties based on which operator-subspaces are involved. For example, because intrinsic properties as defined in Definition \ref{def:intrinsicFOGI} do not rely on orthogonality we can describe them in terms of the intrinsic dual map $\cA_g^*$ as

\begin{equation}
\cF_g = \mathrm{kernel}\left(\mathcal{A}_g^*\right)\subseteq \Delta^*_g.
\end{equation}

Unfortunately, all of the remaining subspaces defined in Sec. \ref{Taxonomy} do not have a satisfying canonical version since they rely on orthogonality with respect to other subspaces. Relational FOGI properties complement intrinsic properties to span the entire FOGI space

\begin{equation}\label{eq:partition}
    \cF_\cG = \left(\bigoplus_{g\in\cG}\cF_g \right) + \cF^R,
\end{equation}

\noindent but because we can no longer define them as orthogonal to intrinsic properties, $\cF^R$ is not unique. The quotient space $\cF_\cG/\left(\bigoplus_{g\in\cG}\cF_g \right)$ defines the equivalence classes of all relational properties, i.e. every element of this space is a set of properties. Choosing a property to represent each class/element of a quotient space is called choosing a \emph{section}. Any section of $\cF_\cG/\left(\bigoplus_{g\in\cG}\cF_g \right)$ defines a different vector space $\cF^R$ which satisfies Eq. (\ref{eq:partition}). But when an inner product is present, we can naturally define a section of $\cF_\cG/\left(\bigoplus_{g\in\cG}\cF_g \right)$ whose elements are orthogonal to all intrinsic properties. The same analysis is true for all n-wise irreducible FOGI subspaces.

Under the assumption of an inner product, we presented a correspondence between relational FOGI properties and gauge generators. Let us revisit the same derivation canonically. First, recall from Eq. (\ref{eq:tuple}), that any $\varphi$ can be written as a tuple of intrinsic properties

\begin{equation}
\varphi=\left(\varphi_\rho, \varphi_M, \varphi_{G_1}, \ldots,\varphi_{G_N}\right).\notag
\end{equation}

\noindent Because a relational FOGI property is not contained in the span of the intrinsic FOGI subspaces, then some of the intrinsic properties $\varphi_g$ that make up a relational property must not be FOGI themselves ($\varphi_g \notin \mathrm{kernel}\left(\cA_g^*\right)$). By the definition of the kernel, these intrinsic properties satisfy $\cA_g^*(\varphi_g) \neq0$. We collect these non-zero values into a set that represents a property's \emph{inherently relational content}:

\begin{equation}
R(\varphi) = \left\{r_g = \cA_g^*(\varphi_g) \mid r_g \neq 0 \right\}.\end{equation}

\noindent By definition every $r_g \in R(\varphi)$ lives in the range of $\cA_g^*$. We denote the range of $\cA_g^*$ as $\cK_g^*$, since it is a subspace of $\cK^*$. The existence of a non-trivial $\cK_g^*$ is what allows its corresponding operation $g$ to have relational FOGI properties. This is the canonical version of an operation's reference frame introduced in Sec. \ref{sec:referenceframes}, so we call this subspace ($\cK^*_g$) the operation's \emph{canonical reference frame}.

Any FOGI property must satisfy $\mathcal{A}_\mathcal{G}^*(\varphi)=0$.  Plugging this into Eq. (\ref{eq:sumduals}) yields an analogous equation to Eq. (\ref{eq:ZeroSum}):
\begin{equation}\label{eq:canonrelcon}
\sum_{g\in \mathcal{G}}\mathcal{A}_g^*(\varphi_g)
=0.
\end{equation}
As we just observed, every element in this sum must either be 0, or part of the property's relational content, and  thus Eq. (\ref{eq:canonrelcon}) simplifies to 
\begin{equation}\label{eq:ZeroSumCanon}
\sum_{r_g\in R(\varphi)}r_g=0. 
\end{equation}
So the space of relational FOGI properties of $\mathcal{G}$ is isomorphic to the space of solutions to Eq. (\ref{eq:ZeroSumCanon}), subject to $r_g\in\mathcal{K}_g^*$. Unlike the correspondence under an inner product assumption, a set of $r_g$'s that satisfy Eq. (\ref{eq:ZeroSumCanon}) has no canonical mapping to a single relational FOGI property.  What used to be an isomorphism between properties ($\Delta^*$) and gauge generators ($\cK$), is instead (canonically) a many-to-one map between properties ($\Delta^*$) and gauge generator functionals ($\cK^*$).

\section{FOGI Properties with Elementary Error Generators} \label{EEG}
Doing actual calculations of and with FOGI properties requires choosing a concrete basis for error set space $\Delta$.  The basis of \textit{elementary error generators} \cite{Blume-Kohout2022-ln} is a particularly convenient choice. Here, we show how to use the basis of elementary error generators to write down concrete, interpretable FOGI properties/errors for specific gate sets.  Throughout this section we will use the Pauli matrices $\{\Id, \sigma_x, \sigma_y, \sigma_z \}$ to form a basis for Hilbert-Schmidt space for one qubit.

\subsection{The elementary error generator basis}
\label{sec:eeggates}
One way to parameterize gate sets explicitly is to use the individual matrix elements of the PTMs for $\rho$, $M$, and the $\{G_i\}$. But those matrix elements do not typically have a clear physical meaning. It is more useful to represent each operation $g$ by its \emph{error generator} $L_g$, as
\begin{eqnarray}
\label{eq:postgate}
  G_i &=& e^{L_{G_i}} \overline{G}_i \nonumber\\
  \rho &=& e^{L_\rho} \overline{\rho} \label{eq:errorgenprimitives}\\
  M &=&  \overline{M} e^{L_M}. \nonumber
\end{eqnarray}

\noindent If we assume our errors are small (Assumption \ref{ass:epsilon}), and thus constrain noisy operations to be within an $\epsilon$-neighborhood of their targets (see Definition \ref{def:neighborhood}) then
\begin{equation}
    e^{L_g}=\Id + L_g + O\left(\epsilon^2\right).
\end{equation}

\noindent We parameterize each error generator $L_g$ as a linear combination of $d^2(d^2-1)$ \emph{elementary error generators} $\{\hat{L}_i\}$ \cite{Blume-Kohout2022-ln}, 

\begin{equation}
  L_g = \sum_{i=1}^{d^2(d^2-1)} \epsilon_{i}(g) \hat{L}_{i}. \label{eq:errorgenlinearcombo}
\end{equation}

\noindent Each $\epsilon_i(g)$ is the coefficient or \textit{error rate} of the elementary error generator $\hat{L}_i$ on operation $g$.

The elementary error generators are categorized into three classes of distinct error mechanisms, spanned by four distinct sectors/subspaces $(\mathbb{H}, \mathbb{S}, \mathbb{C} \textrm{ and } \mathbb{A})$: 
\begin{enumerate}
    \item \textbf{Hamiltonian} elementary generators ($H_P\in \mathbb{H}$) are indexed by a single Pauli $P$ and generate unitary errors.
    \item \textbf{Stochastic} elementary generators generate random-unitary errors.  They are divided into (i) \textbf{Pauli-stochastic} generators ($S_P\in\mathbb{S}$) indexed by a single Pauli, and (ii) \textbf{Pauli-correlation} generators ($C_{P,Q}\in\mathbb{C}$) indexed by a pair of distinct Paulis.
    \item \textbf{Active} elementary generators ($A_{P,Q} \in \mathbb{A}$) generate error processes that require feedback from the environment, including non-unital affine shifts (as observed in amplitude damping).  They are indexed by a pair of distinct Pauli basis elements.
\end{enumerate}

\noindent Extracting the rate of a specific elementary error generator ($\epsilon_i$) from a gate's error ($L_{G_j}$) is given by the inner product with the corresponding \emph{dual elementary error generator} $\hat{L}^*_{i}$\cite{Blume-Kohout2022-ln}, 

\begin{equation}
\epsilon_i(G_j) = \bvec{\hat{L}}_{i}^{*}\cdot \bvec{L_{G_j}} = \Tr\left[ \hat{L}^{*\dagger}_{i} L_{G_j} \right].
\end{equation}

\noindent We use lowercase characters to differentiate error rates from their corresponding (upper-case) elementary error generator:

\begin{itemize}
    \item $\mathrm{rate}(H_P) = h_P$
    \item $\mathrm{rate}(S_P) = s_P$
    \item $\mathrm{rate}(C_{P,Q}) = c_{P,Q}$
    \item $\mathrm{rate}(A_{P,Q}) = a_{P,Q}$
\end{itemize}
 For example, a single qubit error generator is completely described by a list of 12 rates or coefficients:
\begin{align}
    \epsilon_i \in \{&h_X, h_Y, h_Z, s_X, s_Y, s_Z, c_{X,Y}, c_{X,Z}, c_{Y,Z},\notag \\ 
        &a_{X,Y}, a_{X,Z}, a_{Y,Z}\}.
\end{align}

In Eq. (\ref{eq:error}) we defined gate errors as  $\delta_{G_j} = G_j - \overline{G}_j$. If following Eq. \eqref{eq:postgate}, we represent a gate's error by a post-gate error channel, then $\delta_{G_j}$ becomes
\begin{align}
    \delta_{G_j} &= G_j - \overline{G}_j = e^{L_{G_j}}\overline{G}_j - \overline{G}_j  \\&=L_{G_j}\overline{G}_j + O\left(\epsilon^2\right)
\end{align}
To use basis elements $\hat{L}_i$ to parameterize errors instead of $\hat{L}_i\overline{G}_j$, we perform a change of coordinates where all gate errors $\delta_{G_j} \in \Delta_{G_j}$ become

\begin{align}\label{eq:coordchange1}
    &\delta_{G_j} \rightarrow \delta_{G_j}\overline{G}_j^\dagger = L_{G_j},
\end{align}
and similarly properties $\Phi_j \in \Delta^*_{G_j}$ become
 \begin{equation}\label{eq:coordchange2}
     \Phi_{j} \rightarrow\overline{G_j}\Phi_{j}.
 \end{equation}

This preserves the value of any linear functional $\varphi(\delta)$. To see this, recall that in Eq. (\ref{eq:PTMproperty}) we showed that any property intrinsic to a gate $G_i$ can be written as $\Tr(\Phi_iG_i)$ for some $\Phi_i$, and thus

\begin{align}
    \varphi(\delta_{G_j})=\Tr\left(\Phi_j\delta_{G_j}\right) \rightarrow \Tr\left(\overline{G}_j\Phi_j\delta_{G_j}\overline{G}_j^\dagger\right).
\end{align}

With this change of coordinates we are able to use the elements $\hat{\bvec{L}}_i$ as a basis for the \textit{gate} portions of error set space $\Delta$, wherein the elements of $\bvec{\delta}_{G_j} \in \Delta_{G_j}$ are the $d^2(d^2-1)$ coefficients $\epsilon_i(G_j)$ (error rates); and linear functionals $\varphi(\delta_{G_j})$ are linear combinations of those coefficients (error rates).

Furthermore, the gauge action map for gates $\mathcal{A}_{G_j}$, defined in Eq. \eqref{eq:ALinMap} as $[K,\overline{G}_j]$, in these coordinates is given by 

\begin{align}
        \mathcal{A}_{G_j}(K)=\left[K,\overline{G}_j\right]\overline{G}_j^\dagger= K - \overline{G}_jK\overline{G}_j^\dagger\label{eq:eeggaugeaction},
\end{align}

\noindent where $K$ is a gauge generator as defined in \eqref{eq:gaugegenerator}. Any gauge generator can be written as a linear combination of elementary error generators with coefficients $k_i$,

\begin{equation}
    K = \sum_i{k_i \hat{L}_i}.
\end{equation}

\noindent The entries of the gauge action map from input gauge generators to output gauge directions are given by 

\begin{align}
    (A_{G_k})_{i,j} &= \Tr\left(\hat{L}_j^{*\dag} A_{G_k}\left(\hat{L}_i\right)\right),\\
    & = \Tr\left(\hat{L}_j^{*\dag}\left( \hat{L}_i-\overline{G}_k\hat{L}_i\overline{G}_k^\dagger \right)\right).
    \label{eq:eeggaugeactionelements}
\end{align}

\subsection{SPAM-independent FOGI properties can be isolated to classes}
\label{sec:gateonlyFOGI}

A major advantage of the elementary error generator representation is that it divides errors into three \textit{classes} -- Hamiltonian $(\mathbb{H})$, stochastic ($\mathbb{S} \oplus \mathbb{C}$), and active $(\mathbb{A})$ -- that represent distinct kinds of physical error process, and can often be analyzed separately.  The error associated with a gate $g$ generally spans all classes, but each element of the \textit{basis} of elementary error generators belongs to one and only one \emph{sector} ($\mathbb{H},\mathbb{S},\mathbb{C}$ or $\mathbb{A}$). The error generator classes are closed under unitary conjugation, so $\overline{G} L \overline{G}^\dagger$ lies in the same class as $L$, and the same is true of the corresponding classes of the dual elementary error generators \cite{Blume-Kohout2022-ln}.

Happily, FOGI properties of the \textit{gates} can be constructed to respect these classes, because the gauge action on gates does not couple them.  Recall from Eq. (\ref{eq:eeggaugeaction}) that a small gauge transformation $T = e^{\epsilon K}$ simply adds
\begin{equation}
    L_K = \epsilon\left( K - \overline{G}_i K \overline{G}_i^\dagger \right)
\end{equation}
to gate $G_i$'s error generator.  Now decompose the gauge generator by classes as $K = K_\mathbb{H} + K_\mathbb{SC} + K_\mathbb{A}$. Each component is itself a gauge generator, since the elementary error generators span $\cK$.  $L_K$ is linear in $K$, and each component's contribution $L_{K_X} = \epsilon\left( K_X - \overline{G}_i K_X \overline{G}_i^\dagger \right)$ is the difference between $K_X$ and a unitarily conjugated $K_X$, so it lies entirely in class $X$.  Therefore gauge transformations shift each class of the gates' error rates independently, so the $\mathbb{H}$ rates are shifted only by $K_\mathbb{H}$, and similarly for $\mathbb{S}\oplus\mathbb{C}$ and $\mathbb{A}$.

This block structure isolates FOGI properties by class. A property $\varphi$ supported only on the gates is a linear combination of the gates' error rates, and so splits by class as $\varphi = \varphi_\mathbb{H} + \varphi_\mathbb{SC} + \varphi_\mathbb{A}$. Since gauge shifts in one class change only that class' rates, $\varphi_Y\left(\mathcal{A}_\cG(K_X)\right) = 0$ whenever $Y \neq X$. Therefore, for any given FOGI property $\varphi$,
\begin{align}
    &\varphi(\cA(K_X))=0, \forall K_X \\&\implies \varphi_X(\cA(K_X))=0, \forall K_X\\
    &\implies \varphi_X(\cA(K))=0 ,\forall K\\
    &\implies \varphi_X \text{ is FOGI.}
\end{align}

\begin{lemma}\label{lem:sectorDecomposition}
    Any FOGI property whose support solely contains gate operations can be decomposed into FOGI properties which are each isolated to a single class $\mathbb{H}$, $\mathbb{S} \oplus \mathbb{C}$, or $\mathbb{A}$.
\end{lemma}

\noindent Consequently, a FOGI basis for the gate-only properties can be chosen with every element confined to a single sector, and the analysis of the gauge action splits into three smaller, independent linear problems.  This is valuable not so much because it reduces the size of the linear systems to be solved, but because it preserves the physical intuition behind the elementary error generators. Unfortunately, SPAM operations break this symmetry, and relational properties that involve SPAM operations generally cannot be restricted to a single sector.

We note that the $\mathbb{S}$ and $\mathbb{C}$ classes, while not closed under general unitary conjugation, are separately closed under Clifford conjugation \cite{Blume-Kohout2022-ln}. So for Clifford gate sets the same argument yields a finer decomposition:

\begin{corollary}\label{cor:cliffordSectors}
Any FOGI property whose support solely contains Clifford gate operations can be decomposed into FOGI properties which are each isolated to a single class of elementary error generators $\mathbb{H}$, $\mathbb{S}$, $\mathbb{C}$, or $\mathbb{A}$.
\end{corollary}
\noindent These results mean that FOGI properties of \textit{gates}, especially for Clifford gates, can be made relatively simple and elegant.  

\subsection{Error generators for SPAM operations}
\label{sec:eegSPAM}
For reversible logic gates $G_j$ described by $d^2\times d^2$ TP superoperators, the $d^2(d^2-1)$ elementary error generator rates $\epsilon_i(G_j)$ provide a convenient parameterization of $\delta_{G_j} = G_j - \overline{G}_j$.  But for the state preparation ($\rho$) and measurement ($M$) operations, $L_\rho$ and $L_M$ have too many free parameters.  Some linear combinations of elementary error generators in $L_\rho$ and $L_M$ have no effect on $\rho$ and $M$. For example, consider the pure state $\sket{0}$ which lies aligned with the $Z$-axis. $\sket{0}$ is invariant under rotations about the $Z$-axis, and thus the state preparation operation is the same for all values of the $h_Z$ error rate in $L_\rho$. Moreover, different error generators can produce identical observable effects on $\rho$. A depolarizing channel results in the shrinking of its $\rho_z$ component . Similarly, an amplitude damping error channel along the Z-axis can transform the state in the exact same way by also only shrinking its $\rho_z$ component. These two error channels are generated by \emph{distinct} elementary error generators -- $S_X+S_Y+S_Z \text{ and } A_{X,Y} $ -- yet they both yield the same state. So, any given $\overline{\rho}$ is sensitive to a linear combination of some generators, but completely insensitive to others. The space spanned by generators our operations are insensitive to is called the \emph{irrelevant} space of errors.

We eliminate the irrelevant space by restricting each $L_g$ to the \textit{support} of the linear maps
\begin{align}
L_\rho \to \sket{\delta_\rho} &= L_\rho\sket{\overline{\rho}} + O\left(\epsilon^2\right),\\
L_M \to \delta_M &= \overline{M} L_M + O\left(\epsilon^2\right).
\end{align}
We call the support of this mapping the \textit{relevant} space for $\rho$ or $M$.  Relevant error generator rates uniquely parameterize noisy SPAM operations.

Unfortunately, the relevant spaces for $\rho$ and $M$ generally do not inherit the basis of elementary error generators, because each elementary error generator is a linear combination of relevant and irrelevant generators.  Thus, we need to construct new bases for $\rho$ and $M$'s relevant spaces.  To do so, we construct one basis element for each matrix element of the operation's PTM, as the linear combination of error generators that vary that matrix element only.

This is straightforward for $\rho$, but to parameterize $M$ most effectively, we need to change its basis slightly.  We introduced $M$ as a matrix whose rows are the Pauli-basis effects $\sbra{E}$ of a POVM.  In this representation, $M$ transforms vectors in the Pauli-basis to probability vectors over the measurement's classical output. For simplicity we focus the rest of our analysis on single qubit gate sets, yet everything presented here can be generalized to any number of qudits.

First, to make the analysis more elegant, we identify these probability vectors with diagonal density matrices, e.g.

\begin{equation}
    \left(\begin{array}{c} p_0 \\ p_1 \\ p_2 \\ p_3 \end{array}\right)
    \to
    \left(\begin{array}{cccc} p_0 & 0 & 0 & 0 \\ 0 & p_1 & 0 & 0 \\ 0 & 0 & p_2 & 0 \\ 0 & 0 & 0 & p_3 \end{array}\right),
\end{equation}

\noindent and then (assuming the number of measurement outcomes is a power of 2), represent this ``classical'' density matrix in the basis of Pauli $Z$-type operators (e.g. for two qubits $\{\Id, \Id\otimes Z, Z \otimes \Id, Z \otimes Z\}$).  In this representation, $M$ is a true PTM that maps vectors in the Pauli-basis to $Z$-type Pauli-basis vectors.  

To illustrate this change of basis, observe that a single-qubit $Z$ measurement would have originally been represented by
\begin{equation}
    \overline{M }_Z=\begin{pmatrix}
        \sbra{0}\\ \sbra{1}
    \end{pmatrix} =\left(\begin{array}{cccc}1 & 0 & 0 & 1 \\ 1 & 0 & 0 & -1\end{array}\right),
\end{equation}
but in the new basis becomes
\begin{equation}
    \overline{M}_Z = \begin{pmatrix}
        \sbra{0}+ \sbra{1}\\
        \sbra{0}- \sbra{1}
    \end{pmatrix}= \left(\begin{array}{cccc}2 & 0 & 0 & 0 \\ 0 & 0 & 0 & 2\end{array}\right).
\end{equation}

As an example, consider a target gate set with single-qubit SPAM operations $\overline{\rho} = \sket{0}$ and POVM effects $\sbra{\overline{E}_0} = \sbra{0}, \text{ and }\sbra{\overline{E}_1} = \sbra{1}$.  The noisy operations, as Pauli transfer matrices, are
\begin{equation}\label{eq:deltarho}
\sket{\rho} = \sket{\overline{\rho}} + \sket{\delta_\rho} = \frac12\left(\begin{array}{c}1\\0\\0\\1\end{array}\right) + \left(\begin{array}{c}0\\\rho_x\\\rho_y\\\rho_z\end{array}\right)
\end{equation}
and
\begin{equation}\label{eq:deltaM}
M = \overline{M} + \delta_M = \left(\begin{array}{cccc}2 & 0 & 0 & 0 \\ 0 & 0 & 0 & 2\end{array}\right) + \left(\begin{array}{ccccc}0&0&0&0\\M_1&M_x&M_y&M_z\end{array}\right).
\end{equation}
Both noisy operations have their top rows fixed due to the probability constraints $\sbraket{\Id}{\rho} = 1/2$ and $\sbra{E_0} + \sbra{E_1} = \sbra{\Id}$. Consequently, $\sket{\rho}$ has 3 free matrix elements that can be parameterized by $\rho_x,\rho_y,\rho_z$, and $M$ has 4 free matrix elements that can be parameterized by $M_1,M_x,M_y,M_z$.  We will work out the error generators that correspond to these matrix elements. Since $\rho$ has three degrees of freedom, it is sensitive to a 3-dimensional space of relevant error generators. 
This implies the existence of subspaces of the error generators that form bases for $M_Z$ and $\rho$ with 4 and 3 elements respectively;

\begin{align}
    &L_{\rho} = \epsilon_x(\rho)\hat{L}^\rho_x+ \epsilon_y(\rho)\hat{L}^\rho_y + \epsilon_z(\rho)\hat{L}^\rho_z, \text{and}\\
    &L_M = \epsilon_1(M)\hat{L}^M_1 + \epsilon_x(M)\hat{L}^M_x + \epsilon_y(M)\hat{L}^M_y+ \epsilon_z(M)\hat{L}^M_z.
\end{align}
As previously motivated, we choose these basis elements such that they vary a single vector or matrix element of their target operations. In other words,

\begin{align}
    &\epsilon_i(\rho)\hat{L}^\rho_i\sket{\overline{\rho}} = \epsilon_i(\rho)\sket{\sigma_i},
\end{align}

\noindent where $\{\Id, \sigma_x, \sigma_y, \sigma_z \}$ are the single-qubit Pauli matrices. In this convention, the elements of $\sket{\delta_\rho}$ are the coefficients $\epsilon_i$ of the relevant error generators, and thus constitute a basis for state preparation errors
\begin{align}
    \sket{\delta_\rho} &= \left(\epsilon_x(\rho)\hat{L}^\rho_x + \epsilon_y(\rho)\hat{L}^\rho_y + \epsilon_z(\rho)\hat{L}^\rho_z\right)\sket{\overline{\rho}}\\ &=\begin{pmatrix}0\\\epsilon_x(\rho)\\\epsilon_y(\rho)\\ \epsilon_z(\rho)\end{pmatrix}.
\end{align}
With some linear algebra we find these elements to be 
\begin{eqnarray}
\hat{L}^\rho_x &=& {H}_Y + {C}_{X,Z} - 2{A}_{Y,Z}\label{rhobasisstart} \\
\hat{L}^\rho_y &=& -{H}_X + {C}_{Y,Z} + 2{A}_{X,Z} \\
\hat{L}^\rho_z &=& -{S}_X - {S}_Y - 2{A}_{X,Y},\label{rhobasisend}
\end{eqnarray}
The same analysis for $\delta_M$ yields a 4-dimensional space of relevant error generators,
\begin{eqnarray}
\hat{L}^M_1 &=& -8{A}_{X,Y} \label{Mbasisstart}\\
\hat{L}^M_x &=&  4({C}_{X,Z} - {H}_Y)\\
\hat{L}^M_y &=&  4({C}_{Y,Z} + {H}_X)\\
\hat{L}^M_z &=& -4({S}_X + {S}_Y).\label{Mbasisend}
\end{eqnarray}

We can therefore write noisy SPAM operations as
\begin{align}
    \rho &= \sket{\overline{\rho}} + L^\rho\sket{\overline{\rho}} + O\left(\epsilon^2\right), \\
    M_Z &= \overline{M}_Z + \overline{M}L^M + O\left(\epsilon^2\right),
\end{align}
where $L^{\rho}=\sum_i\epsilon_i(\rho)\hat{L}^{\rho} _i$ and $L^M=\sum_i\epsilon_i(M)\hat{L}^{M} _i$.

Due to our choice of basis, the dual generator basis is simply the dual Pauli basis, 
\begin{align}\label{eq:spameegduals1}
    & \sbra{\hat{\rho}_i^*}\in\{\sbra{\sigma_x},\sbra{\sigma_y}, \sbra{\sigma_z}\}, \\
    & \hat{M}_i^* \in\{\ket{1}\sbra{\Id}, \ket{1}\sbra{\sigma_x}, \ket{1}\sbra{\sigma_y}, \ket{1}\sbra{\sigma_z} \}.\label{eq:spameegduals2}
\end{align}

The rate of each elementary error generator on a SPAM operation $\rho/M$ is given by the inner product between the error $\sket{\delta_\rho}/\delta_M$ and a dual generator $\sbra{\hat{\rho}_i}/\hat{M}_i^*$, as
\begin{align}
\epsilon_i(\rho) &= \bvec{\hat{\rho}}_i^*\cdot \bvec{\delta_\rho} =\sbraket{\sigma_i}{\delta_\rho} \\
&=\sbra{\sigma_i}L_{\rho}\sket{\overline{\rho}}, \text{ and}\\
\epsilon_i(M) &= \bvec{\hat{M}}_{i}^*\cdot \bvec{\delta_M} = \Tr[ \hat{M}_i^{*\dagger}\delta_M ]\\
& = \bra{1}\overline{M}L_{M}\sket{\sigma_i}.
\end{align}

\noindent This defines a basis for the \textit{SPAM} portions of error set space $\Delta$, wherein the elements of $\bvec{\delta}_{\rho} \in \Delta_{\rho}$ are the $d^2-1$ coefficients $\epsilon_i(\rho)$, and the elements of $\bvec{\delta}_M \in \Delta_M$ are the $d^2(m-1)$ coefficients $\epsilon_i(M)$.

Where appropriate we take advantage of two notational changes to improve readability:  (1) Gate error rates $\epsilon(g)$ are composed of two elements, its error class $\epsilon$, and the gate the error is attributed to $(g)$. If the error rates are unambiguously attributed to a gate $g$, we will often ignore writing the corresponding gate, e.g. an error intrinsic to $G$, $S_y(G) + S_z(G)$ can be simplified to $S_y + S_z$. (2) Properties can be expressed by their covector $\varphi$, or equivalently, by their action on an arbitrary error $\delta$. For example, consider the property
\begin{equation}\label{eq:covec}
\varphi = \left(a\sbra{\sigma_x},\ 0,\ b\,H_y^*,\ 0\right),
\end{equation}
where the blocks correspond to \(\rho\), \(M\), \(G_x\), and \(G_y\), respectively.  
Its action on an error vector

\begin{equation}
    \delta = \left(\sket{\delta_\rho}, \delta_M,L_{G_x}, L_{G_y} \right)
\end{equation}
is
\begin{align}
\varphi(\delta)&=a\sbraket{\sigma_x}{\delta_\rho} +b \Tr(H_y^{*\dagger}L_{G_x}) \\&= a\,\rho_x + b\,h_y(G_x)\label{eq:linearcomb}.
\end{align}
Thus the same property may be represented either as a covector in the stacked error space (Eq. (\ref{eq:covec})), or equivalently by the linear combination of error rates that it extracts (Eq. (\ref{eq:linearcomb})).

\subsection{Intrinsic FOGI properties of single qubit gates}

As shown in Sec. \ref{sec:commutant}, the intrinsic FOGI space for a gate $G_i$ is directly related to the commutant of its target gate, $\mathrm{comm}(\overline{G}_i)$. This is a particularly useful relationship to determine the size of the intrinsic FOGI space of any unitary single-qubit gate $G_\theta$. Every single-qubit unitary gate is a Bloch sphere rotation by some angle $\theta$, and the size of $\overline{G}_\theta$'s commutant is completely determined by $\theta$. This yields the following single-qubit intrinsic FOGI space dimensions (see Appendix \ref{app:intrinsicFOGIdim} for details):

\begin{equation}
    \dim(\cF_{G_\theta})=\begin{cases}
        12 & (\theta=0)\\
        6 & (\theta=\pi)\\
        4 & \text{otherwise},
    \end{cases}
    \label{eqn:rotation_classes}
\end{equation}
The case $\theta=0$ describes an idle gate. Because an idle gate commutes with every error, all of its intrinsic properties are FOGI. For the other two cases, consider the intrinsic FOGI spaces of a $G_x(\pi)$ gate and a $G_x\left(\frac{\pi}{2}\right)$ which are spanned by the following properties:

\begin{table}[H]
  \centering
  \begin{tabular}{@{} c @{\hspace{1.5cm}} c @{}}
    \begin{tabular}{|c|}
      
      \multicolumn{1}{c}{$\cF_{G_x(\pi)}$} \\ \hline
      $h_X$     \\ \hline
      $s_X$     \\ \hline
      $s_Y$     \\ \hline
      $s_Z$     \\ \hline
      $a_{Y,Z}$ \\\hline
      $c_{Y,Z}$     \\ \hline
    \end{tabular}
  &%
    \begin{tabular}{|c|}
      
       \multicolumn{1}{c}{$\cF_{G_x(\pi/2)}$} \\ \hline
      $h_X$      \\ \hline
      $s_X$      \\ \hline
      $s_Y + {s}_Z$ \\ \hline
      $a_{Y,Z}$\\ \hline
    \end{tabular}
  \end{tabular}
\end{table}\noindent $h_X$ quantifies a coherent rotation error that is aligned with the target gate. This corresponds to systematic over- or under-rotations. The stochastic properties (S and C) quantify the dephasing errors towards different axes of the Bloch sphere. For example, $s_Y\left(G_x\left(\frac{\pi}{2}\right)\right)+s_Z\left(G_x\left(\frac{\pi}{2}\right)\right)$ quantifies a dephasing error aligned with the gate's rotation axis. This is a \emph{bit flip error} with respect to the target gate eigenstates $\ket{+}$ and $\ket{-}$. Finally, the $a_{Y,Z}$ property quantifies a biased error along the $X$-axis, this is associated with a \emph{$T_1$ decay error} with respect to $\ket{-}$. 

Elementary-error-generator sectors (Hamiltonian, stochastic, and active) are preserved under unitary changes of basis. Consequently, all single-qubit \(\pi/2\) gates have the same \emph{qualitative} intrinsic FOGI error types, differing only by a rotation of axes. To see this, recall from Eq.~(\ref{eq:gateFOGIcomm}) that a property \(\Phi\) is intrinsic to  \(G_x(\pi/2)\)  and FOGI if and only if
\[
[\Phi, G_x(\pi/2)] = 0.
\]
Now let \(G\) be any other single-qubit \(\pi/2\) rotation gate. There exists a unitary change of basis \(U\) such that
\[
G = U G_x(\pi/2) U^\dagger.
\]
Next, define the property
\[
 \Phi' = U \Phi U^\dagger.
\]
Then
\begin{align}
[\Phi',G]
&= [U\Phi U^\dagger,\, U G_x(\pi/2) U^\dagger] \\
&= U[\Phi,G_x(\pi/2)]U^\dagger.
\end{align}
Therefore, if \([\Phi,G_x(\pi/2)]=0\), then \([\Phi',G]=0\). Thus, conjugation by \(U\) maps intrinsic FOGI properties of \(G_x(\pi/2)\) to intrinsic FOGI properties of \(G\). Since this map is invertible, the intrinsic FOGI spaces of all single-qubit \(\pi/2\) gates are isomorphic.

This establishes that the intrinsic FOGI structure is independent of the rotation axis for fixed angle \(\pi/2\). More generally, because this argument did not require that $\theta=\pi/2$, any single-qubit gate with rotation angle \(\theta \notin \{0,\pi\}\) will have its intrinsic FOGI space composed of the same four qualitative error types, as stated in the following lemma:
\begin{lemma}\label{lem:intrinsicsNature}
Let \(G\) be a single-qubit gate whose target action is a rotation by angle
\(\theta \notin \{0,\pi\}\) about any axis.
Then the intrinsic FOGI space \(\cF_G\) is 4-dimensional.
The properties that span this space quantify the following errors:
\begin{enumerate}
\item over/under-rotation about the target axis,
\item dephasing in the target gate's eigenbasis,
\item symmetric mixing of the two target-gate eigenstates (bit-flip error),
\item biased relaxation toward one target-gate eigenstate ($T_1$ decay error).
\end{enumerate}
\end{lemma}

\subsection{Pairwise relational FOGI properties}
Pairwise relational FOGI properties quantify a variety of qualitatively different errors. For example, consider the discrepancy in the difference between outcome probabilities for a noisy prepare-and-measure circuit,

\begin{align}
    &\left(P\left(0|M_Z,\rho\right) - P\left(1|M_Z,\rho\right)\right) - \left(P\left(0| \overline{M}_Z, \overline{\rho}\right) - P\left(1| \overline{M}_Z, \overline{\rho}\right)\right) \\&=  \sbraket{E_0-E_1}{\rho} - 1,
\end{align}

\noindent where the target operations are computational basis elements: $\sbra{\overline{E}_0} = \sbra{0}, \sbra{\overline{E}_1} = \sbra{1}$, and $\sket{\overline{\rho}}=\sket{0}$. By parameterizing their errors using the relevant error generators as shown in Eqs. \eqref{eq:deltarho} and \eqref{eq:deltaM} we find that

\begin{align}
    &\sbraket{E_0-E_1}{\rho} - 1 \\&=\sbraket{\overline{E}_0-\overline{E}_1}{\overline{\rho}} \\&+ \sbraket{\delta_{E_0-E_1}}{\overline{\rho}} + \sbraket{\overline{E}_0-\overline{E}_1}{\delta_{\rho}} - 1 + O\left(\epsilon^2\right)\\
    &= \frac{1}{2}(M_1+M_z) + 2\rho_z + O\left(\epsilon^2\right).\label{eq:SPAMbias}
\end{align}

\noindent Notice that the FOGI basis for a gate set with the same SPAM operations shown in Fig. \ref{fig:XYfogis}(d) contains the FOGI property (number 14)

\begin{equation}
    \varphi_{\mathrm{SPAM}}(\delta) = \rho_z + \frac{1}{4}(M_1 + M_z)
\end{equation}

\noindent which is directly proportional (up to $O(\epsilon^2)$) to Eq. (\ref{eq:SPAMbias}). Therefore this FOGI property directly quantifies the Z-bias between the SPAM operations in $\cG$. Some gauge choices may attribute Z-bias errors to $\rho$ or $M$ separately, but a non-zero value of $\varphi_{\mathrm{SPAM}}$ means the error is inherently relational between $\rho$ and $M$.

Another example of a common relational property is \emph{axis misalignment error}. 
Consider two noisy $\pi/2$ rotations $G_x$ and $G_y$ around axes $\hat{n}_x$ and $\hat{n}_y$, respectively. $\hat{n}_x$ and $\hat{n}_y$ should ideally be separated by an angle of $\overline{\theta}=\pi/2$. The axis misalignment error is given by

\begin{equation}
    \overline{\theta}-\theta=\frac{\pi}{2} - \theta = \frac{\pi}{2} - \cos^{-1}(\hat{n}_x\cdot \hat{n}_y).
\end{equation}

\noindent this can be expressed in terms of the Hamiltonian error generator rates for the two gates as

\begin{equation}
    \cos^{-1}(\hat{n}_x\cdot \hat{n}_y) = \frac{\pi}{2} - (h_Y(G_x)+h_Z(G_x) + h_X(G_y) - h_Z(G_y)) + O(h^2),
\end{equation}

\noindent where $O(h^2)$ captures all squared Hamiltonian error rates.

\noindent This error is directly quantified by FOGI property $9$ shown in Fig. \ref{fig:XYfogis}, panel d):
\begin{equation}
    \varphi_{\mathrm{axis}}(\delta) = h_Y(G_x) + h_Z(G_x) + h_X(G_y) - h_Z(G_y).
\end{equation}

\subsection{Average anisotropy of bit-flip:  An example of $3$-wise relational errors}\label{sec:anisotropy}

In Sec. \ref{sec:n-wise-relational} we introduced higher-order irreducible relational FOGI properties. To construct a concrete example of a 3-wise property let us consider the gate set $\{\rho, M_Z, G_x, G_y\}$ where the gates are $\pi/2$ rotations around the $X$ and $Y$ axes respectively. From Lemma \ref{lem:intrinsicsNature}, $G_x$ has an intrinsic FOGI property which quantifies its bit-flip error
\begin{equation}
    \varphi_{\mathrm{bf}}(L_{G_x}) = s_y(G_x) + s_z(G_x).
\end{equation}

On the other hand, the \emph{anisotropy} of this bit-flip error along the gate's transversal plane ($Y-Z$) is not quantified by any FOGI properties.

\begin{figure}
\centering
\includegraphics[width=8.5cm]{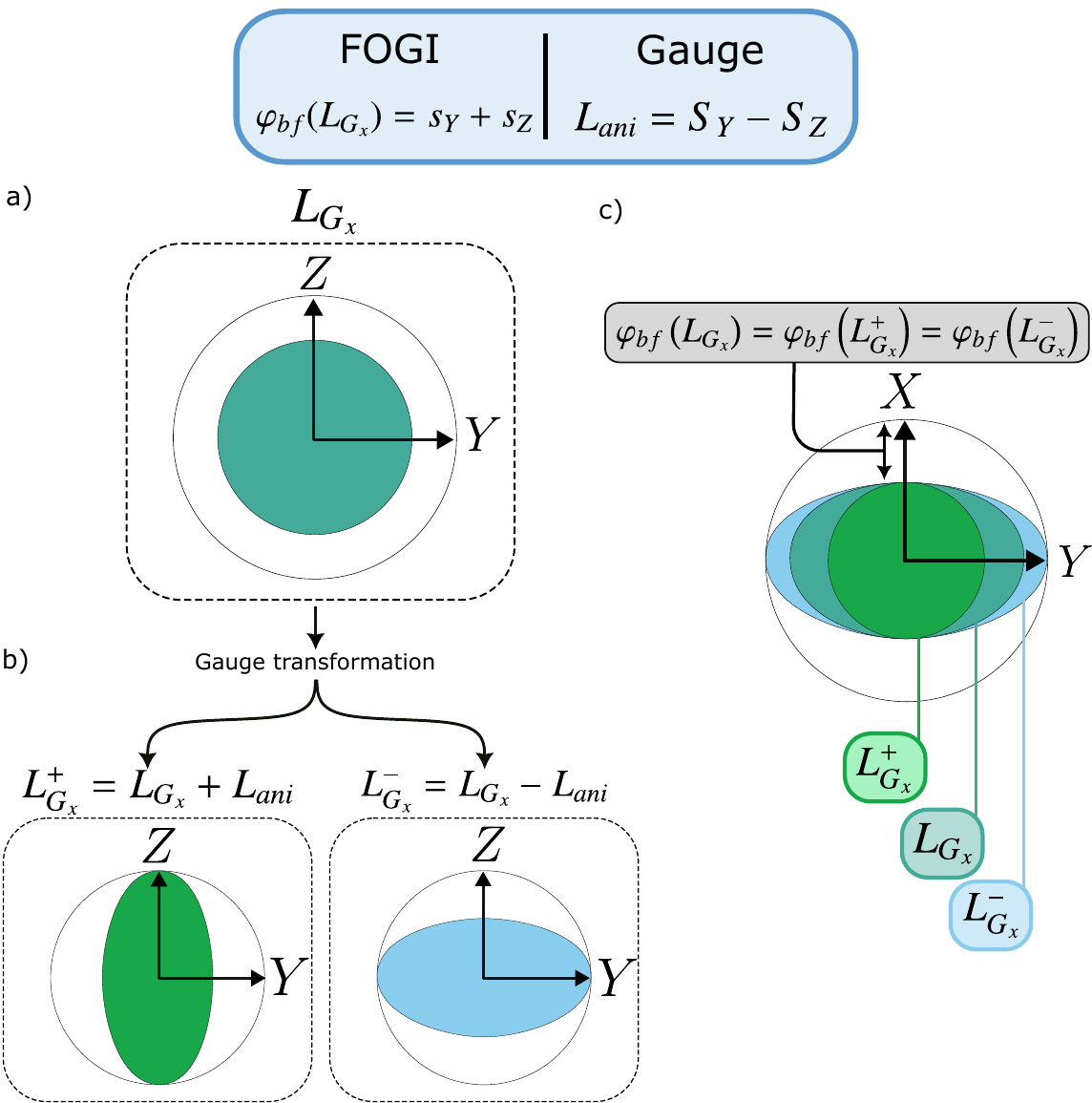}
\caption{Within a simple gate set with $\pi/2$ rotation gates $\cG=\{\sket{0}, M_z, G_x, G_y\}$, the gate $G_x$ has an intrinsic FOGI property $\varphi_{\mathrm{bf}}(L_{G_x}) = s_y+s_z$. This FOGI property corresponds to the bit-flip error of the gate with respect to its eigenstates from Lemma \ref{lem:intrinsicsNature}, i.e. the decoherence of a state's $X$ component. The anisotropy of this process along $Y$ and $Z$ can be varied by shifting $G_x$'s error generator in the direction $L_{ani} = S_y - S_z$. \textbf{a)} Consider an error channel for the $G_x$ gate, denoted $L_{G_x}$, with a bit-flip error rate $\varphi_{\mathrm{bf}}(\delta) = a$ and is isotropic in the $Y-Z$ plane. $\mathbf{b)}$ Within this gate set the anisotropy of bit-flip direction is a gauge direction. We perform two distinct gauge transformations in this direction resulting in $L_{G_x}^+=L_{G_x}+L_{ani}$ and $L_{G_x}^-=L_{G_x}-L_{ani}$ to obtain two gauge equivalent errors which only differ by the anisotropy along the $Y-Z$ plane. The direction on which we vary by $L_{ani}$ determines the squeeze and elongation of the $Z$ and $Y$ axes of the bit-flip error channel. In $\mathbf{c)}$ we superimpose all gauge equivalent errors $L_{G_x}, L_{G_x}^+, \text{ and } L_{G_x}^-$.  Because they are gauge equivalent, the FOGI property $\varphi_{\mathrm{bf}}$ remains unchanged, and thus they all have the same decoherence along the X-axis.}
\label{fig:anisotropy}
\end{figure}

This implies the existence of a gauge direction which allows for the variation of the anisotropy of the bit-flip error in the $Y-Z$ plane. This direction is:

\begin{equation}\label{eq:anisotropy}
    L_{ani} = S_Y(G_x) - S_Z(G_x).
\end{equation}

Consider an error channel for the $G_x$ gate, denoted $L_{G_x}$, which is isotropic in the $Y-Z$ plane as depicted in Fig. \ref{fig:anisotropy}a) and whose bit-flip FOGI property is some non-zero value $\varphi_{\mathrm{bf}}(L_{G_x})=a$. In Fig. \ref{fig:anisotropy}b)
we perform two gauge transformations by shifting $L_{G_x}$ in the $\pm L_{ani}$ directions. The positive shift results in a (gauge equivalent) error whose $Y$--$Z$ plane is squeezed along the $Y$ axis and stretched along the $Z$ axis, while a negative shift results in the opposite stretching and squeezing. Because all three of these error channels are gauge equivalent, they must have the same value for all FOGI properties, including $\varphi_{\mathrm{bf}}$. Therefore, they all have the same decoherence along the X-axis as shown in Fig. \ref{fig:anisotropy}c).

All of these statements above also apply for $G_y$ but on different axes of the Bloch sphere. However, if we modify our gate set by adding another non-commuting $\pi/2$ gate, e.g.~$G_z(\pi/2)$, the situation drastically changes. Within this gate set $\{\rho, M_Z, G_x, G_y, G_z \}$, the anisotropy of bit-flip errors for all gates are no longer gauge directions. For example, the outcome probabilities of the circuit 

\begin{equation*}
    M_ZG_yG_zG_x\sket{0}
\end{equation*}

\noindent change if we vary the gate set in the $S_y(G_x)-S_z(G_x)$ direction. As a consequence, a new FOGI property emerges that is linearly dependent on these former gauge directions and quantifies this new observable error, which we call the \emph{average anisotropy of bit-flip} error:

\begin{align}
\varphi_{\mathrm{aa}}(\delta) = s_Z(G_x) - s_Y(G_x)\notag\\ +  s_X(G_y) - s_Z(G_y) \\+ s_Y(G_z) - s_X(G_z),\notag 
\label{eqn:assymetry_of_bit-flip}
\end{align}

\noindent This error spans the entire space of irreducible three-wise gate FOGI properties for this gate set.

\subsection{Example construction of a FOGI basis for a one-qubit gate set} \label{sec:Detailed}
\begin{figure*}
    \centering
    \includegraphics[width=1\linewidth]{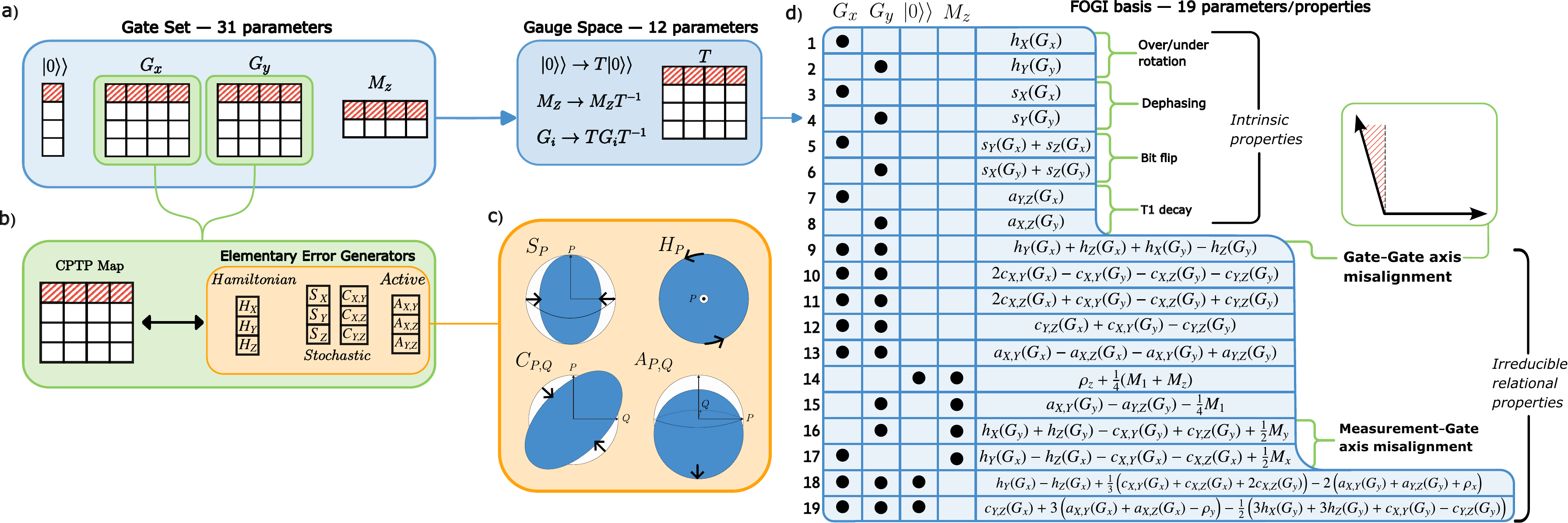}
    \caption{\textbf{A first-order gauge-invariant representation for a one-qubit gate set using the elementary error generators (a)} A noisy gate set composed of standard computational SPAM operations plus two $\pi/2$ rotations around the $X$ and $Y$-axis has a total of 31 free parameters, where the top row elements of the Pauli transfer matrices (and vectors) are constrained for trace preservation. \textbf{b)} Each operation can be decomposed into the \emph{elementary error generators (EEG)\cite{Blume-Kohout2022-ln}} representation, which categorizes errors into: (1) Hamiltonian (coherent), (2) stochastic (dephasing), and (3) active ($T_1$-like processes and other non-stochastic incoherent noise processes). \textbf{ c)} The effect of different elementary error generators on the Bloch sphere. \textbf{ d) }Gauge transformations span a 12 dimensional space given by a single TP matrix which acts on all operations in the gate set. A complete FOGI representation (basis) lacks these degrees of freedom, and thus has a total of 31-12=19 parameters, each of which quantifies an observable error in the gate set. This basis only contains \emph{irreducible} FOGI properties which are more amenable to interpretation. The first 8 properties are intrinsic, meaning that in the original EEG representation they are associated with same operation in all gauges, while the rest are relational and in the EEG representation may be attributed to different operations depending on the gauge choice.}
    \label{fig:XYfogis}
\end{figure*}

 In this section we present an instantiation of Algorithm \ref{algo:basis} to construct a basis of irreducible FOGI properties for a gate set with two $\pi/2$ rotations around the $X$ and $Y$ axes and standard computational basis SPAM operations

\begin{equation*}
    \cG = \{\sket{\rho}, M_Z, G_x, G_y\},
\end{equation*} whose errors are represented with the elementary error generators. 

Aside from the description of the target operations of $\cG $, all their corresponding gauge action matrices $A_g$ are needed as input. The requisite gauge action matrices are constructed using Eq. \eqref{eq:eeggaugeaction} for gate operations and Eq. \eqref{eq:ALinMap} for SPAM. This yields the combined gauge action for all operations 

\begin{equation}
\mathcal{A}_g(K)=\begin{cases}
  K\sket{0}      & \text{if } g = \rho,\\
  -\overline{M}_ZK        & \text{if } g = M,\\
  K - \overline{G}_jK\overline{G}_j^{-1}    & \text{if } g = G_x \text{ or } G_y.
\end{cases}
\end{equation}

\noindent The entries of the gauge action matrices for $G_x$ and $G_y$ are obtained from Eq. \eqref{eq:eeggaugeactionelements}, while the SPAM gauge action matrices are given by

\begin{align}
    \left(A_{\sket{0}}\right)_{i,j} &= \sbra{\hat{\rho}_j}\mathcal{A}_{\sket{0}}\left(\hat{L}_i^\rho\right) = \sbra{\sigma_j}\hat{L}_i^\rho\sket{\overline{0}}, \text{and} \\
    \left(A_{M_Z}\right)_{i,j} &= \Tr\left(\mathcal{A}_{M_Z}\left(\hat{L}_i^M\right)\sket{\sigma_j}\bra{1}\right)=-\bra{1}\overline{M}_Z\hat{L}_i^M\sket{\sigma_j} \\&=-\begin{pmatrix}0 & 0 & 0 & 2\end{pmatrix}\hat{L}_i^M\sket{\sigma_j}.
\end{align}
\noindent Now we are ready to begin building a basis for $\cF_\cG$. First, we find spanning FOGI properties for each operation's intrinsic space $\cF_g$ by computing the kernel of their corresponding gauge action matrix $A_g^\top$. This yields an empty set for the SPAM operations -- as expected from Lemma \ref{lem:stateFOGI} and Lemma \ref{lem:MFOGIs} -- and four vectors for each of $G_x$ and $G_y$, -- as expected from Lemma \ref{lem:intrinsicsNature}-- which we denote $\bvec{\phi}_i^x$ and $\bvec{\phi}_i^y$, respectively.

Next, we construct purely pairwise relational FOGI properties. First, we build each relational gauge action matrix  $A_{\sket{0},M_z}^\top$, $A_{\sket{0},G_x}^\top$, $A_{\sket{0},G_y}^\top$, $A_{M_z,G_x}^\top$, $A_{M_z,G_y}^\top$, and $A_{G_y,G_x}^\top$, where 

\begin{equation}
    A_{g_1,g_2} = A_{g_1} \stack A_{g_2}.
\end{equation} The kernel of each of these matrices is equivalent to the  FOGI spaces $\mathrm{ker}\left(A_{g_1,g_2}^\top\right)=\cF_{g_1,g_2}$. The intrinsic FOGI spaces we just found live within these spaces, and thus properties that span any $\cF_{g_1,g_2}$ will not generally be pairwise irreducible. To overcome this, we make use of \emph{blocking matrices} -- built out of the intrinsic FOGI properties we just found -- as described in Sec.\ref{sec:pairwise},
\begin{align}
    B_{\cF_{G_k}} = \bvec{\phi}_0^k \stack \bvec{\phi}_1^k
    \stack \bvec{\phi}_2^k 
    \stack \bvec{\phi}_3^k.
\end{align}
Stacking these blocking matrices with every relational gauge action matrix allows us to find pairwise irreducible FOGI properties by computing their kernel. For example, the space of purely pairwise relational FOGI properties between $G_x$ and $G_y$ is given by\[\mathrm{ker}\left(A_{G_x,G_y}^\top\stack B_{\cF_{G_x}} \stack B_{\cF_{G_y}}\right) = \cF_{G_x,G_y}\cap(\cF_{G_x}\oplus\cF_{G_y})^\perp.\]
For a more detailed discussion about blocking matrices and the subspaces they isolate see Sec. \ref{sec:pairwise}. 

Finding a spanning set for each purely pairwise relational FOGI space as described above yields a slightly over-complete set for the entire FOGI space $\cF_\cG$. Which is why, as suggested in Algorithm \ref{algo:basis}, after obtaining a spanning set of FOGI properties for any subspace, each new property should only be accepted if it is linearly independent from those previously found. Figure \ref{fig:XYfogis}(d) shows a collection of 19 FOGI properties obtained from this procedure which form a basis for $\cF_\cG$ \footnote{For the construction of this basis we took an additional step to ensure that gate-only FOGI properties are restricted to a single error generator class as described in Sec. \ref{sec:gateonlyFOGI}}.

\section*{Acknowledgments}
We thank R. Murray for insightful discussions about dual maps and their matrix representations. We thank W. Brown, R. Bilotta, C. Wilson, D. Cherney, N. Manning and D. Clarke for insightful discussions and providing careful feedback on an earlier draft of this work, including motivating the need for the canonical presentation of Section \ref{canonical}. We  thank C. Kelson-Packer for insightful discussions and proposing a simpler proof for Lemma~\ref{lem:MFOGIs}.

This material was funded in part by the U.S. Department of Energy, Office of Science, Office of Advanced Scientific Computing Research and by the Laboratory Directed Research and Development program at Sandia National Laboratories.

Sandia National Laboratories is a multimission laboratory managed and operated by National Technology and Engineering Solutions of Sandia, LLC, a wholly owned subsidiary of Honeywell International Inc., for the U.S. Department of Energy’s National Nuclear Security Administration under contract DE-NA0003525. This paper describes objective technical results and analysis. Any subjective views or opinions that might be expressed in the paper do not necessarily represent the views of the U.S. Department of Energy or the United States Government.

\appendix
\section{Intrinsic FOGI space dimensions for single-qubit gates}
\label{app:intrinsicFOGIdim}

In this appendix we prove Eq.~(\ref{eqn:rotation_classes}).

\begin{equation}
    \dim(\cF_{G_\theta})=\begin{cases}
        12 & (\theta=0)\\
        6 & (\theta=\pi)\\
        4 & \text{otherwise},
    \end{cases}
\end{equation}

Let $\overline G$ be the PTM of a single-qubit unitary gate. Since $\overline G$ is trace preserving and unital, it has block form
\begin{equation}
    \overline G =
    \begin{pmatrix}
    1 & 0\\
    0 & R
    \end{pmatrix},
\end{equation}

\noindent where $R\in SO(3)$ corresponds to the unitary's Bloch-sphere rotation. Let $\theta$ be the rotation angle of $R$.

Gate errors have their top row constrained to $[0,0,0,0]$ due to trace preservation. Because gate properties $\Phi_{G_i}$ act on errors through the trace inner product $Tr(\Phi_{i}\delta_{G_i})$, their left-most column entries act trivially on errors. With this in mind, any gate property may be represented by a matrix of the form
\begin{equation}
    \Phi=\begin{pmatrix}
        0&v^\top\\0&H
    \end{pmatrix}
\end{equation}
where $v\in\mathbb R^3$ and $H\in M_{3\times3}(\mathbb R)$. By Eq.~(\ref{eq:gateFOGIcomm}), $\Phi$ is intrinsic FOGI if and only if it commutes with $\overline G$:
\begin{equation}
[\overline G,\Phi]=0.
\end{equation}
A direct calculation gives
\begin{equation}
[\overline G,\Phi]
=
\begin{pmatrix}
0 & v^\top(\Id-R)\\
 0& RH-HR
\end{pmatrix}.
\end{equation}
Therefore
\[
[\overline G,\Phi]=0
\]
if and only if
\begin{equation}
v^\top(\Id-R)=0
\qquad\text{and}\qquad
RH=HR.
\end{equation}
\noindent Because $R$ is an $SO(3)$ rotation, $R^\top R=\Id$ we can re-write the first condition as $(\Id - R)v=0$. Since $v$ and $H$ are independent blocks, the intrinsic FOGI space decomposes as a direct sum
\begin{equation}
\cF_{G}\cong \ker(\Id - R)\oplus \mathrm{comm}(R),
\end{equation}
and hence
\begin{equation}
\dim(\cF_{G})
=
\dim(\ker(\Id - R))
+
\dim(\mathrm{comm}(R)).
\end{equation}
The dimension of $\mathrm{kernel}(\Id-R)$ is the multiplicity of the eigenvalue $+1$ of R.

Any $R$ written in its eigenbasis has the form
\begin{equation}
    R = 
        \begin{pmatrix}
            1&0&0\\
            0&e^{i\theta}&0\\
            0&0&e^{-i\theta}
        \end{pmatrix}.
\end{equation}

The dimension of $\ker(\Id-R)$ changes only when the multiplicity of the eigenvalue $1$ changes. Since the eigenvalues of $R$ are
\[
\lambda \in \{1,\ e^{i\theta},\ e^{-i\theta},
\}\]
the eigenvalue $1$ has multiplicity $3$ only when $e^{i\theta}=e^{-i\theta}=1$, i.e. when $\theta=0$ modulo $2\pi$. Otherwise, the only $+1$-eigenvector is the rotation axis, so
\[
\dim\ker(\Id-R)=1
\]
for $\theta\neq 0$ modulo $2\pi$.

Now we focus on the dimension of $\mathrm{comm}(R)$. Because this quantity is agnostic of the basis we work in, we write the commutator of $R$ with an arbitrary matrix in $R$'s eigenbasis:

\begin{align}
    [R,B] =  &\begin{pmatrix}
            1&0&0\\
            0&e^{i\theta}&0\\
            0&0&e^{-i\theta}
        \end{pmatrix}\begin{pmatrix}
            b_{11}&b_{12}&b_{13}\\
            b_{21}&b_{22}&b_{23}\\
            b_{31}&b_{32}&b_{33}\\
        \end{pmatrix}-\\&\begin{pmatrix}
            b_{11}&b_{12}&b_{13}\\
            b_{21}&b_{22}&b_{23}\\
            b_{31}&b_{32}&b_{33}\\
        \end{pmatrix}\begin{pmatrix}
            1&0&0\\
            0&e^{i\theta}&0\\
            0&0&e^{-i\theta}
        \end{pmatrix}\\
        &= \begin{pmatrix}
            0&(1-e^{i\theta})b_{12}&(1-e^{-i\theta})b_{13}\\
            (e^{i\theta} -1)b_{21} & 0 & (e^{i\theta}-e^{-i\theta})b_{23}\\
            (e^{-i\theta}-1)b_{31}&(e^{-i\theta}-e^{i\theta})b_{32}&0
        \end{pmatrix}=0
\end{align}

The equation $[R,B]=0$ is linear in the entries $b_{ij}$. Each off-diagonal entry appears multiplied by an eigenvalue difference,
\[
(\lambda_i-\lambda_j)b_{ij}.
\]
If $\lambda_i\neq \lambda_j$, then $b_{ij}=0$. If $\lambda_i=\lambda_j$, then the equation becomes $0=0$, so $b_{ij}$ is unconstrained.

Therefore, every diagonal entry $b_{ii}$ is always free, contributing three dimensions. In addition, every degenerate pair $\lambda_i=\lambda_j$ with $i\neq j$ contributes two extra free parameters, namely $b_{ij}$ and $b_{ji}$.

There are three cases.

If $\theta=0$, then all three eigenvalues are equal:
\[
\lambda_1=\lambda_2=\lambda_3=1.
\]
All matrix entries $b_{ij}$ are free, so
\[
\dim(\mathrm{comm}(R))=9.
\]

If $\theta=\pi$, then
\[
\lambda_1=1,\qquad \lambda_2=\lambda_3=-1.
\]
The three diagonal entries are free, and the degeneracy $\lambda_2=\lambda_3$ contributes the two additional free entries $b_{23}$ and $b_{32}$. Hence
\[
\dim(\mathrm{comm}(R))=3+2=5.
\]

If $\theta\notin\{0,\pi\}$, then
\[
1,\ e^{i\theta},\ e^{-i\theta}
\]
are all distinct. Thus all off-diagonal entries are forced to be zero, and only the three diagonal entries remain free. Therefore
\[
\dim(\mathrm{comm}(R))=3.
\]
Putting all of these results together gives
\[
\dim\left(\mathcal F_{\overline G(\theta)}\right)=
\begin{cases}
12 & \theta=0,\\
6 & \theta=\pi,\\
4 & \text{otherwise}.
\end{cases}
\]

All steps in this proof can be generalized to multi-qubit gates.

\section{Gauge equivalent gate sets within an $\epsilon$-neighborhood}\label{app:smallgauge}
The assumption of an informationally complete (IC) set of operations is ubiquitous in quantum tomography protocols\cite{Nielsen2021-nu,DiMatteo2020operationalgauge, QCVV}. In this appendix we justify the following lemma, which is an underlying assumption regarding small gauge transformations (Sec. \ref{sec:smallgauge}), and is a fundamental building block in this paper:

\begin{lemma}
    Any two gauge-equivalent and informationally-complete gate sets $\cG=\{ \sket{\rho}, M, G_1,\ldots G_N\}$ and $\cG'=\{ \sket{\rho'}, M', G_1',\ldots G_N'\}$ that are within an $\epsilon$-neighborhood of a target gate set $\overline{\cG}$ are related by a small standard gauge transformation. In other words,
    \begin{eqnarray}
  G_i' &=& T G_i T^{-1} \nonumber\\
  \sket{\rho'} &=& T \sket{\rho} \label{eq:gaugerelation}\\
  M' &=& M T^{-1}, \nonumber
\end{eqnarray}
where $T$ is an invertible matrix that is close to the identity which can be generated by a matrix $K$ such that
\begin{equation}
    T=e^{\epsilon K} = \Id + \epsilon K + O\left(\epsilon^2\right).
\end{equation}
\end{lemma}

For the purposes of Lemma 8 we define an informationally-complete gate set to be one such that a full-rank and well-conditioned basis of fiducial SPAM operations can be constructed using polynomially-sized circuits. 

First, we observe the fact that if two IC gate sets are gauge equivalent, then there must exist a gauge transformation as defined in Eq. \ref{eq:gaugerelation} that relates them. This follows as a natural consequence of linear-inversion gate set tomography's (LGST) ability to, given an informationally complete gate set, reproduce that gate set up to the standard gauge freedom given in Eq. \ref{eq:gaugerelation} \cite{Nielsen2021-nu}.

By definition, an IC gate set allows for the construction of fiducial states ${\sket{f_i}}$ that form a basis for all states. Here, a fiducial state means a state obtained from preparing the native state $\sket{\rho}$ and applying some combination of the gates available in the gate set $\{G_1,\ldots G_N\}$.  We construct a matrix whose columns are the fiducial states
\begin{equation}
    F(\cG) = \begin{pmatrix}
        \sket{f_1} , \sket{f_2}, \cdots, \sket{f_{d^2}}
    \end{pmatrix}.
\end{equation}

Because $\cG$ and $\cG'$ are related by a standard gauge transformation $T$, their fiducial states are also related by $T$ as

\begin{equation}
    F(\cG')=TF(\cG).
\end{equation}
Furthermore, since these states form a basis, $F$ is invertible, and thus

\begin{equation}
    T = F(\cG')F(\cG)^{-1}.
\end{equation}

Now consider the norm 
\begin{align}
    \| T-\Id\| &= \| T - F(\cG)F^{-1}(\cG)\|\\
    &= \|F(\cG')F(\cG)^{-1} - F(\cG)F(\cG)^{-1}\| \\&=\|\left(F(\cG')-F(\cG)\right)F(\cG)^{-1}\|\\&\leq \|\left(F(\cG')-F(\cG)\right)\| \hspace{.1em} \|F(\cG)^{-1}\|.\label{eq:appendixnorm}
\end{align}
Let us address the two terms on the right-hand side separately. First, with some algebraic manipulation and the triangle inequality we find that
\begin{align}
    \|F(\cG') - F(\cG)\| &= \| F(\cG') - F(\overline{\cG}) -\left(F(\cG)-F(\overline{\cG}) \right) \|\\
    &\leq \|F(\cG') - F(\overline{\cG})\|+\|F(\cG) - F(\overline{\cG})\|\label{eq:ineq}.
\end{align}

\noindent Both $\cG$ and $\cG'$ are within an $\epsilon$-neighborhood of $\overline{\cG}$ and thus
\begin{align}
    &\|F(\cG')-F(\overline{\cG})\|=O(\epsilon),\\
    &\|F(\cG)-F(\overline{\cG})\|=O(\epsilon).
\end{align}

\noindent Plugging this into Eq. (\ref{eq:ineq}) yields
\begin{equation}
    \|F(\cG') - F(\cG)\| \leq O(\epsilon) + O(\epsilon)=O(\epsilon).
\end{equation}

Second, since the target gate set is informationally complete we choose a fixed set of fiducials for which $F(\bar G)$ is full-rank and well-conditioned. By continuity, any sufficiently small perturbation \(G\) of \(\bar G\) has \(F(G)\) nonsingular with \(\|F(G)^{-1}\| = O(1)\), where the implicit constant depends only on the chosen fiducials and the conditioning of \(F(\bar G)\), not on \(\epsilon\). Therefore its inverse has the same property: $\|F(\cG)^{-1}\|=O(1).$ Putting these two facts together, Eq. (\ref{eq:appendixnorm}) simplifies to

\begin{equation}
    \|T - \Id \|\leq O(\epsilon).\label{eq:gaugesmall}
\end{equation}

The fact that these gauge transformations $T$ are close to the identity means its logarithm is well defined. Therefore, any $T$ can be uniquely described by a generator matrix $K$. Finally, to make sure that Eq. (\ref{eq:gaugesmall}) is satisfied, we constrain our generator to be $K=O(1)$, and define

\begin{equation}
    T = e^{\epsilon K} = \Id + \epsilon K + O\left(\epsilon^2 \right). \qed
\end{equation}
\section{Unobservable FOGI Properties}
\label{app:unobservableFOGIs}

In this work, we've implicitly treated FOGI errors/directions and observable features of a gate set as synonymous. This is the case for most FOGI directions, yet under the right circumstances some FOGI directions as produced by our construction may not be observable. Consider the gate set $\cG=\{\sket{0},M_z, G_z\}$, where $Gz$ is a single qubit rotation about the Z axis of the Bloch sphere. From Lemma \ref{lem:intrinsicsNature}, one of the FOGI errors for $G_z$ is Z-axis over-rotation. Every possible circuit outcome probability for this gate set takes the form 

\begin{equation}
    P_{0/1}=\sbra{E_{0/1}} G_z^k \sket{0}
\end{equation}

for some non-negative integer $k$. Notice that increasing the magnitude of $Z$-axis over-rotation on $G_z$ leaves the outcome unchanged. In other words, the $Z$-axis over-rotation FOGI error is unobservable for such a gate set.

This pathology is a consequence of the non-informational-completeness of the gate set in question. For informationally-complete gate sets FOGI errors \emph{are} all observable. This follows as a natural consequence of linear-inversion gate set tomography's (LGST) ability to, given an informationally complete gate set, reproduce that gate set up to the standard gauge freedom given in Eq. (\ref{eq:gaugetransform1}) \cite{Nielsen2021-nu}. As FOGI errors are, by construction, linearly independent from the shifts in error set space induced by standard gauge transformations, LGST implies that these errors must all be observable.


\begin{thebibliography}{26}%
\makeatletter
\providecommand \@ifxundefined [1]{%
 \@ifx{#1\undefined}
}%
\providecommand \@ifnum [1]{%
 \ifnum #1\expandafter \@firstoftwo
 \else \expandafter \@secondoftwo
 \fi
}%
\providecommand \@ifx [1]{%
 \ifx #1\expandafter \@firstoftwo
 \else \expandafter \@secondoftwo
 \fi
}%
\providecommand \natexlab [1]{#1}%
\providecommand \enquote  [1]{``#1''}%
\providecommand \bibnamefont  [1]{#1}%
\providecommand \bibfnamefont [1]{#1}%
\providecommand \citenamefont [1]{#1}%
\providecommand \href@noop [0]{\@secondoftwo}%
\providecommand \href [0]{\begingroup \@sanitize@url \@href}%
\providecommand \@href[1]{\@@startlink{#1}\@@href}%
\providecommand \@@href[1]{\endgroup#1\@@endlink}%
\providecommand \@sanitize@url [0]{\catcode `\\12\catcode `\$12\catcode
  `\&12\catcode `\#12\catcode `\^12\catcode `\_12\catcode `\%12\relax}%
\providecommand \@@startlink[1]{}%
\providecommand \@@endlink[0]{}%
\providecommand \url  [0]{\begingroup\@sanitize@url \@url }%
\providecommand \@url [1]{\endgroup\@href {#1}{\urlprefix }}%
\providecommand \urlprefix  [0]{URL }%
\providecommand \Eprint [0]{\href }%
\providecommand \doibase [0]{http://dx.doi.org/}%
\providecommand \selectlanguage [0]{\@gobble}%
\providecommand \bibinfo  [0]{\@secondoftwo}%
\providecommand \bibfield  [0]{\@secondoftwo}%
\providecommand \translation [1]{[#1]}%
\providecommand \BibitemOpen [0]{}%
\providecommand \bibitemStop [0]{}%
\providecommand \bibitemNoStop [0]{.\EOS\space}%
\providecommand \EOS [0]{\spacefactor3000\relax}%
\providecommand \BibitemShut  [1]{\csname bibitem#1\endcsname}%
\let\auto@bib@innerbib\@empty
\bibitem [{\citenamefont {Nielsen}\ \emph {et~al.}(2021)\citenamefont
  {Nielsen}, \citenamefont {Gamble}, \citenamefont {Rudinger}, \citenamefont
  {Scholten}, \citenamefont {Young},\ and\ \citenamefont
  {Blume-Kohout}}]{Nielsen2021-nu}%
  \BibitemOpen
  \bibfield  {author} {\bibinfo {author} {\bibfnamefont {Erik}\ \bibnamefont
  {Nielsen}}, \bibinfo {author} {\bibfnamefont {John~King}\ \bibnamefont
  {Gamble}}, \bibinfo {author} {\bibfnamefont {Kenneth}\ \bibnamefont
  {Rudinger}}, \bibinfo {author} {\bibfnamefont {Travis}\ \bibnamefont
  {Scholten}}, \bibinfo {author} {\bibfnamefont {Kevin}\ \bibnamefont {Young}},
  \ and\ \bibinfo {author} {\bibfnamefont {Robin}\ \bibnamefont
  {Blume-Kohout}},\ }\bibfield  {title} {\enquote {\bibinfo {title} {Gate set
  tomography},}\ }\href {\doibase 10.22331/q-2021-10-05-557} {\bibfield
  {journal} {\bibinfo  {journal} {Quantum}\ }\textbf {\bibinfo {volume} {5}},\
  \bibinfo {pages} {557} (\bibinfo {year} {2021})}\BibitemShut {NoStop}%
\bibitem [{\citenamefont {Hashim}\ \emph {et~al.}(2025)\citenamefont {Hashim},
  \citenamefont {Nguyen}, \citenamefont {Goss}, \citenamefont {Marinelli},
  \citenamefont {Naik}, \citenamefont {Chistolini}, \citenamefont {Hines},
  \citenamefont {Marceaux}, \citenamefont {Kim}, \citenamefont {Gokhale},
  \citenamefont {Tomesh}, \citenamefont {Chen}, \citenamefont {Jiang},
  \citenamefont {Ferracin}, \citenamefont {Rudinger}, \citenamefont {Proctor},
  \citenamefont {Young}, \citenamefont {Siddiqi},\ and\ \citenamefont
  {Blume-Kohout}}]{Hashim2025-rz}%
  \BibitemOpen
  \bibfield  {author} {\bibinfo {author} {\bibfnamefont {Akel}\ \bibnamefont
  {Hashim}}, \bibinfo {author} {\bibfnamefont {Long~B}\ \bibnamefont {Nguyen}},
  \bibinfo {author} {\bibfnamefont {Noah}\ \bibnamefont {Goss}}, \bibinfo
  {author} {\bibfnamefont {Brian}\ \bibnamefont {Marinelli}}, \bibinfo {author}
  {\bibfnamefont {Ravi~K}\ \bibnamefont {Naik}}, \bibinfo {author}
  {\bibfnamefont {Trevor}\ \bibnamefont {Chistolini}}, \bibinfo {author}
  {\bibfnamefont {Jordan}\ \bibnamefont {Hines}}, \bibinfo {author}
  {\bibfnamefont {J~P}\ \bibnamefont {Marceaux}}, \bibinfo {author}
  {\bibfnamefont {Yosep}\ \bibnamefont {Kim}}, \bibinfo {author} {\bibfnamefont
  {Pranav}\ \bibnamefont {Gokhale}}, \bibinfo {author} {\bibfnamefont {Teague}\
  \bibnamefont {Tomesh}}, \bibinfo {author} {\bibfnamefont {Senrui}\
  \bibnamefont {Chen}}, \bibinfo {author} {\bibfnamefont {Liang}\ \bibnamefont
  {Jiang}}, \bibinfo {author} {\bibfnamefont {Samuele}\ \bibnamefont
  {Ferracin}}, \bibinfo {author} {\bibfnamefont {Kenneth}\ \bibnamefont
  {Rudinger}}, \bibinfo {author} {\bibfnamefont {Timothy}\ \bibnamefont
  {Proctor}}, \bibinfo {author} {\bibfnamefont {Kevin~C}\ \bibnamefont
  {Young}}, \bibinfo {author} {\bibfnamefont {Irfan}\ \bibnamefont {Siddiqi}},
  \ and\ \bibinfo {author} {\bibfnamefont {Robin}\ \bibnamefont
  {Blume-Kohout}},\ }\bibfield  {title} {\enquote {\bibinfo {title} {Practical
  introduction to benchmarking and characterization of quantum computers},}\
  }\href {\doibase 10.1103/prxquantum.6.030202} {\bibfield  {journal} {\bibinfo
   {journal} {PRX quantum}\ }\textbf {\bibinfo {volume} {6}},\ \bibinfo {pages}
  {030202} (\bibinfo {year} {2025})}\BibitemShut {NoStop}%
\bibitem [{\citenamefont {Blume-Kohout}\ \emph {et~al.}(2025)\citenamefont
  {Blume-Kohout}, \citenamefont {Proctor},\ and\ \citenamefont {Young}}]{QCVV}%
  \BibitemOpen
  \bibfield  {author} {\bibinfo {author} {\bibfnamefont {Robin}\ \bibnamefont
  {Blume-Kohout}}, \bibinfo {author} {\bibfnamefont {Timothy}\ \bibnamefont
  {Proctor}}, \ and\ \bibinfo {author} {\bibfnamefont {Kevin}\ \bibnamefont
  {Young}},\ }\bibfield  {title} {\enquote {\bibinfo {title} {{Quantum
  Characterization, Verification, and Validation}},}\ }\href@noop {} {\
  (\bibinfo {year} {2025})},\ \Eprint {http://arxiv.org/abs/2503.16383}
  {arXiv:2503.16383 [quant-ph]} \BibitemShut {NoStop}%
\bibitem [{\citenamefont {Emerson}\ \emph {et~al.}(2005)\citenamefont
  {Emerson}, \citenamefont {Alicki},\ and\ \citenamefont
  {Życzkowski}}]{RB_emerson}%
  \BibitemOpen
  \bibfield  {author} {\bibinfo {author} {\bibfnamefont {Joseph}\ \bibnamefont
  {Emerson}}, \bibinfo {author} {\bibfnamefont {Robert}\ \bibnamefont
  {Alicki}}, \ and\ \bibinfo {author} {\bibfnamefont {Karol}\ \bibnamefont
  {Życzkowski}},\ }\bibfield  {title} {\enquote {\bibinfo {title} {Scalable
  noise estimation with random unitary operators},}\ }\href {\doibase
  10.1088/1464-4266/7/10/021} {\bibfield  {journal} {\bibinfo  {journal}
  {Journal of Optics B: Quantum and Semiclassical Optics}\ }\textbf {\bibinfo
  {volume} {7}},\ \bibinfo {pages} {S347–S352} (\bibinfo {year}
  {2005})}\BibitemShut {NoStop}%
\bibitem [{\citenamefont {Emerson}\ \emph {et~al.}(2007)\citenamefont
  {Emerson}, \citenamefont {Silva}, \citenamefont {Moussa}, \citenamefont
  {Ryan}, \citenamefont {Laforest}, \citenamefont {Baugh}, \citenamefont
  {Cory},\ and\ \citenamefont {Laflamme}}]{RB_Emerson2}%
  \BibitemOpen
  \bibfield  {author} {\bibinfo {author} {\bibfnamefont {Joseph}\ \bibnamefont
  {Emerson}}, \bibinfo {author} {\bibfnamefont {Marcus}\ \bibnamefont {Silva}},
  \bibinfo {author} {\bibfnamefont {Osama}\ \bibnamefont {Moussa}}, \bibinfo
  {author} {\bibfnamefont {Colm}\ \bibnamefont {Ryan}}, \bibinfo {author}
  {\bibfnamefont {Martin}\ \bibnamefont {Laforest}}, \bibinfo {author}
  {\bibfnamefont {Jonathan}\ \bibnamefont {Baugh}}, \bibinfo {author}
  {\bibfnamefont {David~G.}\ \bibnamefont {Cory}}, \ and\ \bibinfo {author}
  {\bibfnamefont {Raymond}\ \bibnamefont {Laflamme}},\ }\bibfield  {title}
  {\enquote {\bibinfo {title} {Symmetrized characterization of noisy quantum
  processes},}\ }\href {\doibase 10.1126/science.1145699} {\bibfield  {journal}
  {\bibinfo  {journal} {Science}\ }\textbf {\bibinfo {volume} {317}},\ \bibinfo
  {pages} {1893–1896} (\bibinfo {year} {2007})}\BibitemShut {NoStop}%
\bibitem [{\citenamefont {Knill}\ \emph {et~al.}(2008)\citenamefont {Knill},
  \citenamefont {Leibfried}, \citenamefont {Reichle}, \citenamefont {Britton},
  \citenamefont {Blakestad}, \citenamefont {Jost}, \citenamefont {Langer},
  \citenamefont {Ozeri}, \citenamefont {Seidelin},\ and\ \citenamefont
  {Wineland}}]{RBKnill_2008}%
  \BibitemOpen
  \bibfield  {author} {\bibinfo {author} {\bibfnamefont {E.}~\bibnamefont
  {Knill}}, \bibinfo {author} {\bibfnamefont {D.}~\bibnamefont {Leibfried}},
  \bibinfo {author} {\bibfnamefont {R.}~\bibnamefont {Reichle}}, \bibinfo
  {author} {\bibfnamefont {J.}~\bibnamefont {Britton}}, \bibinfo {author}
  {\bibfnamefont {R.~B.}\ \bibnamefont {Blakestad}}, \bibinfo {author}
  {\bibfnamefont {J.~D.}\ \bibnamefont {Jost}}, \bibinfo {author}
  {\bibfnamefont {C.}~\bibnamefont {Langer}}, \bibinfo {author} {\bibfnamefont
  {R.}~\bibnamefont {Ozeri}}, \bibinfo {author} {\bibfnamefont
  {S.}~\bibnamefont {Seidelin}}, \ and\ \bibinfo {author} {\bibfnamefont
  {D.~J.}\ \bibnamefont {Wineland}},\ }\bibfield  {title} {\enquote {\bibinfo
  {title} {Randomized benchmarking of quantum gates},}\ }\href {\doibase
  10.1103/physreva.77.012307} {\bibfield  {journal} {\bibinfo  {journal}
  {Physical Review A}\ }\textbf {\bibinfo {volume} {77}} (\bibinfo {year}
  {2008}),\ 10.1103/physreva.77.012307}\BibitemShut {NoStop}%
\bibitem [{\citenamefont {Carignan-Dugas}\ \emph {et~al.}(2023)\citenamefont
  {Carignan-Dugas}, \citenamefont {Dahlen}, \citenamefont {Hincks},
  \citenamefont {Ospadov}, \citenamefont {Beale}, \citenamefont {Ferracin},
  \citenamefont {Skanes-Norman}, \citenamefont {Emerson},\ and\ \citenamefont
  {Wallman}}]{errorreconstructioncompiledcalibration}%
  \BibitemOpen
  \bibfield  {author} {\bibinfo {author} {\bibfnamefont {Arnaud}\ \bibnamefont
  {Carignan-Dugas}}, \bibinfo {author} {\bibfnamefont {Dar}\ \bibnamefont
  {Dahlen}}, \bibinfo {author} {\bibfnamefont {Ian}\ \bibnamefont {Hincks}},
  \bibinfo {author} {\bibfnamefont {Egor}\ \bibnamefont {Ospadov}}, \bibinfo
  {author} {\bibfnamefont {Stefanie~J.}\ \bibnamefont {Beale}}, \bibinfo
  {author} {\bibfnamefont {Samuele}\ \bibnamefont {Ferracin}}, \bibinfo
  {author} {\bibfnamefont {Joshua}\ \bibnamefont {Skanes-Norman}}, \bibinfo
  {author} {\bibfnamefont {Joseph}\ \bibnamefont {Emerson}}, \ and\ \bibinfo
  {author} {\bibfnamefont {Joel~J.}\ \bibnamefont {Wallman}},\ }\href
  {https://arxiv.org/abs/2303.17714} {\enquote {\bibinfo {title} {The error
  reconstruction and compiled calibration of quantum computing cycles},}\ }
  (\bibinfo {year} {2023}),\ \Eprint {http://arxiv.org/abs/2303.17714}
  {arXiv:2303.17714 [quant-ph]} \BibitemShut {NoStop}%
\bibitem [{\citenamefont {Proctor}\ \emph {et~al.}(2017)\citenamefont
  {Proctor}, \citenamefont {Rudinger}, \citenamefont {Young}, \citenamefont
  {Sarovar},\ and\ \citenamefont {Blume-Kohout}}]{Proctor2017-wc}%
  \BibitemOpen
  \bibfield  {author} {\bibinfo {author} {\bibfnamefont {Timothy}\ \bibnamefont
  {Proctor}}, \bibinfo {author} {\bibfnamefont {Kenneth}\ \bibnamefont
  {Rudinger}}, \bibinfo {author} {\bibfnamefont {Kevin}\ \bibnamefont {Young}},
  \bibinfo {author} {\bibfnamefont {Mohan}\ \bibnamefont {Sarovar}}, \ and\
  \bibinfo {author} {\bibfnamefont {Robin}\ \bibnamefont {Blume-Kohout}},\
  }\bibfield  {title} {\enquote {\bibinfo {title} {What randomized benchmarking
  actually measures},}\ }\href {\doibase 10.1103/PhysRevLett.119.130502}
  {\bibfield  {journal} {\bibinfo  {journal} {Phys. Rev. Lett.}\ }\textbf
  {\bibinfo {volume} {119}},\ \bibinfo {pages} {130502} (\bibinfo {year}
  {2017})}\BibitemShut {NoStop}%
\bibitem [{\citenamefont {Di~Matteo}\ \emph {et~al.}(2020)\citenamefont
  {Di~Matteo}, \citenamefont {Gamble}, \citenamefont {Granade}, \citenamefont
  {Rudinger},\ and\ \citenamefont {Wiebe}}]{DiMatteo2020operationalgauge}%
  \BibitemOpen
  \bibfield  {author} {\bibinfo {author} {\bibfnamefont {Olivia}\ \bibnamefont
  {Di~Matteo}}, \bibinfo {author} {\bibfnamefont {John}\ \bibnamefont
  {Gamble}}, \bibinfo {author} {\bibfnamefont {Chris}\ \bibnamefont {Granade}},
  \bibinfo {author} {\bibfnamefont {Kenneth}\ \bibnamefont {Rudinger}}, \ and\
  \bibinfo {author} {\bibfnamefont {Nathan}\ \bibnamefont {Wiebe}},\ }\bibfield
   {title} {\enquote {\bibinfo {title} {Operational, gauge-free quantum
  tomography},}\ }\href {\doibase 10.22331/q-2020-11-17-364} {\bibfield
  {journal} {\bibinfo  {journal} {{Quantum}}\ }\textbf {\bibinfo {volume}
  {4}},\ \bibinfo {pages} {364} (\bibinfo {year} {2020})}\BibitemShut {NoStop}%
\bibitem [{\citenamefont {Chen}\ \emph {et~al.}(2022)\citenamefont {Chen},
  \citenamefont {Ding},\ and\ \citenamefont {Huang}}]{RBBeyond}%
  \BibitemOpen
  \bibfield  {author} {\bibinfo {author} {\bibfnamefont {Jianxin}\ \bibnamefont
  {Chen}}, \bibinfo {author} {\bibfnamefont {Dawei}\ \bibnamefont {Ding}}, \
  and\ \bibinfo {author} {\bibfnamefont {Cupjin}\ \bibnamefont {Huang}},\
  }\bibfield  {title} {\enquote {\bibinfo {title} {Randomized benchmarking
  beyond groups},}\ }\href {\doibase 10.1103/PRXQuantum.3.030320} {\bibfield
  {journal} {\bibinfo  {journal} {PRX Quantum}\ }\textbf {\bibinfo {volume}
  {3}},\ \bibinfo {pages} {030320} (\bibinfo {year} {2022})}\BibitemShut
  {NoStop}%
\bibitem [{\citenamefont {Merkel}\ \emph {et~al.}(2021)\citenamefont {Merkel},
  \citenamefont {Pritchett},\ and\ \citenamefont {Fong}}]{RBConvolution}%
  \BibitemOpen
  \bibfield  {author} {\bibinfo {author} {\bibfnamefont {Seth~T.}\ \bibnamefont
  {Merkel}}, \bibinfo {author} {\bibfnamefont {Emily~J.}\ \bibnamefont
  {Pritchett}}, \ and\ \bibinfo {author} {\bibfnamefont {Bryan~H.}\
  \bibnamefont {Fong}},\ }\bibfield  {title} {\enquote {\bibinfo {title}
  {Randomized {B}enchmarking as {C}onvolution: {F}ourier {A}nalysis of {G}ate
  {D}ependent {E}rrors},}\ }\href {\doibase 10.22331/q-2021-11-16-581}
  {\bibfield  {journal} {\bibinfo  {journal} {{Quantum}}\ }\textbf {\bibinfo
  {volume} {5}},\ \bibinfo {pages} {581} (\bibinfo {year} {2021})}\BibitemShut
  {NoStop}%
\bibitem [{\citenamefont {Helsen}\ \emph {et~al.}(2022)\citenamefont {Helsen},
  \citenamefont {Roth}, \citenamefont {Onorati}, \citenamefont {Werner},\ and\
  \citenamefont {Eisert}}]{RBFramework}%
  \BibitemOpen
  \bibfield  {author} {\bibinfo {author} {\bibfnamefont {J.}~\bibnamefont
  {Helsen}}, \bibinfo {author} {\bibfnamefont {I.}~\bibnamefont {Roth}},
  \bibinfo {author} {\bibfnamefont {E.}~\bibnamefont {Onorati}}, \bibinfo
  {author} {\bibfnamefont {A.H.}\ \bibnamefont {Werner}}, \ and\ \bibinfo
  {author} {\bibfnamefont {J.}~\bibnamefont {Eisert}},\ }\bibfield  {title}
  {\enquote {\bibinfo {title} {General framework for randomized
  benchmarking},}\ }\href {\doibase 10.1103/PRXQuantum.3.020357} {\bibfield
  {journal} {\bibinfo  {journal} {PRX Quantum}\ }\textbf {\bibinfo {volume}
  {3}},\ \bibinfo {pages} {020357} (\bibinfo {year} {2022})}\BibitemShut
  {NoStop}%
\bibitem [{\citenamefont {Chen}\ \emph {et~al.}(2023)\citenamefont {Chen},
  \citenamefont {Liu}, \citenamefont {Otten}, \citenamefont {Seif},
  \citenamefont {Fefferman},\ and\ \citenamefont
  {Jiang}}]{chen_learnability_2023}%
  \BibitemOpen
  \bibfield  {author} {\bibinfo {author} {\bibfnamefont {Senrui}\ \bibnamefont
  {Chen}}, \bibinfo {author} {\bibfnamefont {Yunchao}\ \bibnamefont {Liu}},
  \bibinfo {author} {\bibfnamefont {Matthew}\ \bibnamefont {Otten}}, \bibinfo
  {author} {\bibfnamefont {Alireza}\ \bibnamefont {Seif}}, \bibinfo {author}
  {\bibfnamefont {Bill}\ \bibnamefont {Fefferman}}, \ and\ \bibinfo {author}
  {\bibfnamefont {Liang}\ \bibnamefont {Jiang}},\ }\bibfield  {title} {\enquote
  {\bibinfo {title} {The learnability of {Pauli} noise},}\ }\href {\doibase
  10.1038/s41467-022-35759-4} {\bibfield  {journal} {\bibinfo  {journal}
  {Nature Communications}\ }\textbf {\bibinfo {volume} {14}},\ \bibinfo {pages}
  {52} (\bibinfo {year} {2023})}\BibitemShut {NoStop}%
\bibitem [{\citenamefont {Chen}\ \emph
  {et~al.}(2026{\natexlab{a}})\citenamefont {Chen}, \citenamefont {Zhang},
  \citenamefont {Jiang},\ and\ \citenamefont {Flammia}}]{senrui_efficient}%
  \BibitemOpen
  \bibfield  {author} {\bibinfo {author} {\bibfnamefont {Senrui}\ \bibnamefont
  {Chen}}, \bibinfo {author} {\bibfnamefont {Zhihan}\ \bibnamefont {Zhang}},
  \bibinfo {author} {\bibfnamefont {Liang}\ \bibnamefont {Jiang}}, \ and\
  \bibinfo {author} {\bibfnamefont {Steven~T.}\ \bibnamefont {Flammia}},\
  }\bibfield  {title} {\enquote {\bibinfo {title} {Efficient self-consistent
  learning of gate set pauli noise},}\ }\href {\doibase 10.1103/1pnv-t9px}
  {\bibfield  {journal} {\bibinfo  {journal} {PRX Quantum}\ }\textbf {\bibinfo
  {volume} {7}},\ \bibinfo {pages} {010305} (\bibinfo {year}
  {2026}{\natexlab{a}})}\BibitemShut {NoStop}%
\bibitem [{\citenamefont {Chen}\ \emph
  {et~al.}(2026{\natexlab{b}})\citenamefont {Chen}, \citenamefont {Chen},
  \citenamefont {Fischer}, \citenamefont {Eddins}, \citenamefont {Govia},
  \citenamefont {Mitchell}, \citenamefont {He}, \citenamefont {Kim},
  \citenamefont {Jiang},\ and\ \citenamefont
  {Seif}}]{chen2026disambiguatingpaulinoisequantum}%
  \BibitemOpen
  \bibfield  {author} {\bibinfo {author} {\bibfnamefont {Edward~H.}\
  \bibnamefont {Chen}}, \bibinfo {author} {\bibfnamefont {Senrui}\ \bibnamefont
  {Chen}}, \bibinfo {author} {\bibfnamefont {Laurin~E.}\ \bibnamefont
  {Fischer}}, \bibinfo {author} {\bibfnamefont {Andrew}\ \bibnamefont
  {Eddins}}, \bibinfo {author} {\bibfnamefont {Luke C.~G.}\ \bibnamefont
  {Govia}}, \bibinfo {author} {\bibfnamefont {Brad}\ \bibnamefont {Mitchell}},
  \bibinfo {author} {\bibfnamefont {Andre}\ \bibnamefont {He}}, \bibinfo
  {author} {\bibfnamefont {Youngseok}\ \bibnamefont {Kim}}, \bibinfo {author}
  {\bibfnamefont {Liang}\ \bibnamefont {Jiang}}, \ and\ \bibinfo {author}
  {\bibfnamefont {Alireza}\ \bibnamefont {Seif}},\ }\href
  {https://arxiv.org/abs/2505.22629} {\enquote {\bibinfo {title}
  {Disambiguating pauli noise in quantum computers},}\ } (\bibinfo {year}
  {2026}{\natexlab{b}}),\ \Eprint {http://arxiv.org/abs/2505.22629}
  {arXiv:2505.22629 [quant-ph]} \BibitemShut {NoStop}%
\bibitem [{\citenamefont {McLaren}\ \emph {et~al.}(2025)\citenamefont
  {McLaren}, \citenamefont {Graydon}, \citenamefont {Mahmoud},\ and\
  \citenamefont {Wallman}}]{BenchQI}%
  \BibitemOpen
  \bibfield  {author} {\bibinfo {author} {\bibfnamefont {Darian}\ \bibnamefont
  {McLaren}}, \bibinfo {author} {\bibfnamefont {Matthew~A.}\ \bibnamefont
  {Graydon}}, \bibinfo {author} {\bibfnamefont {Ali~Assem}\ \bibnamefont
  {Mahmoud}}, \ and\ \bibinfo {author} {\bibfnamefont {Joel~J.}\ \bibnamefont
  {Wallman}},\ }\href {https://arxiv.org/abs/2502.00179} {\enquote {\bibinfo
  {title} {Benchmarking quantum instruments},}\ } (\bibinfo {year} {2025}),\
  \Eprint {http://arxiv.org/abs/2502.00179} {arXiv:2502.00179 [quant-ph]}
  \BibitemShut {NoStop}%
\bibitem [{\citenamefont {Zhang}\ \emph {et~al.}(2025)\citenamefont {Zhang},
  \citenamefont {Chen}, \citenamefont {Liu},\ and\ \citenamefont
  {Jiang}}]{learnable_MCM}%
  \BibitemOpen
  \bibfield  {author} {\bibinfo {author} {\bibfnamefont {Zhihan}\ \bibnamefont
  {Zhang}}, \bibinfo {author} {\bibfnamefont {Senrui}\ \bibnamefont {Chen}},
  \bibinfo {author} {\bibfnamefont {Yunchao}\ \bibnamefont {Liu}}, \ and\
  \bibinfo {author} {\bibfnamefont {Liang}\ \bibnamefont {Jiang}},\ }\bibfield
  {title} {\enquote {\bibinfo {title} {Generalized cycle benchmarking algorithm
  for characterizing midcircuit measurements},}\ }\href {\doibase
  10.1103/PRXQuantum.6.010310} {\bibfield  {journal} {\bibinfo  {journal} {PRX
  Quantum}\ }\textbf {\bibinfo {volume} {6}},\ \bibinfo {pages} {010310}
  (\bibinfo {year} {2025})}\BibitemShut {NoStop}%
\bibitem [{\citenamefont {Hines}\ and\ \citenamefont
  {Proctor}(2025)}]{jordanmcmcycle}%
  \BibitemOpen
  \bibfield  {author} {\bibinfo {author} {\bibfnamefont {Jordan}\ \bibnamefont
  {Hines}}\ and\ \bibinfo {author} {\bibfnamefont {Timothy}\ \bibnamefont
  {Proctor}},\ }\bibfield  {title} {\enquote {\bibinfo {title} {Pauli noise
  learning for mid-circuit measurements},}\ }\href {\doibase
  10.1103/PhysRevLett.134.020602} {\bibfield  {journal} {\bibinfo  {journal}
  {Phys. Rev. Lett.}\ }\textbf {\bibinfo {volume} {134}},\ \bibinfo {pages}
  {020602} (\bibinfo {year} {2025})}\BibitemShut {NoStop}%
\bibitem [{\citenamefont {Carignan-Dugas}\ \emph {et~al.}(2018)\citenamefont
  {Carignan-Dugas}, \citenamefont {Boone}, \citenamefont {Wallman},\ and\
  \citenamefont {Emerson}}]{Carignan-Dugas2018-np}%
  \BibitemOpen
  \bibfield  {author} {\bibinfo {author} {\bibfnamefont {Arnaud}\ \bibnamefont
  {Carignan-Dugas}}, \bibinfo {author} {\bibfnamefont {Kristine}\ \bibnamefont
  {Boone}}, \bibinfo {author} {\bibfnamefont {Joel~J}\ \bibnamefont {Wallman}},
  \ and\ \bibinfo {author} {\bibfnamefont {Joseph}\ \bibnamefont {Emerson}},\
  }\bibfield  {title} {\enquote {\bibinfo {title} {From randomized benchmarking
  experiments to gate-set circuit fidelity: how to interpret randomized
  benchmarking decay parameters},}\ }\href {\doibase 10.1088/1367-2630/aadcc7}
  {\bibfield  {journal} {\bibinfo  {journal} {New J. Phys.}\ }\textbf {\bibinfo
  {volume} {20}},\ \bibinfo {pages} {092001} (\bibinfo {year}
  {2018})}\BibitemShut {NoStop}%
\bibitem [{\citenamefont {Lin}\ \emph {et~al.}(2019)\citenamefont {Lin},
  \citenamefont {Buonacorsi}, \citenamefont {Laflamme},\ and\ \citenamefont
  {Wallman}}]{Lin2019-qx}%
  \BibitemOpen
  \bibfield  {author} {\bibinfo {author} {\bibfnamefont {Junan}\ \bibnamefont
  {Lin}}, \bibinfo {author} {\bibfnamefont {Brandon}\ \bibnamefont
  {Buonacorsi}}, \bibinfo {author} {\bibfnamefont {Raymond}\ \bibnamefont
  {Laflamme}}, \ and\ \bibinfo {author} {\bibfnamefont {Joel~J}\ \bibnamefont
  {Wallman}},\ }\bibfield  {title} {\enquote {\bibinfo {title} {On the freedom
  in representing quantum operations},}\ }\href {\doibase
  10.1088/1367-2630/ab075a} {\bibfield  {journal} {\bibinfo  {journal} {New J.
  Phys.}\ }\textbf {\bibinfo {volume} {21}},\ \bibinfo {pages} {023006}
  (\bibinfo {year} {2019})}\BibitemShut {NoStop}%
\bibitem [{\citenamefont {Flammia}(2022)}]{ACES}%
  \BibitemOpen
  \bibfield  {author} {\bibinfo {author} {\bibfnamefont {Steven~T.}\
  \bibnamefont {Flammia}},\ }\bibfield  {title} {\enquote {\bibinfo {title}
  {Averaged circuit eigenvalue sampling},}\ \ }(\bibinfo  {publisher} {Schloss
  Dagstuhl – Leibniz-Zentrum für Informatik},\ \bibinfo {year} {2022})\ pp.\
  \bibinfo {pages} {4:1--4:10}\BibitemShut {NoStop}%
\bibitem [{\citenamefont {Marceaux}\ and\ \citenamefont {Young}(2023)}]{JP}%
  \BibitemOpen
  \bibfield  {author} {\bibinfo {author} {\bibfnamefont {J.~P.}\ \bibnamefont
  {Marceaux}}\ and\ \bibinfo {author} {\bibfnamefont {Kevin}\ \bibnamefont
  {Young}},\ }\bibfield  {title} {\enquote {\bibinfo {title} {Streaming quantum
  gate set tomography using the extended kalman filter},}\ }\bibfield
  {booktitle} {\emph {\bibinfo {booktitle} {2023 IEEE International Conference
  on Quantum Computing and Engineering (QCE)}},\ }\href {\doibase
  10.1109/QCE57702.2023.00159} {\ \textbf {\bibinfo {volume} {01}},\ \bibinfo
  {pages} {1401--1411} (\bibinfo {year} {2023})}\BibitemShut {NoStop}%
\bibitem [{\citenamefont {Madzik}\ \emph {et~al.}(2022)\citenamefont {Madzik},
  \citenamefont {Asaad}, \citenamefont {Youssry}, \citenamefont {Joecker},
  \citenamefont {Rudinger}, \citenamefont {Nielsen}, \citenamefont {Young},
  \citenamefont {Proctor}, \citenamefont {Baczewski}, \citenamefont {Laucht},
  \citenamefont {Schmitt}, \citenamefont {Hudson}, \citenamefont {Itoh},
  \citenamefont {Jakob}, \citenamefont {Johnson}, \citenamefont {Jamieson},
  \citenamefont {Dzurak}, \citenamefont {Ferrie}, \citenamefont
  {Blume-Kohout},\ and\ \citenamefont {Morello}}]{UNSWreference}%
  \BibitemOpen
  \bibfield  {author} {\bibinfo {author} {\bibfnamefont {Mateusz~T.}\
  \bibnamefont {Madzik}}, \bibinfo {author} {\bibfnamefont {Serwan}\
  \bibnamefont {Asaad}}, \bibinfo {author} {\bibfnamefont {Akram}\ \bibnamefont
  {Youssry}}, \bibinfo {author} {\bibfnamefont {Benjamin}\ \bibnamefont
  {Joecker}}, \bibinfo {author} {\bibfnamefont {Kenneth~M.}\ \bibnamefont
  {Rudinger}}, \bibinfo {author} {\bibfnamefont {Erik}\ \bibnamefont
  {Nielsen}}, \bibinfo {author} {\bibfnamefont {Kevin~C.}\ \bibnamefont
  {Young}}, \bibinfo {author} {\bibfnamefont {Timothy~J.}\ \bibnamefont
  {Proctor}}, \bibinfo {author} {\bibfnamefont {Andrew~D.}\ \bibnamefont
  {Baczewski}}, \bibinfo {author} {\bibfnamefont {Arne}\ \bibnamefont
  {Laucht}}, \bibinfo {author} {\bibfnamefont {Vivien}\ \bibnamefont
  {Schmitt}}, \bibinfo {author} {\bibfnamefont {Fay~E.}\ \bibnamefont
  {Hudson}}, \bibinfo {author} {\bibfnamefont {Kohei~M.}\ \bibnamefont {Itoh}},
  \bibinfo {author} {\bibfnamefont {Alexander~M.}\ \bibnamefont {Jakob}},
  \bibinfo {author} {\bibfnamefont {Brett~C.}\ \bibnamefont {Johnson}},
  \bibinfo {author} {\bibfnamefont {David~N.}\ \bibnamefont {Jamieson}},
  \bibinfo {author} {\bibfnamefont {Andrew~S.}\ \bibnamefont {Dzurak}},
  \bibinfo {author} {\bibfnamefont {Christopher}\ \bibnamefont {Ferrie}},
  \bibinfo {author} {\bibfnamefont {Robin}\ \bibnamefont {Blume-Kohout}}, \
  and\ \bibinfo {author} {\bibfnamefont {Andrea}\ \bibnamefont {Morello}},\
  }\bibfield  {title} {\enquote {\bibinfo {title} {Precision tomography of a
  three-qubit donor quantum processor in silicon},}\ }\href {\doibase
  10.1038/s41586-021-04292-7} {\bibfield  {journal} {\bibinfo  {journal}
  {Nature}\ }\textbf {\bibinfo {volume} {601}},\ \bibinfo {pages} {348–353}
  (\bibinfo {year} {2022})}\BibitemShut {NoStop}%
\bibitem [{\citenamefont {Stemp}\ \emph {et~al.}(2024)\citenamefont {Stemp},
  \citenamefont {Asaad}, \citenamefont {Blankenstein}, \citenamefont
  {Vaartjes}, \citenamefont {Johnson}, \citenamefont {Madzik}, \citenamefont
  {Heskes}, \citenamefont {Firgau}, \citenamefont {Su}, \citenamefont {Yang},
  \citenamefont {Laucht}, \citenamefont {Ostrove}, \citenamefont {Rudinger},
  \citenamefont {Young}, \citenamefont {Blume-Kohout}, \citenamefont {Hudson},
  \citenamefont {Dzurak}, \citenamefont {Itoh}, \citenamefont {Jakob},
  \citenamefont {Johnson}, \citenamefont {Jamieson},\ and\ \citenamefont
  {Morello}}]{Stemp_2024}%
  \BibitemOpen
  \bibfield  {author} {\bibinfo {author} {\bibfnamefont {Holly~G.}\
  \bibnamefont {Stemp}}, \bibinfo {author} {\bibfnamefont {Serwan}\
  \bibnamefont {Asaad}}, \bibinfo {author} {\bibfnamefont {Mark R.~van}\
  \bibnamefont {Blankenstein}}, \bibinfo {author} {\bibfnamefont {Arjen}\
  \bibnamefont {Vaartjes}}, \bibinfo {author} {\bibfnamefont {Mark A.~I.}\
  \bibnamefont {Johnson}}, \bibinfo {author} {\bibfnamefont {Mateusz~T.}\
  \bibnamefont {Madzik}}, \bibinfo {author} {\bibfnamefont {Amber J.~A.}\
  \bibnamefont {Heskes}}, \bibinfo {author} {\bibfnamefont {Hannes~R.}\
  \bibnamefont {Firgau}}, \bibinfo {author} {\bibfnamefont {Rocky~Y.}\
  \bibnamefont {Su}}, \bibinfo {author} {\bibfnamefont {Chih~Hwan}\
  \bibnamefont {Yang}}, \bibinfo {author} {\bibfnamefont {Arne}\ \bibnamefont
  {Laucht}}, \bibinfo {author} {\bibfnamefont {Corey~I.}\ \bibnamefont
  {Ostrove}}, \bibinfo {author} {\bibfnamefont {Kenneth~M.}\ \bibnamefont
  {Rudinger}}, \bibinfo {author} {\bibfnamefont {Kevin}\ \bibnamefont {Young}},
  \bibinfo {author} {\bibfnamefont {Robin}\ \bibnamefont {Blume-Kohout}},
  \bibinfo {author} {\bibfnamefont {Fay~E.}\ \bibnamefont {Hudson}}, \bibinfo
  {author} {\bibfnamefont {Andrew~S.}\ \bibnamefont {Dzurak}}, \bibinfo
  {author} {\bibfnamefont {Kohei~M.}\ \bibnamefont {Itoh}}, \bibinfo {author}
  {\bibfnamefont {Alexander~M.}\ \bibnamefont {Jakob}}, \bibinfo {author}
  {\bibfnamefont {Brett~C.}\ \bibnamefont {Johnson}}, \bibinfo {author}
  {\bibfnamefont {David~N.}\ \bibnamefont {Jamieson}}, \ and\ \bibinfo {author}
  {\bibfnamefont {Andrea}\ \bibnamefont {Morello}},\ }\bibfield  {title}
  {\enquote {\bibinfo {title} {Tomography of entangling two-qubit logic
  operations in exchange-coupled donor electron spin qubits},}\ }\href
  {\doibase 10.1038/s41467-024-52795-4} {\bibfield  {journal} {\bibinfo
  {journal} {Nature Communications}\ }\textbf {\bibinfo {volume} {15}}
  (\bibinfo {year} {2024}),\ 10.1038/s41467-024-52795-4}\BibitemShut {NoStop}%
\bibitem [{\citenamefont {Blume-Kohout}\ \emph {et~al.}(2022)\citenamefont
  {Blume-Kohout}, \citenamefont {da~Silva}, \citenamefont {Nielsen},
  \citenamefont {Proctor}, \citenamefont {Rudinger}, \citenamefont {Sarovar},\
  and\ \citenamefont {Young}}]{Blume-Kohout2022-ln}%
  \BibitemOpen
  \bibfield  {author} {\bibinfo {author} {\bibfnamefont {Robin}\ \bibnamefont
  {Blume-Kohout}}, \bibinfo {author} {\bibfnamefont {Marcus~P}\ \bibnamefont
  {da~Silva}}, \bibinfo {author} {\bibfnamefont {Erik}\ \bibnamefont
  {Nielsen}}, \bibinfo {author} {\bibfnamefont {Timothy}\ \bibnamefont
  {Proctor}}, \bibinfo {author} {\bibfnamefont {Kenneth}\ \bibnamefont
  {Rudinger}}, \bibinfo {author} {\bibfnamefont {Mohan}\ \bibnamefont
  {Sarovar}}, \ and\ \bibinfo {author} {\bibfnamefont {Kevin}\ \bibnamefont
  {Young}},\ }\bibfield  {title} {\enquote {\bibinfo {title} {A taxonomy of
  small markovian errors},}\ }\href {\doibase 10.1103/PRXQuantum.3.020335}
  {\bibfield  {journal} {\bibinfo  {journal} {PRX Quantum}\ }\textbf {\bibinfo
  {volume} {3}},\ \bibinfo {pages} {020335} (\bibinfo {year}
  {2022})}\BibitemShut {NoStop}%
\bibitem [{\citenamefont {Bartlett}\ \emph {et~al.}(2007)\citenamefont
  {Bartlett}, \citenamefont {Rudolph},\ and\ \citenamefont
  {Spekkens}}]{bartlett2007reference}%
  \BibitemOpen
  \bibfield  {author} {\bibinfo {author} {\bibfnamefont {Stephen~D}\
  \bibnamefont {Bartlett}}, \bibinfo {author} {\bibfnamefont {Terry}\
  \bibnamefont {Rudolph}}, \ and\ \bibinfo {author} {\bibfnamefont {Robert~W}\
  \bibnamefont {Spekkens}},\ }\bibfield  {title} {\enquote {\bibinfo {title}
  {Reference frames, superselection rules, and quantum information},}\
  }\href@noop {} {\bibfield  {journal} {\bibinfo  {journal} {Reviews of Modern
  Physics}\ }\textbf {\bibinfo {volume} {79}},\ \bibinfo {pages} {555--609}
  (\bibinfo {year} {2007})}\BibitemShut {NoStop}%
\end{thebibliography}
\end{document}